Roadmap

# Roadmap on quantum nanotechnologies

Arne Laucht[1,*], Frank Hohls[2], Niels Ubbelohde[2], M Fernando Gonzalez-Zalba[3,27], David J Reilly[4,5], Søren Stobbe[6], Tim Schröder[7], Pasquale Scarlino[8], Jonne V Koski[8], Andrew Dzurak[1], Chih-Hwan Yang[1], Jun Yoneda[1], Ferdinand Kuemmeth[9], Hendrik Bluhm[10], Jarryd Pla[1], Charles Hill[11], Joe Salfi[12], Akira Oiwa[13,14,28], Juha T Muhonen[29], Ewold Verhagen[15], M D LaHaye[30,31], Hyun Ho Kim[25,26], Adam W Tsen[25], Dimitrie Culcer[16,17], Attila Geresdi[18], Jan A Mol[19], Varun Mohan[20], Prashant K Jain[21,22,23,24] and Jonathan Baugh[25]

[1] Centre for Quantum Computation and Communication Technology, School of Electrical Engineering and Telecommunications, UNSW Sydney, New South Wales 2052, Australia

[2] Physikalisch-Technische Bundesanstalt, 38116, Braunschweig, Germany

[3] Quantum Motion Technologies, Nexus, Discovery Way, Leeds, LS2 3AA, United Kingdom

[4] School of Physics, University of Sydney, Sydney, NSW 2006, Australia

[5] Microsoft Corporation, Station Q Sydney, University of Sydney, Sydney, NSW 2006, Australia

[6] Department of Photonics Engineering, DTU Fotonik, Technical University of Denmark, Building 343, DK-2800 Kgs. Lyngby, Denmark

[7] Department of Physics, Humboldt-Universität zu Berlin, 12489, Berlin, Germany

[8] Department of Physics, ETH Zürich, CH-8093, Zürich, Switzerland

[9] Niels Bohr Institute, University of Copenhagen, 2100, Copenhagen, Denmark

[10] JARA-FIT Institute for Quantum Information, RWTH Aachen University and Forschungszentrum Jülich, 52074, Aachen, Germany

[11] School of Physics, University of Melbourne, Melbourne, Australia

[12] Department of Electrical and Computer Engineering, The University of British Columbia, Vancouver BC V6T 1Z4, Canada

[13] The Institute of Scientific and Industrial Research, Osaka University, Ibaraki, Osaka 567-0047, Japan

[14] Center for Quantum Information and Quantum Biology, Institute for open and Transdisciplinary Research Initiative, Osaka University, 560-8531, Osaka, Japan

[15] Center for Nanophotonics, AMOLF, 1098 XG, Amsterdam, The Netherlands

[16] School of Physics, The University of New South Wales, Sydney 2052, Australia

[17] Australian Research Council Centre of Excellence in Future Low-Energy Electronics Technologies, UNSW Node, The University of New South Wales, Sydney 2052, Australia

[18] QuTech and Kavli Institute of Nanoscience, Delft University of Technology, 2600 GA Delft, The Netherlands

[19] School of Physics and Astronomy, Queen Mary University of London, E1 4NS, United Kingdom

[20] Department of Materials Science and Engineering, University of Illinois at Urbana-Champaign, Urbana, IL 61801, United States of America

[21] Department of Chemistry, University of Illinois at Urbana-Champaign, Urbana, IL 61801, United States of America

[22] Materials Research Laboratory, University of Illinois at Urbana-Champaign, Urbana, IL 61801, United States of America

[23] Department of Physics, University of Illinois at Urbana-Champaign, Urbana, IL 61801, United States of America

[24] Beckman Institute for Advanced Science and Technology, University of Illinois at Urbana-Champaign, Urbana, IL 61801, United States of America

[25] Institute for Quantum Computing, University of Waterloo, Waterloo, Ontario N2L 3G1, Canada

[26] School of Materials Science and Engineering & Department of Energy Engineering Convergence, Kumoh National Institute of Technology, Gumi 39177, Korea

[27] Present address: Quantum Motion Technologies, Windsor House, Cornwall Road, Harrogate, HG1 2PW, United Kingdom

[28] Center for Spintronics Research Network (CSRN), Graduate School of Engineering Science, Osaka University, Osaka 560-8531, Japan

[29] Department of Physics and Nanoscience Center, University of Jyväskylä, FI-40014 University of Jyväskylä, Finland

[30] Department of Physics, Syracuse University, Syracuse, NY 13244-1130, United States of America

[31] Present Address: United States Air Force Research Laboratory, Rome, NY 13441, United States of America

[*] Author to whom any correspondence should be addressed.

E-mail: a.laucht@unsw.edu.au

**Abstract**

Quantum phenomena are typically observable at length and time scales smaller than those of our everyday experience, often involving individual particles or excitations. The past few decades have seen a revolution in the ability to structure matter at the nanoscale, and experiments at the single particle level have become commonplace. This has opened wide new avenues for exploring and harnessing quantum mechanical effects in condensed matter. These quantum phenomena, in turn, have the potential to revolutionize the way we communicate, compute and probe the nanoscale world. Here, we review developments in key areas of quantum research in light of the nanotechnologies that enable them, with a view to what the future holds. Materials and devices with nanoscale features are used for quantum metrology and sensing, as building blocks for quantum computing, and as sources and detectors for quantum communication. They enable explorations of quantum behaviour and unconventional states in nano- and opto-mechanical systems, low-dimensional systems, molecular devices, nano-plasmonics, quantum electrodynamics, scanning tunnelling microscopy, and more. This rapidly expanding intersection of nanotechnology and quantum science/technology is mutually beneficial to both fields, laying claim to some of the most exciting scientific leaps of the last decade, with more on the horizon.

(Some figures may appear in colour only in the online journal)

## Contents

## Introduction

The year 2019 marks the 60th anniversary of Richard Feynman's seminal lecture 'There's Plenty of Room at the Bottom: An Invitation to Enter a New Field of Physics' at the California Institute of Technology on 29 December 1959. In his lecture, Feynman considered the possibilities of directly manipulating individual atoms, designing microscopes with atomistic resolution, and building nanoscale machines. His talk was very visionary and sparked many ideas that can be classified as nanotechnology, although the actual term was not coined until some 20 years later and the field of nanotechnology did not emerge as a research direction until the 1980s. Around this time, in 1989, it was also when the IOP journal Nanotechnology was founded, and thus the year 2019 coincidentally also marks the journal's 30th anniversary. All of this is more than enough reason to reflect on some of the contributions that nanotechnology has made to science in recent years. In this Roadmap article, we would like to take a closer look at the importance of nanotechnology to shape the field of quantum systems by reviewing the state-of-the-art in a number of different subfields from metrology, quantum communication, quantum computation, to low-dimensional systems.

This Roadmap is structured similarly to other Roadmaps that have been published on e.g. magnetism [1] and plasma physics [2]. Each section is written by experts in their fields and tries to capture the state-of-the-art as well as outline some of the future challenges and research directions.

In section 1 on Metrology and sensing we have two contributions. The first one by Frank Hohls and Niels Ubbelohde details how semiconductor nanodevices can be used to deterministically shuttle single electrons with high frequency to redefine the ampere in units of time and the elementary charge. In some sense this is a perfect example of a nanomachine as Feynman might have envisioned it—it is an apparatus that can move individual electrons. The second contribution by M Fernando Gonzalez-Zalba and David J Reilly then shows how proper use of classical microwave engineering allows the detection of a single charge movement in a nanodevice, which in combination with Pauli spin blockade even allows the detection of a single spin state.

Section 2 on Quantum light sources, cavities and detectors takes a closer look at the interaction of quantum systems with photons. Søren Stobbe and Tim Schröder review quantum light sources, their properties, and their ability to emit various quantum states of light. Quantum light sources play an important role for quantum communication and complex photonic quantum systems. Akira Oiwa then extends the discussion to long-distance quantum networks that rely on quantum repeaters based on coherent quantum interfaces between static qubits (solid-state spins) and flying qubits (photons). Long-distance entanglement is required for distributed quantum computing as well as a secure quantum internet. The third contribution by Pasquale Scarlino and Jonne V Koski discusses the coupling of charges and spins to photons in superconducting microwave resonators, for the purpose of spin readout, remote spin–spin coupling, and semiconductor–superconductor hybrid quantum devices.

In section 3 we are discussing quantum computing with spins in the solid state. Electronic spins have shown to be highly-coherent quantum systems that promise to be scalable to a large number of qubits in very little space when integrated in a semiconductor platform. The four contributions in this section cover different implementations. The first two contributions make use of quantum dots (QDs) in the electrostatically-defined potential of nanostructures in gallium-arsenide (GaAs)-based material systems by Ferdinand Kuemmeth and Hendrik Bluhm, and in silicon-based material systems by Andrew Dzurak, Chih-Hwan Yang and Jun Yoneda. The other two contributions look at more naturally-confined spin systems, namely donor spins in silicon by Jarryd J Pla and Charles Hill, and acceptor spins in silicon by Joe Salfi.

Section 4 covers the topic of Nano- and opto-mechanics. Here, Juha Muhonen and Ewold Verhagen discuss the coupling of mechanical resonators to electromagnetic fields, and illuminate how mechanical resonators can be used as a transducer to convert quantum signals from one electromagnetic mode to another, as e.g. from microwave frequencies to optical frequencies. Matthew D LaHaye then writes about quantum nanomechanical systems that allow fundamental explorations of motion and quantum thermodynamics with applications in quantum computation, communication, sensing, and hybrid quantum platforms.

Section 5 is dedicated to low-dimensional systems. In their contribution Hyun Ho Kim and Adam W Tsen discuss CrI3 as an example of a 2D semiconductor with magnetic properties that exhibits strong tunnel magnetoresistance (TMR) when used in a quantum tunneling device, with applications in spin filters and for magnetic memories. Dimitrie Culcer and Attila Geresdi then write about topological states and how they can be exploited for dissipationless transport at low temperatures, spin–orbit torque devices at room temperature and topologically protected quantum electronics in general.

In section 6, Jan Mol reports on the progress in molecular devices, where the unique properties of individual molecules are exploited to build functional electronic devices. In particular, molecular designs allow for the engineering of optical, magnetic and quantum effects that are not readily achievable in lithographically defined nanostructures. And finally, in section 7 on nanoplasmonics, Varun Mohan and Prashant K Jain discuss the high-localization of electromagnetic fields using nanoplasmonic structures that allow the spatiotemporal concentration of optical energy far below the diffraction limit of light. These effects can be exploited for highly-efficient single photon sources, enhanced photocatalytic conversion, and all-optical nanoplasmonic circuits for computation.

*Jonathan Baugh* and *Arne Laucht*
Editors of the Roadmap on Quantum Nanotechnology

# 1. Metrology and sensing

## 1.1. Electrical quantum metrology with single electrons

*Frank Hohls and Niels Ubbelohde*

Physikalisch-Technische Bundesanstalt, Braunschweig, Germany

*1.1.1. Status.* The generation of a quantized electric current $I = ef$ by single-electron (SE) control with $e$ the elementary charge and $f$ the repetition frequency was suggested soon after the first demonstrations of SE devices [3, 4]. Initially, in series connected metallic SE transistors (SETs) were pursued, cumulating in a SE current source built of six SETs with error probability $1.5 \times 10^{-8}$, albeit at a current level of only 0.81 pA, limited by the tunnelling barriers with fixed transparencies [3]. SE pumps (SEPs) with higher current levels can be realized using semiconductor QDs, where the SE tunnelling rates can be varied [5]. These tunable-barrier SEPs are also less complex to operate, requiring only two control gates and a single time-dependent drive signal, and operate up to several GHz driving frequency [6]. The research on SEPs also pushed the development of improved current measurement capabilities [5, 7]. This allowed in recent years to confirm the quantization accuracy for several GaAs and silicon based SEPs operated at $f \geqslant 0.5$ GHz at sub-ppm relative uncertainty [7, 8] (example in figure 1).

Since 20 May 2019 the International System of Units (SI) is fully defined by a set of fundamental constants with fixed values [9], among them the elementary charge $e = 1.602176634 \times 10^{-19}$ As. In this new SI SEPs are now the shortest path to a representation of the unit Ampere, using only $e$ and a frequency $f$ derived from the hyperfine splitting of Caesium. The alternative path combines two quantum effects, the quantum Hall effect (QHE) and the Josephson effect, which both incorporate an additional fundamental constant, the Planck constant $h$. The realization of a suitable SE based primary current standard would impact both the metrological practice and the fundamentals of metrology: Firstly, it would allow to improve the measurement accuracy for small currents, relevant e.g. in semiconductor technology and environmental sensing. Secondly, a comparison between the quantized currents generated along the two mentioned paths ('Quantum Metrology Triangle' [3, 4], figure 2(b)) with sufficiently increased accuracy would test the fundament of electrical quantum metrology.

*1.1.2. Current and future challenges.* The most important task in the development of a primary quantum standard based on SEPs is to establish the universality of the current to frequency relation, which relies on the manipulation of tunnelling rates over many orders of magnitude. The robustness of the operating principle in the presence of disorder and potential fluctuators and a fundamental understanding of the dynamics of electronic transport under high frequency excitation, including the role of electron spin and magnetic field, are both experimentally and theoretically very challenging questions [4, 5].

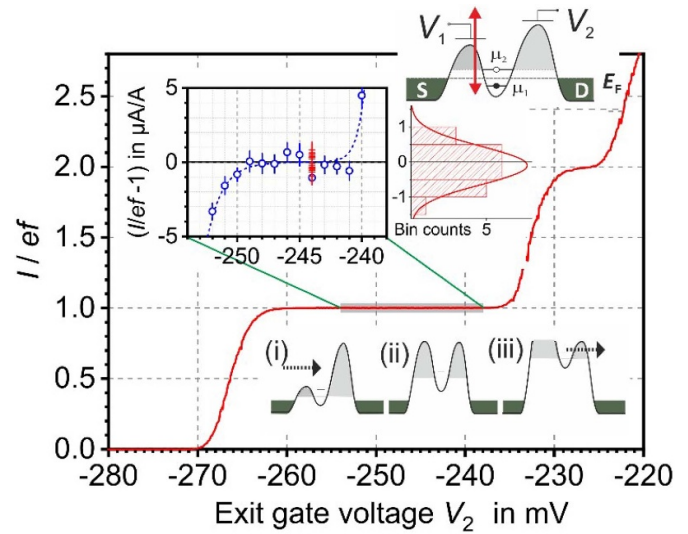

**Figure 1.** Example of quantized single-electron current generated by a quantum dot SEP as function of control voltage ($f = 600$ MHz, $B = 9.2$ T, $ef = 96.130600$). Schematic upper right: QD with gate controlled tunnelling barriers. Lower schematic: pumping cycles: (i) electron loading, (ii) isolation and (iii) ejection to drain. Inset plot left: high accuracy measurement of current deviation from $ef$; right: histogram of 1 h measurement points shown in red. Average $-0.1$ ppm agrees with $ef$ within uncertainty $1.6 \times 10^{-7}$. Graph adopted from [8].

The short time scales of charge transfer at frequencies in the regime of $\sim 1$ GHz makes it difficult to directly resolve the success rate of isolating and subsequently transferring single electrons in QDs and the rarity of errors in this residually still stochastic process necessitates new concepts and improvements to the sensitivity of charge and current detection.

A further increase in the current level towards the nanoampere regime is required for metrological applications at accuracies better than $10^{-7}$ and poses another challenge, which can be addressed by an increase in excitation frequency or device parallelization. However, all realizations of SEPs, that have so far shown good (sub-ppm) accuracy, were prone to strong degradation when operated beyond 1 GHz [7]. Similarly, the reproducibility of tunnel coupled dynamic QDs as needed for parallelization is not yet understood.

The above challenges have to be solved to provide the single electron path for the test of the fundaments of electrical quantum metrology at the desired uncertainty level of $10^{-8}$. In addition, this test requires also large improvements for a current comparison based on the second leg, where the QHE and the Josephson effect are used to generate resp. measure a quantized current of only $\sim 1$ nA or less. This requires at least one order of magnitude improvement compared to the best present techniques.

*1.1.3. Advances in science and technology to meet challenges.* To demonstrate the universality of SEPs advances will necessitate the validation of current quantization at accuracies better than $10^{-8}$. Comparative measurement of multiple SEPs could be realized by either a null measurement of the differential current or an *in-situ* validation detecting the charge trapped on an island between two current sources (figure 2(a)).

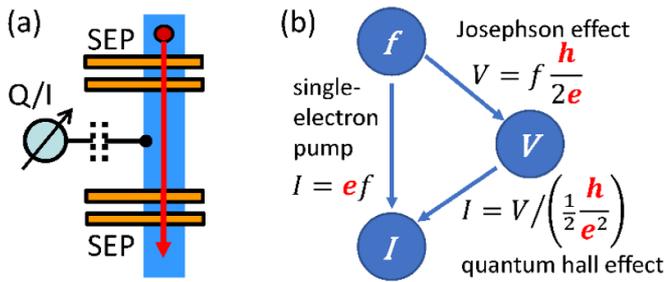

**Figure 2.** (a) Schematic for *in-situ* validation of SEP universality. Two SEPs are connected in series and operated at the same frequency. Any difference in the generated SEP currents can be measured either by a charge Q built up on an isolated node in between the SEPs or by an error current $I_{err}$ through a lead tapping the node. (b) The two possible primary realizations of a current standard. The left SEP path allows direct transfer from a frequency (derived from a primary frequency standard) to a current. The right path uses Ohms law with a quantized voltage generated by the Josephson effect and a quantized resistance by the quantum Hall effect. A comparison of the currents generated along the two paths realizes the 'Quantum Metrology Triangle'.

Towards this goal multiple technological capabilities have to be achieved: to acquire a statistical basis large enough to verify rare quantization errors would require in the case of charge detection a detector bandwidth in excess of 100 kHz with correspondingly low identification errors or in the case of direct current verification noise levels in the very low fA/rtHz range and a current level in the nanoampere range with considerable demands to the overall stability of the experiment. Similar improvements are needed for the transfer accuracy from QHE and Josephson effect to a current of $\sim 1$ nA, where for this path to a primary current source the low level of current is demanding.

Device technology has to be developed to increase SEP accuracy by maximizing charging energy and sharp transients in the time-dependent tunnelling rates. While GaAs is presently the most reliable technology basis for high accuracy SEPs, silicon based SEPs, especially by utilizing strong SE confinement in trap states [6], have shown the potential for higher frequency. However, a large increase of the presently much too low yield in the fabrication of silicon SEPs is needed. The desired combination of large-bandwidth charge detectors with SEPs into integrated single electron circuits featuring *in-situ* detection sets additional demands on device stability and reproducibility.

Parallelization is very likely to rely on the ability to individually address the QD devices forming the parallel network. Complex connection circuitry and the availability of a high count of individual dc control voltages is therefore required but might benefit by developments towards scalable quantum bit circuits based on semiconductor QDs.

These advances in technology are also necessary for the test of the fundaments of electrical quantum metrology by comparing the SE current to the combination of QHE and Josephson effect at $10^{-8}$ accuracy (figure 2(b)).

*1.1.4. Concluding remarks.* In the recent years large progress has been achieved towards a SE based primary current standard. However, impact and application of SEPs in metrology has been hampered by the missing validation of universality and robustness of the current to frequency relation. Additionally increasing the current will broaden the application range and widen the impact on practical metrology. Finally, adding the connection to QHE and Josephson effect would allow to test and strengthen the fundament of electrical quantum metrology.

**Acknowledgments**

We acknowledge funding under EMPIR project 'SEQUOIA' 17FUN04, co-funded by the EU's Horizon 2020 programme and the EMPIR Participating States.

## 1.2. Fast dispersive readout for solid-state qubits

M Fernando Gonzalez-Zalba[1] and David J Reilly[2]

[1] Hitachi Cambridge Laboratory, United Kingdom
[2] Institute for Microsoft Corporation and University of Sydney

### 1.2.1. Status.

A qubit specific measurement readout protocol is an essential ingredient for all quantum computing technologies. The minimum time required to perform a measurement ($t_{\min}$) is an important characteristic of the method as for high-fidelity qubit detection, readout has to be faster than the relaxation time of the system ($T_1$). Moreover, to implement fast feedback in error correction protocols, the readout must be faster than the intrinsic decoherence time ($T_2$). Another important aspect is that the time needed to determine the qubit state is bounded by quantum mechanics and hence it is always longer or at best equal to half the dephasing time induced by the back-action of the detector ($t_\varphi$).

For superconducting charge qubits, semiconductor-based qubits and Majorana zero modes, one can use charge sensors such as the quantum point contact or the single-electron transistor (SET)—with charge sensitivities of a fraction of an electron charge—to detect the charge, spin or parity state of these qubits which can be achieved either via direct charge readout or via spin- or parity-to-charge conversion, respectively. However, the direct current (DC) versions of these sensors have an upper bandwidth limit of a few tens of kHz. High-frequency techniques have been developed to overcome these limitations. By embedding the sensor in a LC tank circuit, single electron resolution with a bandwidth in excess of 100 MHz has been reached. However, the radiofrequency (RF) SET, the most sensitive of all charge sensors, does not reach the quantum limit for detection due to the induced measurement back-action caused by the randomness of the charge tunnelling processes. The roadmap for the radio frequency RF-SET is well known and hence will not be subject to further discussion [10].

More recently, research has been shifting towards dispersive readout methods in which the qubit to be sensed is coupled non-resonantly to a high-frequency electrical resonator [11, 12]. In this paradigm, the state-dependent reactance of the qubit manifests in a difference in the reflection or transmission coefficient of the resonator. Dispersive readout approaches the quantum limit, yields high-fidelity sub-microsecond measurement times, and does not require additional sensing elements, simplifying the overall qubit architecture. The method is extensively used to read superconducting qubits in a single-shot manner, and more recently to detect the spin parity of singlet-triplet qubits in silicon with up to 98% fidelity in 6 $\mu$s [13]. Finally, there are proposals to extend this methodology to enable parity detection of Majorana bound states by parity-dependent hybridization to a quantum dot (QD) [14].

Given the different relaxation times, the technical requirements to achieve readout fidelities above error-correction thresholds vary across platforms. Although our discussion will be of general applicability, when specific, we tailor our roadmap for silicon spin qubits implemented in double QDs (DQDs) that offer some of the longest coherence times of all solid-state device platforms while being manufacturable at scale using very large-scale integration processes.

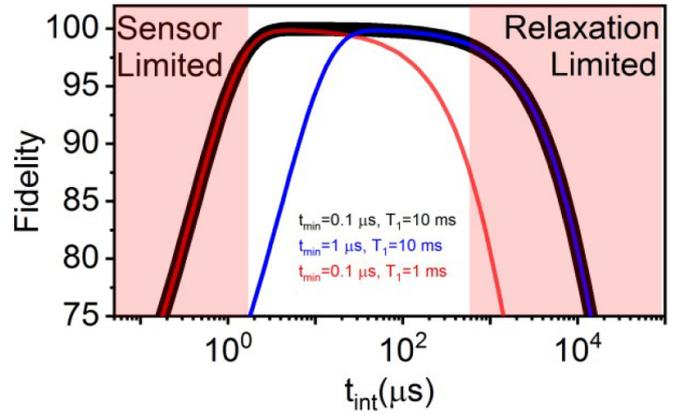

**Figure 3.** Readout fidelity as a function of integration time. The varying parameters are the relaxation time $T_1$ and the minimum integration time $t_{\min}$ defined as the integration time to achieve a signal-to-noise ratio of 1. The red areas indicate integration times for $t_{\min} = 100$ ns and $T_1 = 10$ ms where high fidelity readout (>99%) cannot be obtained.

### 1.2.2. Current and future challenges.

The challenge for dispersive sensing is to increase the readout fidelity well above 99% in timescales shorter than $T_2$ setting a clear target on the integration time ($t_{\text{int}}$) of a measurement. Following the readout model of Barthel *et al* [15], and assuming white noise, we find a useful rule of thumb for dispersive sensor designers: $T_1/25 > t_{\text{int}} > 25\, t_{\min}$ that translates in a necessity to increase the $T_1/t_{\min}$ ratio, see figure 3. In this roadmap, we focus on technological advances to minimize $t_{\min}$. See section 3.1 for a discussion on increasing $T_1$. The minimum measurement time of an impedance matched dispersive sensor, defined here as the integration time to achieve a signal-to-noise ratio of one, can be estimated in the small signal regime using the steady-state approximation [16, 17]:

$$t_{\min} \propto \frac{k_B T_N}{(\alpha e)^2}\left(\frac{C_{\text{r}}}{Q\omega_{\text{r}}}\right) \qquad (1)$$

Equation (1) highlights the different levels where readout fidelity improvements can be accomplished: At the device level, by increasing the coupling $\alpha$ to inter-dot charge transitions. Geometrically, $\alpha$ corresponds to the difference between the ratios of the coupling capacitance of the resonator to each QD and their total capacitance. At the resonator level, by increasing its natural frequency of resonance ($\omega_{\text{r}}$), increasing its loaded quality factor ($Q$) and reducing its capacitance ($C_{\text{r}}$)—or in other words by increasing the resonator impedance $Z_{\text{r}} = \sqrt{L_{\text{r}}/C_{\text{r}}}$ at fixed $\omega_{\text{r}}$. Finally, at the amplification level, by reducing the noise temperature of the first amplifying stage ($T_N$).

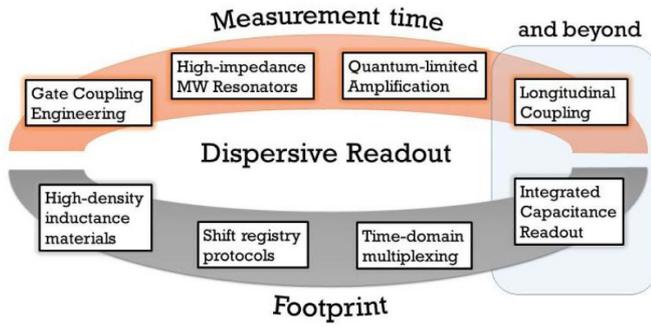

**Figure 4.** Roadmap for dispersive readout highlighting directions for improvement in terms of reduced measurement time and footprint. Strategies beyond dispersive readout are highlighted in the blue rectangle.

Another important timescale is the resonator response time,

$$t_r = \frac{Q}{\omega_r} \quad (2)$$

which should be lower than the timescale of the measurement, setting an upper bound for $Q$. Faster readout can be achieved by increasing the coupling of the resonator to the transmission line but at the expense of increasing $t_{\min}$. Finding a compromise between these two requirements is currently subject of extensive research.

Finally, an important challenge is to minimise the effect of the measurement on the qubit. Enhanced spontaneous emission can occur when the qubit frequency is close to $\omega_r$ via the Purcell effect [18]. Additionally, induced qubit dephasing can occur due to the measurement back-action caused by the photon-noise-induced frequency shift of the resonator. For a thermal population of photons in the resonator $\bar{n} = \left( exp\left( \frac{\hbar \omega_r}{k_B T} \right) - 1 \right)^{-1}$, the induced dephasing time is $t_\varphi = t_\varphi \left( \bar{n}^k \right)$, $k$ being positive and dependent on the qubit-resonator coupling regime. Induced dephasing can be minimised by operating at higher frequencies and cooling the resonator [19].

Another challenge that has received less attention but will impact the prospect for scalability, both for charge and dispersive sensors, is the footprint. As the number of qubits increase, the size of individual resonators will be a limiting factor. Strategies to reduce the size or the number of resonators will need to be put in place.

*1.2.3. Advances in science and technology to meet challenges.* At the device level, for silicon spin qubits, the gate coupling can be increased by using metal-oxide-semiconductor (MOS) structures with small equivalent gate oxide thickness, for example by using high-k dielectrics. However, the density of interface trap charges in these multi-layer oxides will have to be reduced to ensure reproducibility from device to device. Additional enhancements can be obtained by using thin (10 nm) silicon-on-insulator (SOI) and/or using non-planar gate geometries. However, 3D geometries may complicate QD couplings in 2D and therefore fabrication advances and design of novel qubit arrays will have to be proposed. Finally, a large gate coupling to inter-dot transitions can be achieved by minimizing cross coupling capacitance of the sensing gate or by driving the DQD gates in differential mode.

At the resonator level, the field will benefit from moving to on-chip lumped-element MW resonators where, by increasing the operation frequency, $t_{\min}$ and the back action due thermal photons will be minimised. Special care will need to be put in reducing the contribution of Purcell relaxation either by operating at large detuning or introducing Purcell filters, if a small detuning is required. However, the large footprint of these filters will negatively impact scalability. To reduce non-radiative losses in the resonator and boost the internal quality factor, resonators will need to be manufactured on low-loss SOI substrates with high quality interfaces.

At the amplification level, quantum-limited Josephson parameter amplification (JPA) in phase-preserving mode will enable reducing the readout time by an order of magnitude with respect to conventional cryogenic amplifiers given that their noise temperature is set by $T_N = \hbar \omega_r \coth \left( \hbar \omega_r / 2 k_B T \right) / 2 k_B$—where $T$ is the temperature of the amplifier. Furthermore, JPAs enable going beyond the quantum-limit using noise squeezing in the phase sensitive mode. For the large amplification bandwidth necessary for frequency multiplexing, travelling wave amplifiers may be used.

If these technical advances on dispersive (transverse coupling) readout were not sufficient to achieve high-fidelity detection in timescales shorter than the coherence time, longitudinal coupling, by modulation of the resonator-qubit coupling at the frequency of the resonator, could provide even faster readout while being generally quantum-limited [20].

Finally, the footprint. Moving to higher frequencies and lumped-element high-impedance resonators will minimize the size of the sensing resonator and more particularly the inductor which, at the 50 nH level, may occupy a physical area of $100 \times 100$ $\mu$m. Research on industry-compatible high kinetic inductance materials, like TiN, with an estimated kinetic inductance of $L_K > 200$ pH sq$^{-1}$ in 5 nm thin films, could drastically reduce the resonator footprint to sub $\mu$m$^2$. Josephson metamaterials formed by arrays of Josephson junctions may also be a compact alternative.

To reduce the number of resonators, two strategies could be used: a shift registry protocol with dedicated chip areas for sequential readout and/or time-multiplexed readout by sequentially connecting qubits to a readout resonator [21]. Time-multiplexing will have to be further developed to cope with the voltage drifts on the qubit gates associated with charge locking and also be able to manage clock feedthrough effectively by optimizing the control transistors.

*1.2.4. Concluding remarks.* For readout of solid-state qubits, dispersive sensing offers a fast solution with reduced back-action and footprint when compared to RF charge sensors. The different timescales $t_{\min}$, $t_r$, $t_\varphi$, impose strict conditions on the optimal measurement set-up but we find a good compromise when operating the qubit far detuned from the resonant frequency of a cooled high-impedance resonator with moderate $Q$, large capacitive coupling to the qubit and quantum-limited amplification, see figure 4. However, the associated footprint of the resonators will cause a major scalability challenge in the future. The community should think of ways to minimise its impact or even think beyond resonators, adapting concepts for capacitance readout from classical electronics. A compact solution that could be integrated on-chip with a footprint commensurable to the qubit size will be necessary if very large integration quantum computing is to become a reality.

**Acknowledgments** M F G Z acknowledges funding from the European Union's Horizon 2020 research and innovation programme under Grant Agreement No. 951852 and 688539 (http://mos-quito.eu) and support from the Royal Society Industry Pro-gramme and the Winton Programme for the Physics of Sus-tainability.

We thank Lisa A Ibberson and James Haigh for useful discussions.

## 2. Quantum light sources, cavities and detectors

*2.1. Quantum light sources*

Søren Stobbe[1] and Tim Schröder[2]

[1] Department of Photonics Engineering, Technical University of Denmark, Denmark

[2] Department of Physics, Humboldt-Universität zu Berlin, Germany

*2.1.1. Status.* Light consists of electromagnetic waves characterized by their wavelength, propagation direction, spin and orbital angular momentum. Beyond these classical properties, profound quantum mechanical properties of light emerge in the photon statistics and quantum correlations, and light can be categorized into uncorrelated thermal light (light-emitting diodes, the sun), highly correlated coherent light (lasers), and non-classical (quantum) light. Photonic quantum technologies are most often concerned with non-classical quantum states such as single photons, squeezed states, or multiphoton entangled states.

An early motivation for research on quantum light sources was the vision of unconditionally secure data communication systems employing quantum key distribution. It was initially believed that such cryptosystems required single photons but it was later realized that faint laser pulses combined with decoy-state protocols also enable unconditional security using existing technologies.

Contemporary research in quantum light sources extends on these ideas and seeks to address more complex quantum technologies including secure long-distance quantum communication with quantum repeaters and photon-based quantum information processing such as photonic quantum simulators and photonic quantum computers [22]. Meeting these goals will require great theoretical and experimental efforts as no physical system today fulfils the theoretical requirements. The theoretical proposals assume various quantum resources in the form of different quantum light sources as outlined in figure 5. As research in other areas of quantum technologies is also facing steep challenges, many researchers believe that the future of quantum technologies lies in hybrid systems that combine the best of different quantum technologies, e.g. the unprecedented range of optical quantum communication and the state-of-the-art performance of superconducting quantum circuits or ion traps. This has led to the vision of quantum networks [23], i.e. quantum communication links for long-range quantum key distribution or interfacing quantum computers.

Over the past decades, solid-state quantum light sources [23–30], in contrast to trapped atoms and ions, built by carefully engineering the photonic structures surrounding solid-state emitters such as QDs [23–27], color centres in diamond [28, 29], molecules [30], or 2D materials [28] (see table 1) have matured to a level, which makes them the most promising contenders for quantum light-source technologies.

*2.1.2. Current and future challenges.* Photonic quantum technologies encompass devices and visions employing various quantum photonic resources and posing different requirements, which in turn can be implemented with a variety of physical systems that are each more or less well suited and developed (see table 1). While the most important device aspects of quantum light sources have been addressed and at least partially demonstrated experimentally, it remains a significant challenge to meet several or all requirements in the same device. The exact requirements for quantum light sources depend on the particular application but scalable quantum architectures would likely require all figures of merit approaching unity.

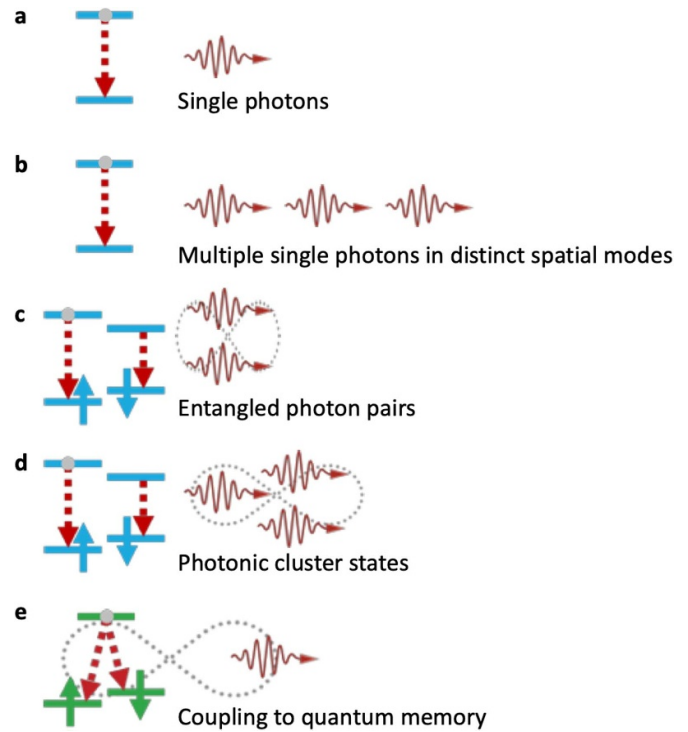

**Figure 5.** Quantum states of light and quantum light sources with wavepackets (red) emitted from optical transitions (dashed red), possibly employing spins (blue and green arrows) and entanglement (grey dotted lines). (a) Single photons emitted from a two-level system such as a quantum dot, a defect center in diamond, a molecule, or an atom. (b) Multiple single photons may be generated by demultiplexing of a single-photon source or multiple single-photon sources. (c) Using more complex level schemes, entangled photon-pair sources may be realized. (d), (e) Combining complex excitation protocols with complex level schemes allows building sources of photonic cluster states or interfacing photons with quantum memories.

For pulsed single-photon sources [23, 24, 27, 29], the key figures of merit are the system efficiency (the probability that there is at least one photon per pulse), purity (the probability that there is no more than one photon per pulse), and coherence (the degree to which two photons in the pulse train are quantum mechanically identical). Sources with unity efficiency are denoted on-demand or deterministic sources but an alternative is heralded sources (the emission time can be accurately measured). The coherence is often characterized by the indistinguishability, which gauges the coherence at short time scales but for scalable quantum technologies, the long-time indistinguishability and ultimately the linewidth is more relevant. Entangled-photon-pair sources [25] have similar figures of merit and in addition, the photon pairs must have a high entanglement fidelity.

**Table 1.** Overview of the most common quantum light sources and their ability to emit various quantum states as well as their most important properties. The black dots indicate experimental demonstrations so far. Notably, several of these quantum states and properties are mutually exclusive and it remains a main challenge in the science and engineering of quantum light sources to combine more functionalities and favourable properties within the same device.

| Quantum light source | Refs. | Quantum state | | | | | Properties | | | |
|---|---|---|---|---|---|---|---|---|---|---|
| | | Single photons | Multiple single photons | Entangled photon pairs | Photonic cluster states | Coupling to quantum memory | High efficiency | High repetition rate | High purity | High coherence |
| Trapped atoms | | ■ | | ■ | | ■ | | | ■ | ■ |
| Trapped ions | | ■ | | | | ■ | | | ■ | ■ |
| Optical quantum dots | [23–27] | ■ | ■ | ■ | ■ | ■ | ■ | | ■ | ■ |
| Defect centres | [28, 29] | ■ | | ■ | ■ | ■ | ■ | | ■ | ■ |
| Molecules | [30] | ■ | | ■ | | | ■ | | ■ | |
| 2D materials | [28] | ■ | | | | | | | | |
| Nonlinear materials | [24] | ■ | | ■ | | | | | ■ | ■ |
| Squeezed laser sources | [31] | | | | | | ■ | ■ | | ■ |

Besides these quantum mechanical parameters, a number of technological aspects are important towards real-world implementations. First, the wavelength should match the application, e.g., the telecom fiber-transmission bands for long-range communication, although most research has focused on shorter wavelengths. Second, optical pumping can lead to unacceptably high costs and complex layout; electrical pumping is preferred. Third, the spectral variation in solid-state emitters known as inhomogeneous broadening remains a major issue for reproducibility and scalability. Fourth, room-temperature operation is desirable whenever possible. Fifth, losses in all components in the optical circuits must be extremely low.

*2.1.3. Advances in science and technology to meet challenges.* Significant scientific progress has been made and many crucial properties of the various physical systems are now well understood but the jump from physics to technology is facing serious challenges. First steps have been taken and it is now time to apply industry-like engineering efforts to achieve efficiency enhancement, scalability, miniaturization, and cost-reduction. Since scientific research is often concerned with reaching the next breakthrough through experimental demonstrations using one or a few working devices, many of the underlying technological challenges are rarely addressed thoroughly and the fabrication yield is seldomly reported in the scientific literature. In many experiments, the fabrication yield is well below one percent and this renders the combination of different experimental techniques highly challenging or even practically impossible without new breakthroughs in nanofabrication and experimental techniques. The needed advances differ for particular quantum light sources and applications, and the the present discussion pertains to most but not necessarily all quantum light sources.

Building high-performance quantum light sources at telecom wavelengths could build on available semiconductor technology such as the indium-phosphide platform, but QDs at these wavelengths are yet to reach the same performance as those at shorter wavelengths, which are based on gallium arsenide. Ultimately, this goal might also require entirely new materials that are unknown or unexplored today. The integration of electrical pumping [25] may be able to replace optical pumping but resonant electrical pumping requires extreme control of tunnelling barriers. The inhomogeneous broadening of quantum emitters [23] remains a major obstacle and its solution appears to be beyond reach of the current generation of nanotechnology but pre-selection of emitters and local electrical tuning methods can at least partly overcome this challenge although it is difficult to combine voltage tunability with current injection. Room-temperature operation of highly efficient and coherent quantum light sources seems impossible within the current state of the art [23] and will likely require entirely new device concepts and/or materials. Scalable technologies with extremely low losses already exist, e.g. in glass-based photonic circuits and fibers and quantum light sources are benefitting tremendously from device concepts developed in data communication, silicon photonics, etc.

*2.1.4. Concluding remarks.* Quantum light sources have developed significantly over the past decades and have seen a shift in the anticipated applications from single-photon emitters for quantum secure communication to more complex photonic quantum systems and networks. Today, research is diversifying and new ideas gain importance. It appears that quantum light sources of the future will not just generate single-photon states but will provide entangled photon pairs, multiple parallel single photons, couple stationary quantum memories to photons, generate multi-photon cluster states, or squeezed laser light for communication or quantum imaging [31]. With sufficient performance, such sources would enable boson-sampling experiments [27] and quantum repeaters for long-distance quantum communication and quantum networks. The requirements will unquestionably change as the theoretical developments of quantum-information protocols progresses because although much of the governing physics has been understood and demonstrated experimentally, radical breakthroughs in the technology of quantum light sources are needed and should be expected.

**Acknowledgments**

S S gratefully acknowledges the Villum Foundation (Young Investigator Programme) and T S the Federal Ministry of Education and Research of Germany (BMBF, project DiNOQuant13N14921) for financial support.

## 2.2. Semiconductor–superconductor hybrid circuit-QED

P Scarlino and J V Koski

Department of Physics, ETH Zürich, CH-8093 Zürich, Switzerland

*2.2.1. Status.* Standard approaches to studying light–matter interaction consist of coupling one atom to one or few electromagnetic modes of a cavity. In the context of circuit quantum electrodynamics (cQED), this concept has been implemented in the microwave domain with an on-chip superconducting resonator coupled to superconducting artificial atoms, providing the means to probe and manipulate their quantum state and to entangle them [32]. Recently, cQED has been explored for hybrid systems, where semiconductor-based qubits are defined by the orbital (charge) or the spin degree of freedom of electrons/holes confined in electrostatically defined QDs, having led to the observation of coherent interaction between a microwave photon and a charge qubit [33–35] or a spin qubit [36–38]. The state-of-the-art hybrid cQED experiments with semiconductor QDs have demonstrated dispersive qubit readout [39] (see figure 6(a)), virtual-photon-mediated interaction between two charge qubits [40] (see figure 6(b)), between a transmon and a charge qubit [41] (see figure 6(c)) and a resonant exchange spin qubit in GaAs [42], and between two spin qubits in SiGe [43].

The cQED architecture is one of the most promising platforms for realizing two-qubit gates between distant qubits in a future quantum processor, providing an interaction range determined by the cavity length (up to a few millimeters). The method would be particularly useful for semiconductor QD platforms where direct qubit-qubit coupling is typically limited to the spatial extent of the wavefunction of the confined particle (up to a few hundred nanometers). To scale up QD-based architectures, small clusters of QD qubits could be coupled by resonators [44] (see figure 7) in contrast to solely relying on technically challenging realizations of dense 1D or 2D arrays of QDs. In addition to the applications in quantum information processing, hybrid cQED can also contribute to exploring more complex mesoscopic systems, such as Majorana fermions [14], Kondo systems, or Luttinger liquids.

A coherent link between semiconductor- and superconductor-based qubits may give access to the best of both device architectures by, for example, providing an interface between fast-operated transmons and long-coherence spin qubits as a quantum memory. Further enhancing the qubit–photon interaction strength allows exploring the fundamental physics of ultra-strong coupling regime (USCR), where the strength of the qubit–photon coupling is comparable to the cavity photon energy. There, the more efficient interactions could provide not only shorter operation times, but also simpler protocols where the natural evolution of a USC system replaces a sequence of quantum gates [47].

*2.2.2. Current and future challenges.* In order to achieve an efficient qubit manipulation via electric means, it is required that the qubit computational states present a finite electric dipole moment. This also subjects the qubit to dephasing by electric noise, which is the dominant contribution to qubit decoherence. While cQED is well established for superconducting qubit platforms, the realization for semiconductor qubits has proven challenging. The comparably small dipole moment of QD qubits leads to a weaker interaction with the zero-point fluctuations of microwave photons in superconducting resonators. Furthermore, typical dephasing rates of an electron charge qubit is of the order of 100 MHz–10 GHz, as measured by using conventional transport or charge detection techniques and confirmed by first generation of hybrid cQED devices. Encoding the quantum information mainly into the electron spin degree of freedom suppresses electric-noise-induced dephasing, however at the price of decreased electric dipole moment and therefore lower qubit–photon coupling strength and increased susceptibility to magnetic noise [36]. Over the past few years, however, the decoherence rates of QD-based qubits embedded in a cQED architecture have been reduced by almost two orders of magnitude, down to a few MHz level, both for the spin and charge degree of freedom. While not yet demonstrated, virtual-photon-mediated spin–spin coupling is within experimental reach [43] with further improvements in resonator quality factors, spin-photon coupling rates, and further suppression of noise-induced dephasing.

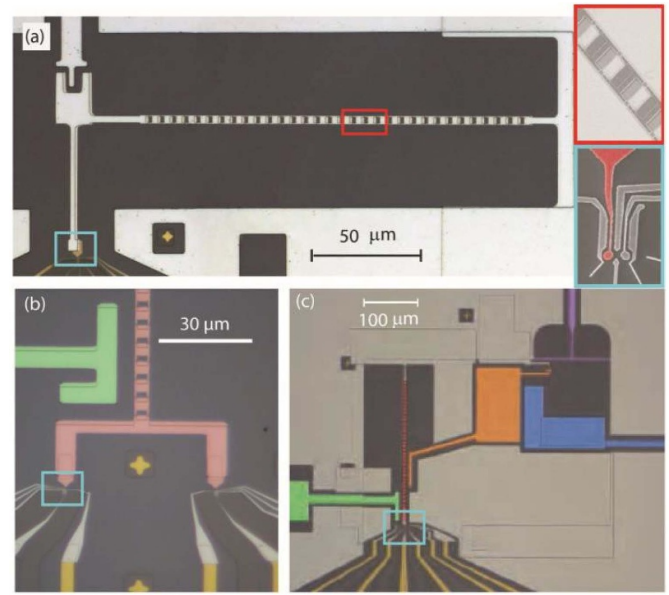

**Figure 6.** False-color optical micrographs of hybrid cQED devices with high impedance SQUID array resonator coupled to charge qubits in GaAs. (a) SQUID array resonator (light gray) coupled to a single charge qubit defined via depletion gates (yellow). Enlarged view of the SQUIDs in the resonator (charge qubit) is shown in the inset enclosed by the red (blue) line [33]. (b) Optical micrograph of a device with two charge qubits coherently coupled by a SQUID array resonator [40]. (c) False-color optical micrograph of the device showing the SQUID array resonator (red) mediating the coherent coupling between a single island transmon (orange) and a charge qubit [41].

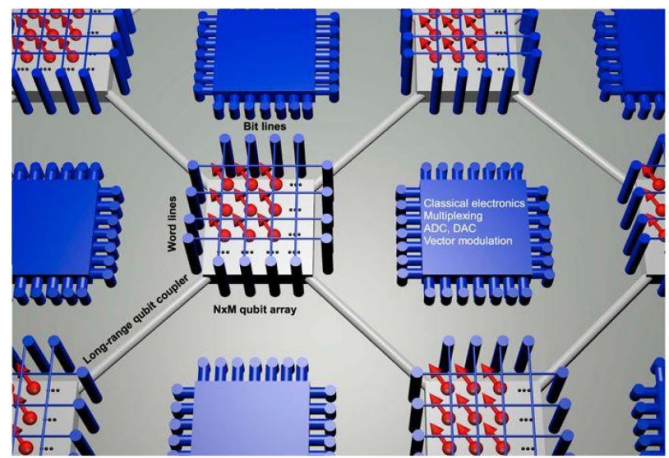

**Figure 7.** A schematic illustrating a potential scaling method of a spin-qubit-based quantum processor. A microwave resonator can provide the long-distance interaction between local clusters of mutually coupled spin qubits [44].

The ultimate goal of practical hybrid cQED based quantum computation with high-fidelity gates and readout requires further improvement of the qubit coherence time while maintaining a high qubit–photon coupling strength for reaching gating times much shorter than those of the qubit coherence. The main challenge is to mitigate the noise-induced decoherence, either by optimizing the qubit design to have noise-insensitive energy dispersion, or by decreasing the noise magnitude. An additional challenge arises from internal relaxation processes, such as qubit energy decay by phononic, or photonic loss channels. Furthermore, when implementing multiple qubits in a practical quantum computer, one of the major objectives, irrespective of the qubit architecture, is solving the wiring and coupling challenge, i.e. the implementation of control lines and electronics for a dense qubit array while realizing a mutual coherent link between the arrays [44]. Implementation of hybrid cQED may be essential for realizing long distance coherent coupling within the qubit network and, ultimately, implementing error correction protocols in these systems, for example with surface code.

*2.2.3. Advances in science and technology to meet challenges.* The qubit–photon coupling is determined by the vacuum fluctuations in voltage $V_0$ that scales with the resonator impedance as $V_0 \propto \sqrt{Z_r}$. In recent experiments, high qubit–photon coupling has been achieved by engineering the resonator to have a high impedance $Z_r$ beyond the typical 50 Ω of conventional coplanar waveguides [33, 36, 38]. This approach is universally applicable to any cQED system striving to maximize the coupling to the charge degree of freedom and is promising for realizing coherent spin–spin coupling. By further increasing the resonator impedance beyond the 1 kΩ of recent experiments, USCR with semiconductor QDs could be reached [47]. High impedance resonators can be fabricated out of high kinetic inductance disordered superconducting thin films. They have shown to preserve a high quality factor even in the presence of a strong (few Tesla) in plane magnetic field [36, 38], characteristic that makes them ideal to explore the spin properties of a mesoscopic system.

The dispersive interaction between a qubit and a microwave resonator provides very high fidelity and fast single shot measurements of the qubit state [32]. This readout technique has been optimized with quantum-limited microwave parametric amplifiers, which enhance the readout signal while introducing a minimal amount of noise. The recent observation of coherent semiconductor–qubit–photon interaction could facilitate the implementation of such a readout technique also for semiconductor qubits, which is orders of magnitude faster than the conventional readout with a charge sensor that currently presents a maximal bandwidth of a few hundred kHz for single shot measurements. Another proposed class of techniques for qubit readout and coupling relies on longitudinal interaction between the qubit and the resonator photons [20]. Longitudinal coupling has been proposed initially as an alternative and more efficient readout and coupling tool for superconducting qubits and recently extended to electron spins and topologically protected states embedded in a cQED architecture.

Recent experiments have explored more complex qubit implementations with energy dispersions that are particularly gate voltage-independent while still maintaining a possibility to manipulate the quantum state electrically [48]. On the other hand, conventional charge qubits have recently shown unexpectedly long coherence times [33, 34, 39]. A particular feature is that they are operated in multi-electron regime, suggesting that Coulomb interactions and decreasing QD charging energy could be relevant for protection from charge noise. These experiments indicate that appropriate engineering of the quantum system may significantly improve the resilience of semiconductor qubits to ubiquitous charge noise.

*2.2.4. Concluding remarks.* Recent progress on hybrid cQED-based approach to semiconductor QDs has led to the observation of coherent charge/spin qubit–photon interaction. Considerable improvement in the qubit coherence time is still necessary to achieve high-fidelity time-domain manipulation and single shot readout. Such technological development could be accessible, however, by engineering the quantum system and optimizing the host material such that the qubits are more resilient and less exposed to electric noise induced decoherence. Achieving long coherence times, combined with enhanced qubit–photon coupling by an optimized design of high impedance resonator, would enable entanglement of distant spin qubits and therefore provide a promising platform for a scalable semiconductor–superconductor hybrid quantum processor. Furthermore, the cQED technology offers a qualitatively new way to investigate the dynamic response of mesoscopic nanocircuits at the fundamental level, allowing direct microwave spectroscopy of the quantum states emerging in more exotic semiconductor and hybrid systems.

## Acknowledgments

This work was supported by the Swiss National Science Foundation through the National Center of Competence in Research (NCCR) Quantum Science and Technology.

## 3. Quantum computing with spins

*3.1. GaAs quantum dots*

*Ferdinand Kuemmeth[1] and Hendrik Bluhm[2]*

[1] Center for Quantum Devices, Niels Bohr Institute, University of Copenhagen, Denmark
[2] JARA-FIT Institute for Quantum Information, RWTH Aachen University and Forschungszentrum Jülich, Germany

*3.1.1. Status.* Gate-defined QDs in GaAs have been used extensively for pioneering spin qubit devices due to the relative simplicity of fabrication and favourable electronic properties such as a single conduction band valley, a small effective mass, and stable dopants. Decades of prior improvements of the growth of III–V heterostructures by molecular beam epitaxy had resulted in the availability of high-quality substrates for various applications, and spin qubits were ultimately first demonstrated in GaAs in 2005, significantly before the first Si qubits in 2012. GaAs spin qubits are now readily produced in many labs, whereas the realization of comparable devices in Si remains challenging. However, a disadvantage is the unavoidable presence of nuclear spins, leading to an intrinsic $T_2^*$ of about 10 ns. Dynamical decoupling can extend the coherence time to the millisecond range [49], and single-qubit control with a fidelity of 99.5% was demonstrated [50]. Nevertheless, these techniques require a significant effort in controlling and suppressing nuclear spin fluctuations, and so far have only been successful for singlet–triplet qubits encoded in two-electron spin states associated with DQDs. GaAs QDs like those in figure 8 are currently used as a testbed for entanglement [45], quantum non-demolition measurements [51], automatic tuning [46, 52], multi-dot arrays [53, 54], coherent exchange coupling [54], teleportation [55] etc, partly because reproducible Si devices are not broadly available yet. Much of the resulting insights can be transferred to Group IV material systems, although specific properties of GaAs are also actively studied. Remarkable recent achievements include the transfer of electrons between QDs using surface acoustic waves (SAW) [56], which could be used to overcome the challenge of connecting distant qubits, and the detection of photo-generated carriers, a precursor to the ability to convert flying photonic qubits into spin states [57]. Last but not least, qubits in GaAs QDs are of interest as a manifestation of quantum many-body physics, such as the central spin problem or itinerant magnetism [58].

*3.1.2. Current and future challenges.* The operation of gate-defined spin qubits relies on voltages—quasistatic voltages for tuning the device to an appropriate operating point, and time-dependent control voltages for the coherent manipulation on nanosecond timescales—which in a modern dilution refrigerator should be practical up to approximately 100 qubits. On the flipside, this makes the quantum processor susceptible to effective electrical noise, requiring a careful trade-off of instrumentation noise and the material's intrinsic charge noise against other engineering constraints. Just like the encoding in specific two-electron spin states makes a singlet–triplet qubit robust to global magnetic field fluctuations, other encodings in three-electron [59] or four-electron [60] spin states have recently been proposed that also mitigate noise in the magnetic gradient between dots (particularly relevant for GaAs) and effective charge noise (relevant also for Si). The role of symmetric operating points [60] in these proposals are being experimentally studied in GaAs multi-dot arrays [61], exposing a new engineering challenge: The large number of physical gate electrodes per QD (facilitated by the relatively large size of GaAs QDs) allows independent tuning of many local degrees of freedom (dot occupation, interdot tunnel barriers, etc.), but ultimately will impose unrealistic wiring requirements. For a processor with more than 1000 spin qubits, a radical change will be needed on how to integrate QDs at cryogenic temperatures with scalable control electronics. Even for current devices, the ultimate limits of coherence and control fidelity are still uncharted, despite the fact that the nature of the hyperfine coupling between electron and nuclear spins is rather well known and many of the resulting effects are now understood in considerable detail. Using appropriate control pulse optimization, substantial improvements in the demonstrated two-qubit gate fidelities can be expected. As for all types of QD qubits, a mechanism for high-fidelity long-range coupling would likely be required for truly scalable quantum circuits, potentially building upon current efforts to couple GaAs dots to superconducting cavities [41] (cf figure 9(d)) or shuttling of electrons (figure 9(e)). Although anecdotal experience in many labs points to a good reproducibility of GaAs QDs, no systematic study supports this evidence, and the limiting factors are unknown. A detailed yield investigation could reveal if the small effective mass is a decisive advantage and could serve as a reference benchmark for Si-based devices.

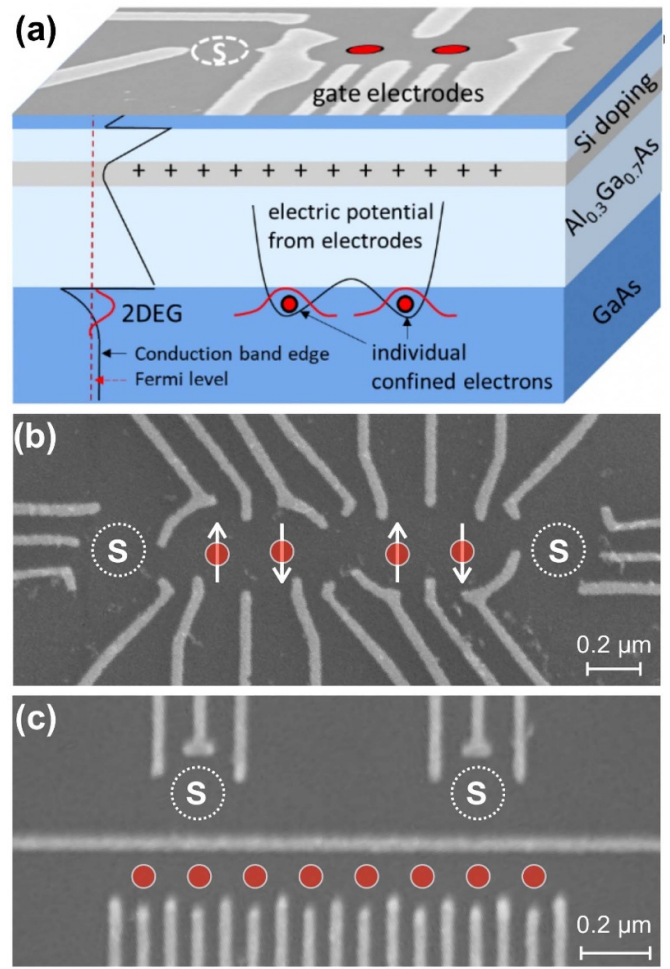

**Figure 8.** Representative GaAs quantum-dot qubit devices—from double dots to linear arrays. (a) Top-gated GaAs heterostructure resulting in controllable one-electron quantum dots with proximal charge sensor (S) for readout. (b) Two proximal double dots to study entanglement between two nearest-neighbour singlet–triplet qubits [45]. (c) Progress towards linear spin chains [46]. Figure credits: Hendrik Bluhm, RWTH Aachen (a), Shannon Harvey, group of Amir Yacoby at Harvard University (b), Christian Volk, group of Lieven Vandersypen, TU Delft (c).

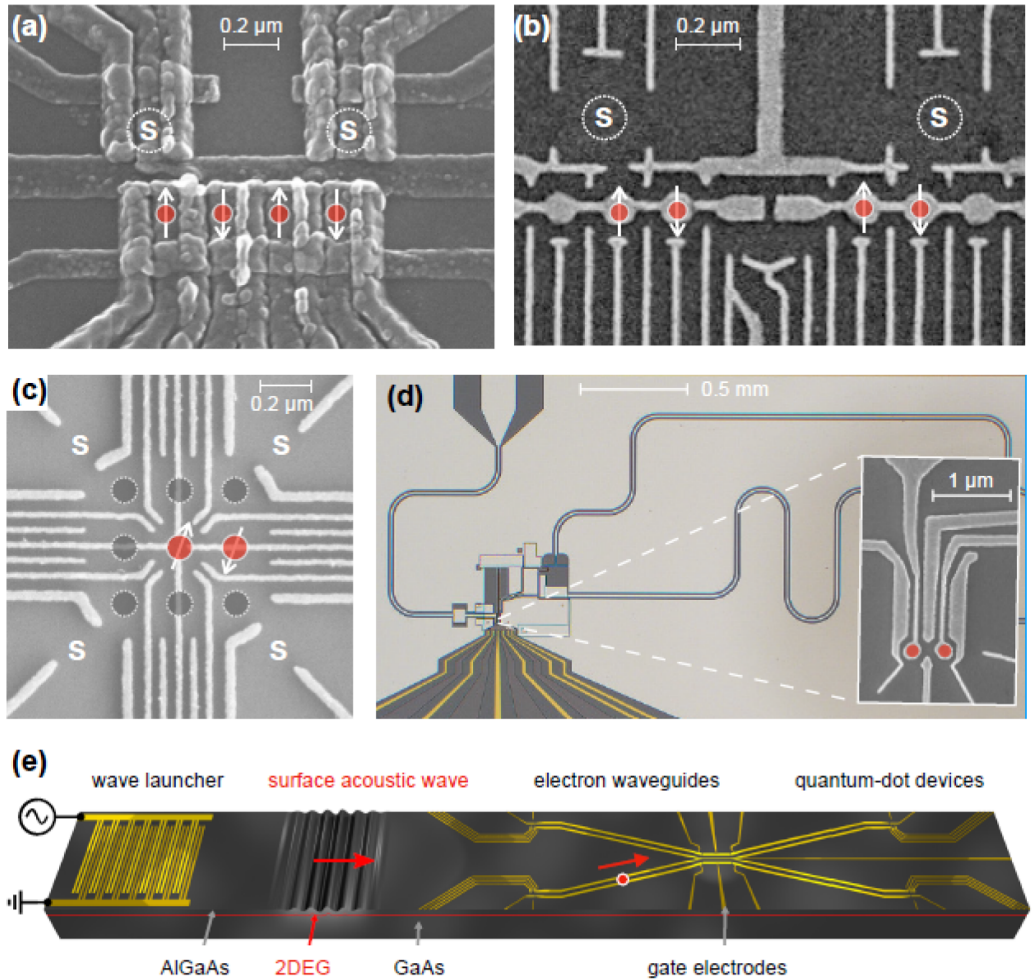

**Figure 9.** Current approaches to larger quantum circuits with GaAs qubits: intermediate distance coupling by (a) Heisenberg teleportation [56] and (b) mediated exchange [54]. First steps towards larger circuits by (c) operating spins within a small 2D array [54], (d) integration with superconducting devices and cQED [41], and by (e) moving electrons via surface acoustic waves (SAW) [56]. Figure credits: John Nichol, University of Rochester (a), Ferdinand Kuemmeth, University of Copenhagen (b), Tristan Meunier, Université Grenoble Alpes (c), Pasquale Scarlino, Group of Andreas Wallraff, ETH Zurich (d), Christopher Bäuerle, Université Grenoble Alpes (e).

*3.1.3. Advances in science and technology to meet challenges.* Further improvement of coherence and control fidelity will benefit from both improved dynamic nuclear polarization procedures to suppress fluctuations of the hyperfine field as well as a reduction of charge noise. Somewhat surprisingly, simulations indicate that charge noise is the more limiting factor. For long range coupling approaches via cavities or electron shuttling, material-specific limitations will have to be understood. Piezoelectricity, spin–orbit coupling, and nuclear spins work against the GaAs material system, whereas the single valley and small mass are advantages. For cavity coupling, current performance metrics are not nearly good enough for high-fidelity entangling gates. From the quantum control point of view, one challenge appears to be that optimal pulses require careful cancelation of errors due to quasi-static noise. Applying simulated pulses in experiments may compromise the desired performance due to imperfect system knowledge and thus require new approaches to gate characterization and calibration.

Regarding device designs, more complicated circuits would greatly benefit from multiple metal layers, as shown in figure 9(a). Yet, the then required dielectrics may be an additional source of charge noise and will sacrifice some of the fabrication simplicity.

An exciting prospect associated with the direct band gap is to convert between spin and photon states (see section 3.5 Quantum interface based on photon-spin coherent transfer), or to entangle them. This capability could be a major advantage over Si by allowing the realization of networks of quantum processors for communication and distributed computing and by opening additional options for long-range on-chip coupling. Much of the fundamental principles have been demonstrated using self-assembled QDs and could be transferred to hybrid devices with some kind of exciton trap coupled to a gate-defined dot [62]. However, such devices yet remain to be realized.

*3.1.4. Concluding remarks.* GaAs-based devices have been crucial for the birth of QD qubits. Much attention is now shifting to Si. While the reasons for this trend are largely compelling, it is not established with scientific rigour that Si is preferable to GaAs when considering all factors. The compatibility of Si with complementary metal-oxide-semiconductor (CMOS) processing is often seen as an advantage. However, one should also keep in mind that process development for the unusual layouts compared to transistors with small feature sizes needed for Si qubits will incur large development costs for foundry fabrication. In any case, GaAs-based devices are likely to remain a workhorse for proof-of-concept quantum information processing and solid-state experiments. Considerable technological and scientific potential may arise from advances in optical coupling.

**Acknowledgments**

F K acknowledges support by the Danish National Research Foundation.

## 3.2. Quantum computing with spins in silicon: dots

*Andrew Dzurak, Chih-Hwan Yang and Jun Yoneda*

School of Electrical Engineering and Telecommunications, The University of New South Wales, Sydney 2052, Australia

### 3.2.1. Status.

Electron spins in QDs exhibit compelling properties for use as qubits. Pioneering studies in GaAs QDs demonstrated basic requirements for spin qubits, including initialization, readout, and coherent control. However, their fidelities are limited by strong hyperfine-mediated dephasing from gallium and arsenic nuclei, which all have non-zero spin. For large-scale quantum computing employing quantum error correction, control fidelities above 99% are required. Silicon QDs have attracted strong interest due to the low natural abundance of nuclear spins in silicon (only 4.7% of atoms contain spin-bearing $^{29}$Si nuclei), which can be further reduced by isotopic enrichment, together with their compatibility with CMOS processes used in industry. These advantages have motivated significant commercial efforts to develop silicon QD-based quantum computing by established semiconductor companies, e.g. Intel and STMicroelectronics in US/Europe, research foundries, e.g. IMEC and CEA-Leti, and new start-ups, e.g. Silicon Quantum Computing in Australia.

Several device technologies have been explored to realise QD spin qubits in silicon-based nanostructures (figures 10(a)–(d)). The long phase coherence time ($T_2^*$) in silicon QDs was first observed using devices based on Si/SiGe heterostructures [63]. Si/SiO$_2$ (or Si-MOS) structures were later used to demonstrate fault-tolerant single-qubit control fidelities ($F_{1Q}$) [64]. Single electrons or holes can be hosted also in nanowires, which can be fabricated much like industry-standard CMOS transistors [65].

Electron spin qubits in silicon QDs have been realized in several operating modes, using different numbers of electrons—notably, Loss-DiVincenzo (LD) qubits based on a single electron (1$e$-), singlet–triplet qubits (2$e$-), hybrid qubits (3$e$-), and exchange only (E-O) qubits (3$e$-). Multi-electron spin states can be efficiently controlled electrically via exchange interactions. For LD qubits, single qubit manipulation can be performed magnetically through electron spin resonance [64] or electrically using electric-dipole spin resonance [66], in each case rotating the spin of a single electron between down and up states (figure 10(e)). For the other qubit modes, controlling multi-electron hybridized states is typically performed via fast voltage pulsing on gate electrodes. Figure 10(f) summarizes demonstrated qubit performance for individual technologies and qubit types.

### 3.2.2. Current and future challenges.

Demonstrating the building blocks required for realisation of a logical qubit is an important next stage of development. This will require a fully functional array of 3–5 qubits with high control and readout fidelities. The large electron effective mass in silicon requires small confining structures to reach the single electron level and fabricating a device with more than three qubits in an academic environment remains a challenge, although advances are being made within the community [71]. Currently, device characteristics and qubit performance (e.g. qubit coherence times, control speed, Landé g-factor, and valley splitting) exhibit device-to-device and dot-to-dot variations.

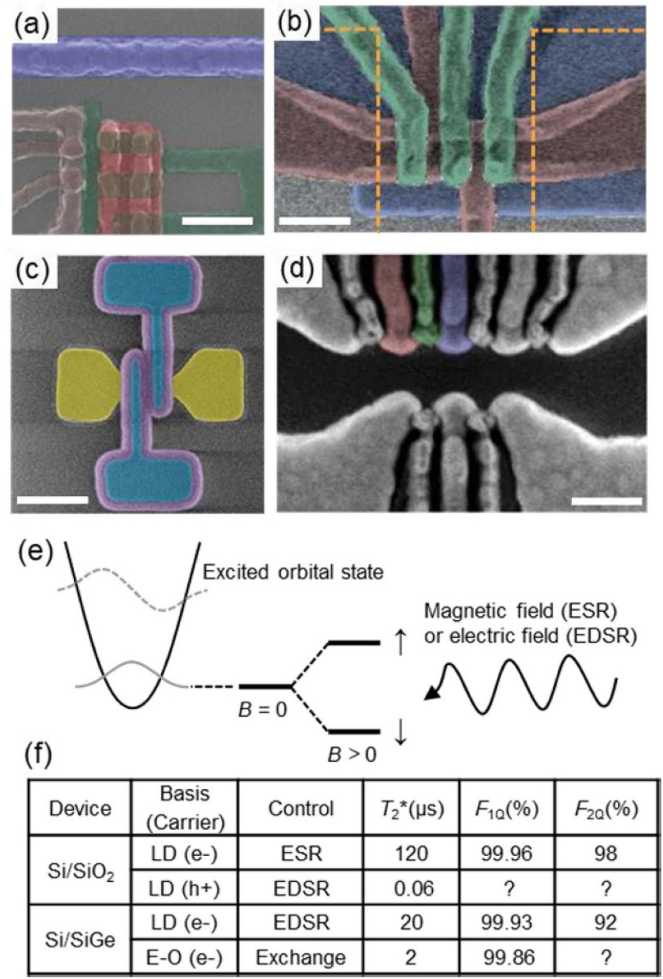

| Device | Basis (Carrier) | Control | $T_2^*$(μs) | $F_{1Q}$(%) | $F_{2Q}$(%) |
|---|---|---|---|---|---|
| Si/SiO$_2$ | LD (e-) | ESR | 120 | 99.96 | 98 |
|  | LD (h+) | EDSR | 0.06 | ? | ? |
| Si/SiGe | LD (e-) | EDSR | 20 | 99.93 | 92 |
|  | E-O (e-) | Exchange | 2 | 99.86 | ? |

**Figure 10.** (a)–(d) Device architectures to implement spin qubits in silicon. Adapted from [37, 64, 65, 67] with permission. All scale bars are 200 nm. Devices (a) and (c) are based on Si/SiO$_2$ structures, whereas those in (b) and (d) employ a Si/SiGe quantum well. (e) Schematic of a LD qubit level structure. A single electron (or hole) in the lowest orbital state exhibits a Zeeman splitting under a magnetic field. Transitions between these spin states can be coherently driven with fields at the resonant frequency. (f) Table summarizing the demonstrated characteristics and metrics of different qubit implementations. Values are quoted from refs. [64–66, 68–70].

Understanding the limiting mechanisms of coherence is crucial for identifying the optimal device technology and qubit basis type. The major sources of decoherence for spin qubits in silicon QDs are charge noise, which impacts via the spin–orbit interaction, and the hyperfine interaction with surrounding $^{29}$Si nuclear spins [37, 63–70]. Characterisation and suppression of these noise mechanisms is an active research topic for many groups, with improved measurement and control techniques being developed.

Control errors need to be further reduced, especially for two-qubit gates, in order to implement error correcting codes.

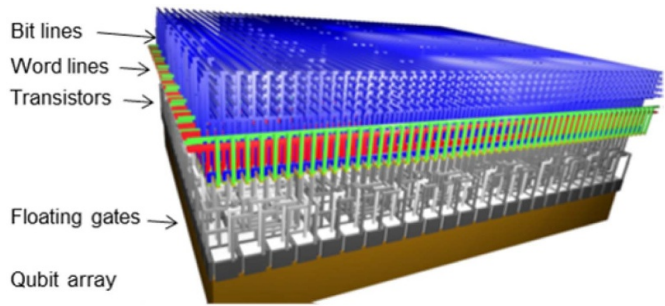

**Figure 11.** Architecture of a 2D spin qubit array using CMOS technology. The qubits are individually addressable via word and bit lines, and the gate biasing is achieved via periodically charging of floating gates. Adapted from [73] with permission.

The two major approaches are to enhance coherence times, and to shorten the gate times. Optimizing the balance between coherence and controllability is a ubiquitous problem in qubit research and will ultimately require thorough characterization of devices with good reproducibility. Readout and initialization should also be performed with sufficiently high fidelities, and the techniques used should be made compatible with multi-qubit device layouts, in terms of physical dimension. Currently, readout is typically achieved with limited sensitivity, and using sensors with a large footprint, such as a SET.

At present, coupling between spin qubits typically relies on the exchange or Coulomb interaction, which are only active at short range. The ability to separate qubits by microns or more would be a useful asset, since it would alleviate the fan-out and wiring problems which will arise when QD architectures are employed in a large-scale quantum processor.

*3.2.3. Advances in science and technology to meet challenges.* One way to overcome the problem of a large footprint and low fidelity of the qubit sensor is to adopt dispersive readout techniques [72, 73] (see also figure 11). These utilize a RF tank circuit connected to a gate electrode that is commonly used for qubit voltage biasing, eliminating the need to have a large sensing component (e.g. SET) nearby. Dispersive readout detects the reflected RF signal to determine the correlation between two spins [72]. Operating at a higher frequency than traditional SET sensing, it can potentially yield a higher signal-to-noise ratio and a larger bandwidth, improving the readout fidelity and reducing the readout time.

Circuit quantum electrodynamics has also recently been used to couple spins to microwave photons in a superconducting cavity, thus opening a path to achieving long-distance spin-photon-spin coupling [37]. Such a long-range qubit coupler would allow the footprint density per qubit to be relaxed, making more room for control/measurement electronics that will be needed for a full-scale quantum processor (figure 7 in section 2.2). Additionally, the qubit spin state can also be read out via the cavity response [37].

Gate fidelity problems can be addressed using software and materials engineering. Implementing advanced software techniques such as dynamical decoupling pulse sequences can prolong the coherence times of the qubits [64, 66], while shaped microwave pulses can be used to improve gate fidelities [74]. Gate fidelities could be further improved by using higher levels of silicon isotopic enrichment, almost eliminating the impact of $^{29}$Si nuclear spin noise, which is presently a major source of two-qubit gate infidelity [68].

To address the important challenge of reproducibility and yield, moving device production from university laboratories to industrial-scale CMOS foundries will be crucial. Growing higher quality oxides and patterning gate electrodes with finer resolution will ensure that QDs are more uniform across a large qubit array. Global CMOS foundries have now begun to invest in the development of silicon QD based quantum processors. Once fabrication processes are established it is envisaged that chips containing large qubit arrays could be mass produced with high yield, opening the prospect of integrating conventional CMOS control electronics with the qubit system, and providing a path to full-scale silicon-based quantum computing in a CMOS-compatible form (figure 11). Furthermore, qubit array modules can be connected through long-range qubit couplers, creating a sparse array (see figure 7 in section 2.2) that reduces the density of the physical qubit control lines that generate heat, and opens up space for classical electronics to be integrated nearby.

*3.2.4. Concluding remarks.* Research in quantum computing based on silicon QDs has grown rapidly over the past few years, due to the availability of isotopically-enriched, nuclear-spin-free silicon, and the potential of utilising silicon foundry manufacture. The primary materials systems used to confine QD qubits include Si/SiGe and Si/SiO$_2$ interfaces and nanowires. For the Si/SiO$_2$ and nanowire structures, dots can contain either electrons or holes depending on the gate bias mode. The reproducibility of silicon QD device parameters remains an important challenge. This includes overcoming differences in qubit resonance frequencies, dot sizes, tunnel and exchange coupling strengths and valley states that may arise from interface disorder and imperfect electrode patterning. Despite the challenges, with the achievement of high single-qubit (99.96%) and two-qubit (98%) gate fidelities, silicon QDs have proven to be a leading candidate for spin-based quantum computing, and research focus is now moving towards the design of large-scale qubit architectures and industrial manufacture.

**Acknowledgments**

We acknowledge support from the US Army Research Office (W911NF-17-1-0198), the Australian Research Council (FL190100167 & CE11E0001017), Silicon Quantum Computing Proprietary Limited, and the NSW Node of the Australian National Fabrication Facility.

### 3.3. Quantum computing with donor spins in silicon

*Jarryd J Pla[1] and Charles Hill[2]*

[1] School of Electrical Engineering and Telecommunications, UNSW Sydney, Australia

[2] School of Physics, University of Melbourne, Melbourne, Australia

*3.3.1. Status.* Of the early proposals to implement a quantum computer in the solid-state, the Kane proposal [75] garnered much attention because of its use of silicon—the workhorse of the multi-trillion-dollar microelectronics industry. The original idea saw information encoded in the nuclear spin states of individual phosphorus donors engineered inside a silicon chip, with interactions between neighbouring nuclei facilitated by their donor-bound electrons (figure 12(a)). In later proposals, the electron was similarly identified as an excellent quantum bit (qubit) candidate. Both the electron and nuclear spins of donors are known to possess exceptionally long coherence times when they are incorporated in a silicon host that has been purified in the nuclear-spin-zero isotope $^{28}$Si [76], reaching seconds to minutes.

The fabrication of nano-electronic devices that isolate individual donors has been achieved by means of two approaches: a conventional MOS fabrication strategy (figure 12(b)) that involves ion-implanting single donors [77], and an 'atom-by-atom' approach (figure 12(c)) that builds devices from the bottom-up using the atomic-precision afforded by a scanning tunnelling microscope (STM) [78]. Critically, several important criteria for establishing a scalable quantum computer have also been met. The abilities to projectively measure the states of donor electron [77] and nuclear spins [79] were achieved in 2010 and 2013, respectively. Measurements were performed electrically using sensitive nano-scale circuits (figure 12(b)) to detect the donor charge and infer the spin states through a process known as spin-to-charge conversion [77]. On-chip broadband antennas have been successfully deployed to deliver microwave and radio-frequency fields for performing coherent control of the electron and nuclear spins [79]. Devices have been made in isotopically-enriched silicon and shown to exhibit the exceptionally-long spin coherence times expected from ensemble measurements [80], as well as led to some of the highest single-qubit control fidelities (>99.95%) for any solid-state qubit. Experiments on multi-donor devices have observed an exchange coupling between two electrons [81]—a basis for two-qubit logic gates. Efforts to demonstrate coherent exchange operations between two qubits are currently underway and producing promising results.

An extensive theory program encompassing device modelling, architectures and control has complemented this experimental work. A metrology technique using atomistic tight binding has been shown to be capable of precisely locating phosphorus donor qubits within a device, up to depths of 5 nm [82].

*3.3.2. Current and future challenges.* Quantum computing with donors in silicon faces several current and future

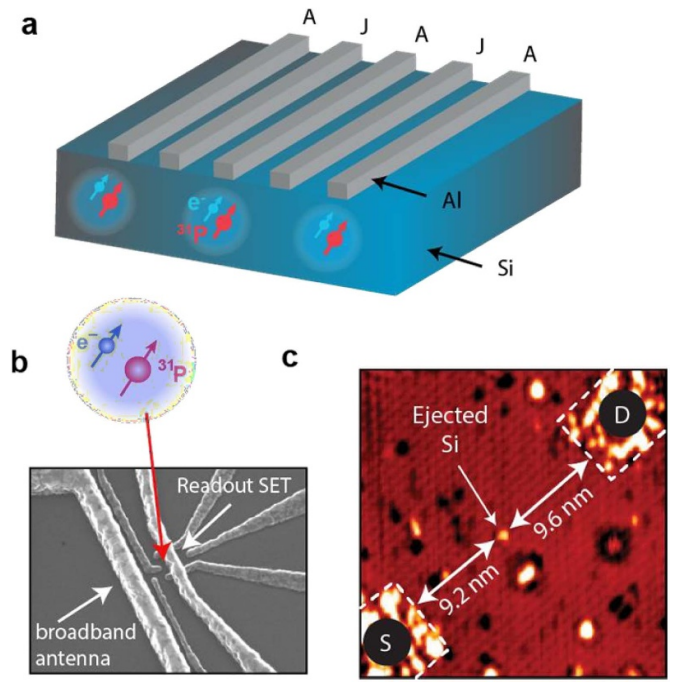

**Figure 12.** (a) The Kane quantum computer proposal. Donor qubits are positioned under electrodes (A-gates) that implement single-qubit operations. Two-qubit gates are controlled using electrodes (J-gates) that produce an exchange interaction between neighbouring electrons. (b) A single-donor device fabricated via ion-implantation of phosphorus into silicon [80]. Aluminium CMOS electronics facilitate qubit control and readout. (b) A precision single-donor transistor device fabricated with STM lithography [78].

challenges relating primarily to the scale up and improvement of the proof-of-principle devices:

*Fabrication of multi-qubit arrays:* In order to demonstrate the large-scale quantum computing envisioned, it is necessary to increase the number of qubits in a device, starting with reliable and controllable couplings in two-qubit devices to devices containing five to ten qubits. These small-scale processors will be critical in the development of manufacturing technology and reaching important milestones such as the demonstration of logical/error-protected qubits. Ultimately, the ability to realize arrays of fabricated qubits will be necessary to implement quantum algorithms and topological quantum error codes [83].

*Control of multi-qubit arrays:* Sophisticated multiplexed control electronics will need to be developed to interface with the qubit array and perform qubit control and readout. Tailored control pulses, which minimize noise in gate operation and measurements, need to be optimized for donor-based quantum computation. Strategies for delivering the microwave and radio-frequency magnetic drive signals used in single qubit gates simultaneously to large numbers of qubits (often referred to as 'global control' [75]) will likely need to be developed to replace the current local methods of generating these fields.

*Long range coupling and transport:* Enhancing the long-range coupling between qubits will allow donors to be placed further apart than is naturally available through the exchange and magnetic dipole–dipole interactions of electrons (which is typically limited to 20–30 nm in silicon), providing space to integrate the classical control and measurement electronics. Similarly, the ability to transport quantum information across a device (over distances of millimetres to centimetres) will need to be established and interconnects between remote processors is also desirable.

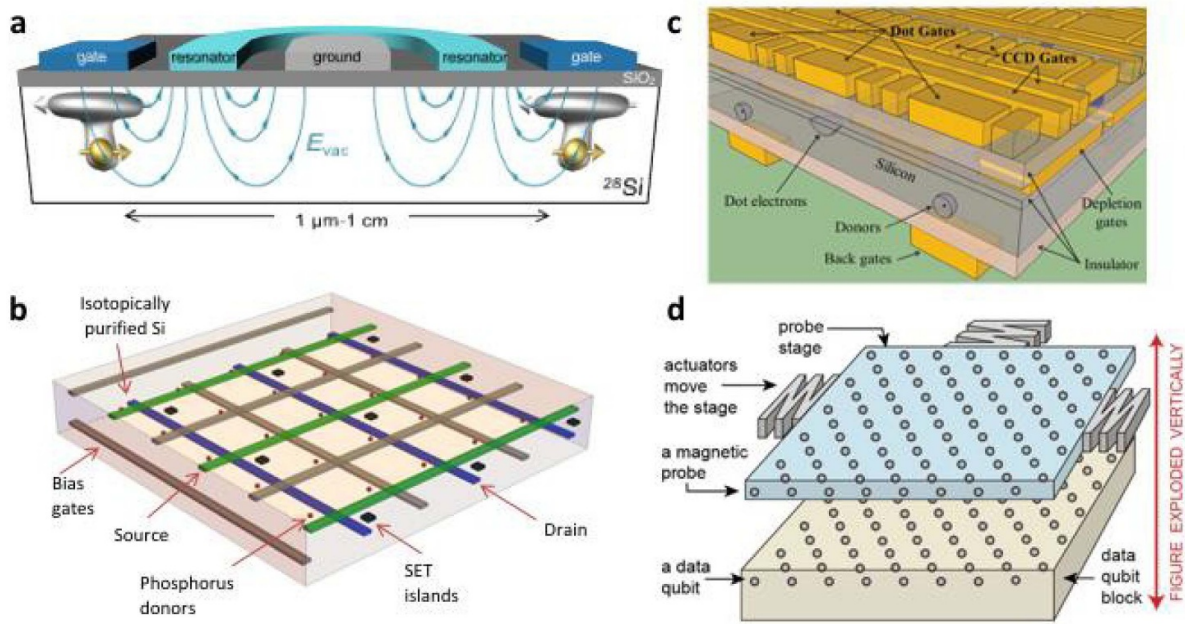

**Figure 13.** (a) The flip-flop qubit [84] promises to extend the range of coupling in a donor-based quantum processor, easing constraints on architecture layout. (b) An architecture by Hill et al [83] hich exploits 3D STM lithography device fabrication and uses a shared control architecture. (c) A proposal by Pica *et al* [85], in which donors are coupled to electrons that can shuttle between adjacent quantum dots. (d) An architecture proposal by O'Gorman *et al* [86], in which mechanically separated 'probe' qubits facilitate parity measurements between data qubits.

*Implementation of QEC*: Ultimately, the goal of a large-scale donor-based quantum computer in silicon is to implement quantum algorithms, such as Shor's factoring algorithm. This will require the application of quantum error correction to combat the detrimental effects of environmental noise. The surface code exhibits some of the highest experimental threshold error rates, lends itself to layout on a silicon surface, and is therefore a natural target for *scalable* designs. As larger numbers of controllable qubits become available, one of the primary goals will be to implement quantum error correction on increasingly large distance codes—ranging from stabilizing a single quantum state (such as a Bell state), and small-scale quantum error codes, to the demonstration of larger-scale codes and QEC primitives such as state injection.

*3.3.3. Advances in science and technology to meet challenges.* In the short term, experiments will focus on characterising the fidelities of two-qubit logic operations. These devices are within the reach of current capabilities.

Moving beyond this to large qubit arrays will require advances in fabrication technology. The STM approach is to extend lithography into 3Ds, moving control gates and readout circuitry into additional planes [83]. The CMOS/ion-implantation devices will target qubit designs that are robust to donor implantation straggle [84], whilst techniques are being developed for rapid deterministic single-ion implantation.

A promising proposal to implement long-range couplings and transport in donor-based quantum processors is the so-called 'flip-flop' qubit [84]. This qubit uses a combination of the electronic and nuclear spin states of a phosphorus donor and can be controlled by microwave *electric* fields—it offers the promise of coupling distances in excess of 150 nm. Due to the strong induced electric dipole, this idea is well suited to exploit cavity QED (cQED) (figure 13(a)), whereby long-range coupling and quantum information transport (over a centimetre) can be achieved through interactions with a high-quality-factor superconducting resonator [84]. Optical interfaces based on rare-earth ions (erbium) or deep chalcogen donors (selenium) are being investigated as ways to achieve quantum information transfer over even longer distances and to connect remote quantum processors.

A number of proposals exist to implement topological quantum error correction with donors in silicon (figures 13(c)–(d)) for realizing scalable quantum computing: Hill *et al* [83] consider a three-level shared control scheme, Pica *et al* propose transferring information by shuttling electrons from donor to donor [85], and O'Gorman *et al* [86] suggest a scheme where data qubits are stationary and coupling is provided by making use of 'probe' spins which are mechanically separated and move to implement the required operations of the surface code.

*3.3.4. Concluding remarks.* Donors in silicon represent a promising pathway to large-scale quantum computation. Single-qubit fidelities are at fault-tolerant levels and the small physical qubit size will allow processors to be fabricated with sufficient numbers of qubits to address difficult tasks such as prime factoring. Challenges facing the field are those of iterative improvement and scale-up: transitioning from the proof-of-principle experiments in the laboratory to larger multi-qubit devices, the development of control electronics, and implementing long-range coupling between qubits.

**Acknowledgments**
The authors would like to thank Lloyd Hollenberg and Andrea Morello for their comments. J J P is supported by an Australian Research Council Discovery Early Career Research Award (DE190101397).

### 3.4. Single-atom qubits: acceptors

*Joe Salfi*

Department of Electrical & Computer Engineering, The University of British Columbia, V6T1Z4, Vancouver, British Columbia, Canada

*3.4.1. Status.* Acceptor dopant atoms have recently been identified as compelling candidates for spin-based quantum technologies. Interest in acceptors ultimately derives from the properties of their acceptor-bound holes (figure 14(A)), where spin–orbit coupling quantizes total angular momentum $J = 3/2$ rather than spin. Under applied magnetic, electric, and elastic fields, different two-level systems can be defined (figure 14(B)) amenable to two-qubit logic over long distances [87–90] and fast single-qubit logic using electric fields [88, 89]. These properties are important to improve the scalability of spin-based technologies, and here derive from spin–orbit coupling, which is comparatively weak for electrons in silicon. Two-qubit operations are predicted to be possible either indirectly, using microwave phonons [87] or microwave photons [88] in CQED or via elastic dipole–dipole [90] or electric dipole–dipole [88] interactions, even while suppressing decoherence from electric field noise [88]. Phonon coupling could enable transducers from microwave to optical photons for quantum networks [87].

Experimental investigation of acceptor-bound holes is underway with B:Si acceptors and is confirming their potential for quantum technologies. The first single atom transistor was demonstrated in an industrially fabricated device [91], exhibiting the $J = 3/2$ Zeeman energy spectrum (figure 14(A)). Readout by spin-to-charge conversion was also demonstrated on an industrially fabricated two-atom device by gate-based reflectometry [92]. Recent materials advances are also paving the way. Isotope purification, which removes random strains in the host, yields narrow linewidths in ensemble continuous wave spin resonance [93]. Recently, pulsed spin resonance has been performed for B acceptors in $^{28}$Si yielding ultra-long 10 ms spin coherence times $T_2$ in a moderate static strain [94], approaching the best results for electron spins. This is highly non-trivial because spin and orbital are mixed in $J = 3/2$ systems, but is a key enabling ingredient making acceptors attractive for quantum technologies.

*3.4.2. Current and future challenges.* The next breakthrough required to establish the suitability of acceptor-based qubits for quantum computing is to couple acceptor qubits in scalable arrays. Two-dopant atom coupling for acceptors has been demonstrated using the exchange mechanism [92, 95]. While useful for two-qubit gates, exchange is short-ranged making it difficult to fabricate 2D qubit arrays and the desired measurement and control devices needed for quantum error correction. Coupling via electric or elastic fields is therefore more attractive. Devices allowing applied strain fields, electric fields, and interaction with interfaces/confinement (figure 14(B)) is predicted to enable control over electric and elastic couplings needed for the long-ranged coupling, via control of energy gaps [87–90] and Rashba-like interactions [88, 89]. Indirect QED-based schemes where interactions are mediated by phonons or photons will require nanomechanical (figure 15(A)) and superconducting (figure 15(B), (C)) cavity design to obtain the desired spin-to-photon or spin-to-phonon coupling, respectively. It also requires integration of acceptor atoms into these cavities, but could allow qubit readout with essentially no overhead.

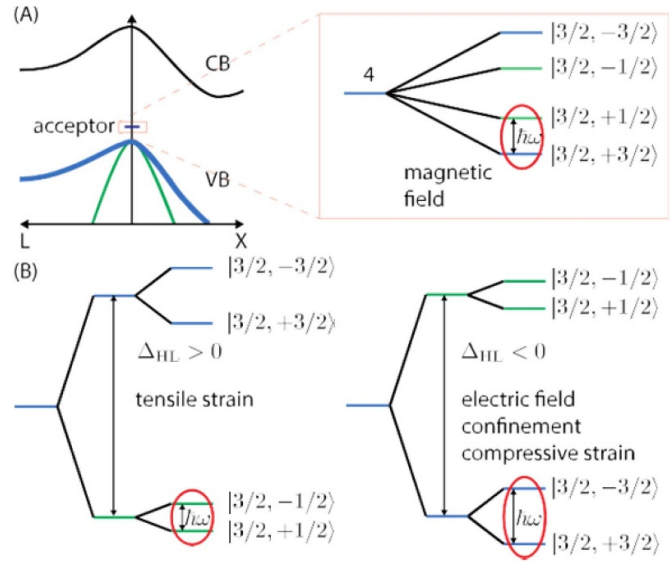

**Figure 14.** (A) Acceptor bound state above the valence band edge. Levels labelled $|J, m_J\rangle$, where $J$ is the total angular momentum and $m_J$ is the projected total angular momentum are split into four in a magnetic field, in a Si crystal. The lowest energy states form a two-level system with a Larmor frequency $\hbar\omega$. (B) Time-reversal symmetric two-level quantum systems are induced when a gap $\Delta_{HL}$ is induced. $|3/2, \pm 1/2\rangle$ is obtained under biaxial tensile strain, and $|3/2, \pm 3/2\rangle$ is obtained under compressive strain, confinement, or electric fields, which couple to electric and elastic fields [87–90].

Another breakthrough would be realizing a complete set of logic gates with fidelities above 99% or higher, as required for large-scale quantum computers that are tolerant to errors. This feat has not yet been accomplished for scalable atomic qubit system; single-qubit gates in other systems have succeeded. The long $T_2$ of acceptors is advantageous for this, because infidelity is bounded by $\tau/T_2$, where $\tau$ is the gate time, which is expected to be very fast [88, 89]. One of the challenges will be to maintain the long $T_2$ times of acceptors when long-range couplings are used. The use of optimal working points ('sweet spots') where $T_2$ is optimized but electric couplings are active have been identified in theoretical work on acceptors is a key enabling concept to achieve this goal [88].

The last key breakthrough mentioned here is to optically interconnect physically separated systems in quantum networks using proposed microwave-to-optical quantum transducers using acceptor dopants [87]. To accomplish this with acceptor atoms, optical structures are needed where the optical modes are sensitive to small mechanical deformations that couple via the spin–phonon interaction.

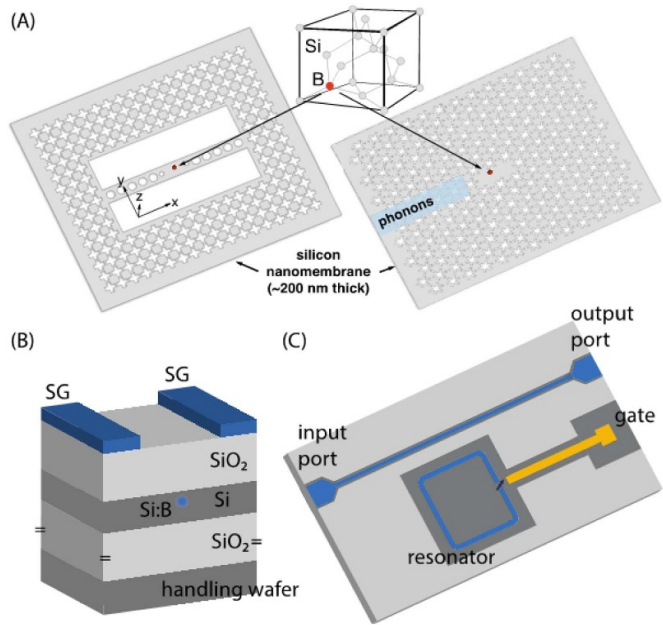

**Figure 15.** (A) Hybrid quantum system of a Si acceptor and a 1D (left) and 2D (right) nanomechanical cavity. An on-chip phonon waveguide could be used to couple to the hybrid system. The elastic field in the nanomechanical resonator couples to the acceptor [87] (B) Si acceptor in a strained silicon-on-insulator structure with side gates SG (left) (C) hybrid quantum system of a Si acceptor and a nanowire ring-based high impedance superconducting resonator. Microwave photons in the resonator couple to the acceptor [88].

*3.4.3. Advances in science and technology to meet challenges.* The above challenges can be met by building acceptor-based qubit systems in a way that leverages the existing process technology and yields devices with the desired characteristics. Like for donors and QDs in silicon, the acceptor platform is backed by the materials and process know-how from the microelectronics industry. Within this context, acceptors are poised to take advantage of recent materials development aimed at extending Moore's law. Indeed, TiN gate materials, which in recent years have been adopted as the gate material in ultra-scaled transistors, have recently been shown to be suitable for building superconducting resonators with high quality factors and high characteristic impedances [96] desirable for QED via electrically mediated interactions with spin qubits. Strain, which enables the enhancement of acceptor $T_2$ to state-of-the-art values, also features in state-of-the-art microelectronic devices. Generally, there are two ways to achieve this, either to use strained SOI (figure 15(a)) or to use the gate material itself, such as TiN, to controllably strain the lattice. There is already an advanced research and industrial fabrication infrastructure in place for silicon-based nanomechanical and nanophotonic structures to build microwave-to-optical quantum transducers for quantum networks with high-quality mechanical and photonic components. Existing silicon-based technologies form a solid basis to investigate scalable quantum information technologies with acceptor qubits.

*3.4.4. Concluding remarks.* Acceptor-based hole spins offer compatibility with established silicon fabrication techniques together with recently demonstrated long spin lifetimes in $^{28}$Si. Their potential for addressable electric spin manipulation and long-distance coupling via electric or elastic fields and makes them compelling new candidates for scalable quantum computers and networks.

**Acknowledgments**

J Salfi acknowledges J T Muhonen for helpful comments and M Khalifa for the schematic of the superconducting microwave resonator. Funding support for this work was provided by the National Science and Engineering Research Council of Canada and the Canadian Foundation for Innovation.

### 3.5. Quantum interface based on photon-spin coherent transfer

*Akira Oiwa*

The Institute Scientific and Industrial Research, Osaka University, Ibaraki, Osaka 567-0047, Japan
Center for Quantum Information and Quantum Biology, Institute for Open and Transdisciplinary Research Initiative, Osaka University, 560-8531, Osaka, Japan
Center for Spintronics Research Network (CSRN), Graduate School of Engineering Science, Osaka University, Osaka 560-8531, Japan

*3.5.1. Status.* Quantum information processing is expected to gain predominance over classical information processing when both quantum computation and quantum communications are involved. Quantum information is transmitted through optical fiber networks, securely connecting to quantum computers and quantum nodes, in order to realize applications that cannot be realized by classical computers and communications. Such quantum networks would provide quantum key distribution, clock synchronization, distributed quantum computation, and other various practical applications, depending on the stage of functionality of the quantum network [97]. However, the optical fiber channel brings loss errors and depolarization errors of single photons, making quantum information transmission imperfect and difficult for long distance quantum communications. Quantum repeaters have been proposed to overcome this difficulty. The quantum repeaters consist of quantum interfaces, which entangle the photons with solid-state qubits, a Bell-measurement scheme and quantum memory at each node. Therefore, the quantum interfaces are indispensable for both basic of quantum physics and applications in quantum information technologies.

Tremendous efforts have been made to realize such quantum interfaces in various physical systems. In solid-state systems, NV centers in diamonds provide a relatively long spin coherence time of electrons and nuclei of the order of milliseconds and seconds, respectively, at room temperature [98, 99].

These are very successful systems for controlling spin states and entangling them with emitted photons [99]. Entanglement of the spin states in remote NV centers over a kilometer has been realized via an optical Bell-measurement [100]. Other quantum systems such as cold atoms, trapped ions, and InAs self-assembled QDs have also been comprehensively studied for use in a quantum interface. However, the ultimate physical system with faithful state controls and highly efficient coupling to photons at an optical communication wavelength is still up for debate. Therefore, further exploration of qubits for quantum networks not restricted to the aforementioned physical systems would create a new research field for novel hybrid quantum systems. Gate-defined QDs, which have been extensively studied as qubits for quantum computing, have great potential for such photon-spin quantum interfaces because of the faithful gate operation and relatively long coherence time.

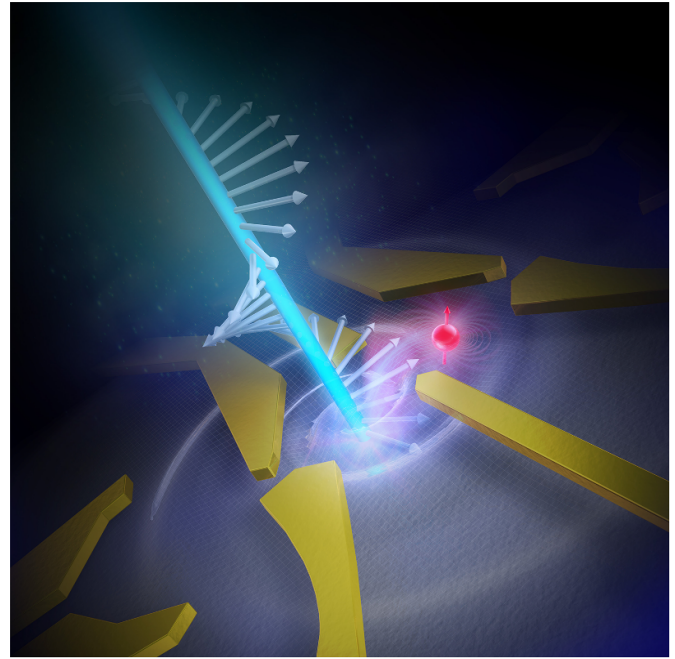

**Figure 16.** Schematic illustration of photon–spin quantum interface using a gate-defined quantum dot.

Importantly, these are also expected to possess compatibility with state-of-the-art optoelectronic devices. In addition, spin-selective interband transition is well established in compound semiconductors, enabling direct coupling to polarization of light.

*3.5.2. Current and future challenges.* As mentioned, spin qubits based on gate-defined QDs have great potential for use in quantum interfaces where they couple to photon polarization states based on the spin-selective Excitation (figure 16) [101]. Toward the quantum interface, the quantum state transfer from the polarization of a single photon to the spin in a gate-defined QD has to be realized at an optical communication wavelength. Sharing entanglement among different quantum computers or end nodes that consist of gate-defined QDs is indispensable in constructing the quantum network. Therefore, the faithful generation of an entanglement between electron spins in two remote QDs has to be achieved using a source of entangled photon pairs, resulting in the quantum teleportation from a photon pair to the electron spin pair, and making an electrical Bell-measurement. Although the idea itself is clear, several technical difficulties must be overcome. For example, the transfer efficiency from photon polarization to electron spin needs to be drastically improved. Otherwise, the system would not be practical to use since the total efficiency of the transfer is the dominant factor affecting the transmission rate of the network. High efficiency is also crucial for proof-of-principle experiments in quantum state and entanglement transfers. To enhance the transfer efficiently, various approaches based on nano-optoelectronic structures are feasible. For example, a photonic nano-cavity consisting of photonic crystals or a distributed Bragg reflector resonator confines photons efficiently and increases the coupling

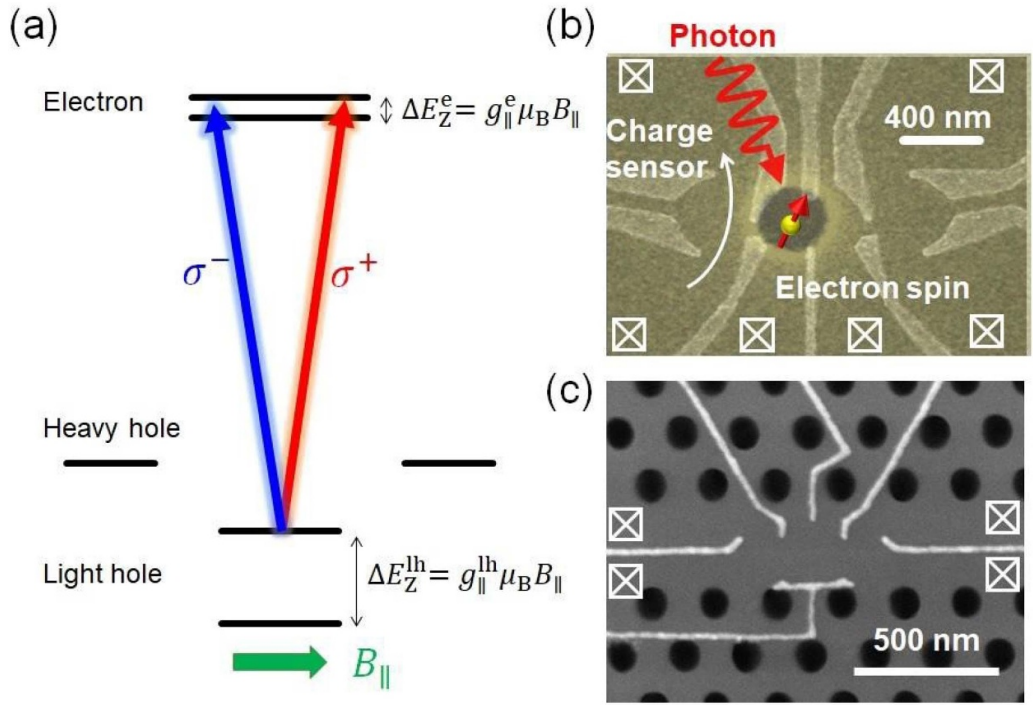

**Figure 17.** (a) Schematic energy diagram of the quantum state conversion via an excitation of a Zeeman–split light hole state in a (001) GaAs quantum well under an applied in-plane magnetic field [101]. Electron and light hole states show Zeeman splitting. (b) Scanning electron micrograph (SEM) picture of a typical gate-defined lateral double QD (DQD) for photon-spin conversion experiments. DQD equips nearby charge sensor QD formed on the left. To prevent excess irradiations on the areas other than QDs, a thick Ti/Au metal mask was fabricated on the surface above the QDs with an aperture. (c) SEM picture of a prototype device of a gate-defined lateral single QD embedded in a nano-cavity formed in a double-hetero type 2D photonic crystal.

between photons and QDs when the QDs are embedded in the cavity.

For the quantum repeaters, faithful Bell-measurement is needed for entanglement swapping at each repeater node. More importantly, quantum memory with a memory time much longer than a millisecond, which is of the order of photon transmission time over 100 km through an optical fiber, is a challenging subject. Development of electrically controllable qubits would lead to an error correction protocol enabling fault-tolerant quantum computing. Introducing a fault-tolerant architecture into quantum networking is one of the goals for future quantum networks [97, 102].

*3.5.3. Advances in science and technology to meet challenges.* Quantum state transfer from single photons to electron (or hole) spins in gate-defined QDs has been progressing for years. The superposition of single photon polarization states can be coherently transferred only to the superposition of single electron spin states in a gate-defined QD by using spin selective interband transition with angular momentum conservation (figure 17(a)) [101, 103, 104]. A very sensitive charge sensor nearby the QD can detect trapping and detrapping as well as spin of a single photoelectron generated in a gate-defined QD as shown in figure 17(b). Indeed, the transfer of photon polarization states to angular momentum and superposition states of electron spin have been demonstrated [105, 116, 116]. Full tomography of a transferred spin state in a QD is within reach. The generation of an entanglement between a photon and a single electron spin in a QD and subsequently between two single electron spins in two remote QDs are needed in order to realize quantum teleportation. Recently, the production of an entangled photon–electron pair in a QD has been achieved using an entangled photon pair, which was created by a spontaneous parametric down conversion [107].

Since gate-defined QDs are based on semiconductor nanodevice technologies, their combination with advanced semiconductor nano-optoelectronics would significantly improve the efficiency for transfers of quantum states and entanglements. Embedding a gate-defined QD in a photonic nano-cavity consisting of photonic crystals shown in figure 17(c) is expected to increase the transfer efficiency from photons to electrons in the QD [108]. The plasmonic effect, which locally enhances the electric field created by light, also provides a route toward higher transfer efficiency [103].

The use of electrically controllable gate-defined QDs offers great benefits in terms of realizing real-world quantum repeaters and quantum networks. Charge sensing of single photoelectrons allows the heralding of the arrival of single photons, enabling heralded entanglement generation. While only one or two of four Bell states are measured in the case of optical Bell-measurement, all four Bell states are expected to be measured when employing electrical means. It has been shown that these features of electrically controllable gate-defined QDs enhance the transmission rate of a spin-based quantum repeater when compared with an optical quantum repeater [103]. Moreover, quantum error correction, which has been comprehensively studied for quantum computing, will also be demanded in order to implement fault-tolerant operations at local nodes in future [97, 108]. To improve the fidelity of qubit operations and maintain a coherence time long enough for quantum memory, which is a crucial constituent of quantum repeaters, the photon-spin quantum interface is likely to be realized in Si- or Ge-based QDs using state-of-the-art Si or Ge nano-photonics. This is because a relatively long coherence time, of the order of milliseconds, has been reported [66].

*3.5.4. Concluding remarks.* Research for quantum networks using spin qubits based on gate-defined QDs has just begun. The aim of a quantum network is to arbitrarily and securely communicate among remote quantum computers in the network and to offer distributed quantum computing as well as other applications that cannot be realized in classical communications. In this article, we have shown that gate-defined QDs have great potential for application in quantum networks. Since almost all proposed quantum networks have assumed purely optical or optically active solid-state qubits, the use of gate-defined QDs provides a novel scheme for constructing quantum repeaters and quantum networks, which is compatible with optoelectronics. However, to reach this ultimate quantum network, the hybridization between different physical systems including superconducting qubits is needed to overcome the disadvantages of each qubit and to improve on their strengths. Finally, we note that not only receivers but also emitters of quantum states would be possible by fully utilizing the benefits of electrical controllability of spin states of the gate-defined QDs.

**Acknowledgments**

Author acknowledges Professor S Tarucha, Professor Y Tokura, Professor S Iwamoto, Dr T Nakajima, Dr T Fujita, Dr H Kiyama, Dr K Kuroyama, Dr S Matsuo, Dr A Ludwig, and Professor A D Wieck for fruitful discussion and cooperation in developing photon-spin quantum interface for years. This work was supported by Grants-in-Aid for Scientific Research S (17H06120), Innovative Areas 'Nano Spin Conversion Science' (Grant No. 26103004), JST CREST (JPMJCR15N2), Dynamic Alliance for Open Innovation Bridging Human, Environment and Materials.

## 4. Nano and opto-mechanics

*4.1. Opto- and electromechanical transduction*

Juha Muhonen[1] and Ewold Verhagen[2]

[1] University of Jyväskylä, Finland
[2] AMOLF, The Netherlands

*4.1.1. Status.* Recent years have seen an explosion of interest towards studying the coupling of electromagnetic fields and mechanical resonators via radiation pressure, down to the quantum level. The extreme sensitivity and control that can be achieved has been put to use in a wide array of applications ranging from detection of gravitational waves to quantum level detection of forces, masses and spins to creation of non-reciprocal optical elements. Often, the coupling is resonantly enhanced by embedding the mechanical element within an optical cavity (creating a cavity optomechanical system). Using a cavity also enables dynamical effects (cooling and amplifications) and versatile control (photon–phonon interaction of swapping and entanglement types) under suitably chosen laser drives [109]. These developments provide the means to usefully incorporate mechanical resonators in quantum technology.

Mechanical resonators are an interesting resource for quantum technology, due to their unique qualities: They can be extremely coherent, allowing significant times over which quantum states can be stored and manipulated, and they couple to a wide variety of other degrees of freedom. A particularly promising aspect is their potential as transducing elements that convert quantum signals from one electromagnetic mode to another, or even to a very different quantum system, such as various types of qubits. Such transduction not only enables their usages as quantum sensors but also as buses, coherent converters, quantum storage elements, and non-reciprocal optical elements capable of bridging and combining with several of the current leading quantum platforms such as superconducting qubits, photonics, trapped atoms or ions and spins in solid state. Some examples of possible quantum applications include using acoustic travelling waves or a mechanical resonator as a quantum link; coherent information transfer from microwaves to optical fields by a mechanical resonator parametrically coupled to both fields; facilitating quantum circuits by coupling two-level systems to the mechanical resonator; storing quantum states in the long-lived phonon states in high-Q mechanical resonators; and creating optical isolators and circulators by introducing non-reciprocity to the optical state transfer. All of these applications are enabled or enhanced by the quantum-level control of mechanical elements allowed by optomechanical interactions.

*4.1.2. Current and future challenges.* Mechanical displacements can strongly alter the response of optical or microwave devices. Thus, micro- or nanoelectromechanical actuation naturally allows for active tuning and switching of components in larger-scale networks [110]. As these systems offer potentially negligible continuous power consumption and cryogenic compatibility, this could be applied in reconfigurable many-mode systems for quantum simulation. Challenges include integration of compliant structures and mitigation of thermo-mechanical noise.

A wider array of possibilities is posed by resonant, coherent control of mechanical systems, which can encode quantum information in their harmonic oscillator modes. Recent studies employing high-quality materials and structural engineering of strain and clamping loss have shown that mechanical resonators can achieve Q-values of the order of $10^{10}$ [111, 112]. Coherence times of seconds seem to be in reach even for GHz-frequency modes, which suffer from small thermal noise at millikelvin temperatures [112]. To create useful quantum memories with these, strong and efficient controlled interaction with other quantum systems is needed. For parametric opto- and electromechanical coupling, a coherent drive field controls and enhances the interaction, supplying the energy difference to convert photons to phonons and back.

If a mechanical mode interacts with two other systems, it can mediate transfer of quantum states from one to the other. Transducers that convert quantum information to optical fields are especially needed, as optical photons are uniquely suited for long-distance transmission of quantum states. Mechanically-mediated microwave-to-optical conversion [113, 114] (see figure 18b) could thus facilitate quantum-coherent communication between e.g. clusters of superconducting qubits.

The control field that assists photon–phonon–photon transfer furthermore controls transfer bandwidth through optomechanical damping [109], and can create synthetic magnetic fields by imprinting control field phase on the transfer (see figure 18a). Indeed, mechanically-mediated optical mode transfer enables useful non-reciprocal functionality including isolation and circulation for optical or microwave photons [115].

Moreover, optomechanical transducers can use direct coupling of mechanical resonators to various qubits, from those based on Josephson junctions to spins of defect centers [116–118] (see figure 18c). Such couplings, if strong enough, leverage quantum non-linearity to create and process non-classical states of acoustic vibrations, making those an inherent part of quantum logic devices. We envision acoustic waves as quantum buses linking different types of qubits on a chip, and using the large Hilbert spaces of mechanical resonators as error correction resources [119].

*4.1.3. Advances in science and technology to meet challenges.* The interaction between a mechanical resonator and a different quantum system is quantified by quantum cooperativity $C_q = 4g^2/(\gamma_i \gamma_m)$, which includes coupling rate $g$ and decoherence rates $\gamma_m$ and $\gamma_i$ of the mechanical and other system, respectively. Achieving cooperativities $C_q \gg 1$ with two other systems simultaneously poses challenges in combining materials and architectures. For optomechanical coupling $g$ is enhanced by laser drive, but this poses a challenge: even small optical absorption can be detrimental in millikelvin environments. Further materials research, pulsing protocols, and thermalization strategies may tackle this problem. For various applications, however, millikelvin temperatures are not a fundamental requirement, as long as $C_q \gg 1$ can be reached with the system in question, as mechanical state initialization could be achieved via optomechanical cooling. Enhancing phonon lifetimes through structural engineering [113, 114] could potentially benefit various mechanical systems.

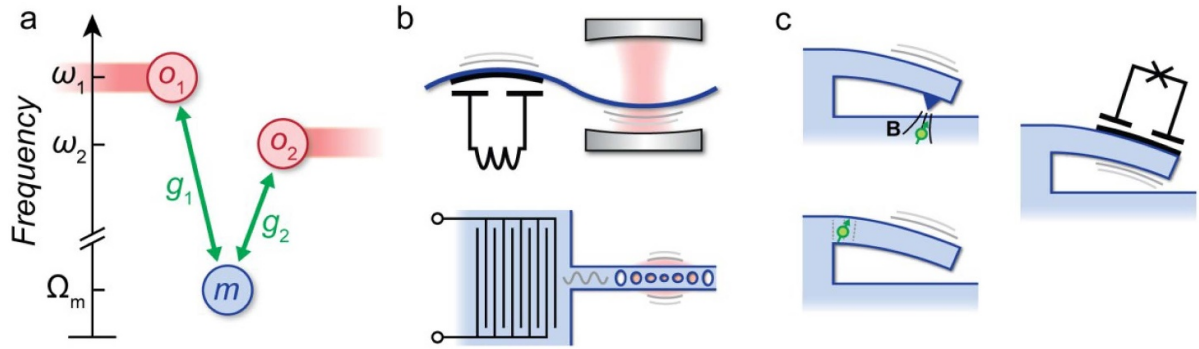

**Figure 18.** (a) Control fields induce coupling (indicated by green arrows) between electromagnetic modes $o_{1,2}$ and mechanical mode *m*, allowing coherent conversion between the em-modes. The conversion can be nonreciprocal as the phase of the coupling rates $g_{1,2}$ is imprinted on the transfer [110]. (b) Schematic of physical systems to implement microwave-to-optical conversion. Top: A vibrating membrane can be coupled parametrically to both an LC oscillator and an optical cavity [111]. Bottom: GHz em fields can be coupled to a nano-optomechanical cavity via surface acoustic waves and an interdigitated transducer. (c) Different coupling mechanisms between a mechanical element and a qubit: magnetic field gradient coupling to a spin qubit, capacitive coupling to a superconducting qubit, and strain-induced coupling to an embedded spin or charge qubit.

Different transducer implementations are being explored: Optical modes that are parametrically coupled to mechanical vibrations of MHz–GHz frequencies in photonic crystal- or larger Fabry-Pérot-like cavities, and GHz fields that are either coupled resonantly to bulk or surface acoustic waves (BAW/SAW) or parametrically to MHz-frequency membrane vibrations [111, 115]. The coupling between qubits and mechanical deformation relies on piezoelectric, magnetic field gradient or strain-induced coupling. Prominent examples are SAW/BAW coupling of superconducting qubits [116] and strain coupling of diamond defect spins [118]. Future advances are expected using different materials and/or qubit systems and geometric design, and there are new efforts to structure and combine materials that are notoriously difficult to process, such as diamond and lithium niobate.

Spatial separation of the different components may help to not let them disturb each other's performance (e.g. metallic circuits and optical cavities). Phononic waveguides can be used to bridge that distance [115]. In this context, developing on-chip acoustic components including mode converters and beam splitters will make on-chip quantum acoustics a versatile technological platform.

The coming years should see a process of natural selection of optimal platforms. Besides bare performance in terms of cooperativity and dissipation, this should be based on factors such as the ability for on-chip and fiber integration, operation bandwidth, and scalability to many reliable components.

*4.1.4. Concluding remarks.* For a long time the goal in the field of optomechanics was to reach the mechanical ground state. Since that was achieved efforts have focussed on improving the quantum coherent properties in order to create truly non-classical mechanical states. We think an especially bright future now lies in combining optomechanical systems with other quantum platforms in order to create new hybrid systems where optomechanics can help to overcome weaknesses in other systems and provide new functionality. The qualities of mechanical resonators make them uniquely powerful as quantum transducers. At the same time the hybrid systems will provide new opportunities to create ultrasensitive sensor devices and study quantum physics and decoherence mechanisms in acoustic systems.

**Acknowledgments**

This work is part of the research programme of the Netherlands Organisation for Scientific Research (NWO), and supported an NWO-Vidi grant, by the European Union's Horizon 2020 research and innovation programme under Grant agreement No. 732894 (FET Proactive HOT), and by Academy of Finland Grant No. 321416

*4.2. Quantum nanomechanics*

M D LaHaye

Syracuse University
Present Affiliation: Air Force Research Lab (Rome, NY)

*4.2.1. Status.* Twenty years ago pioneering nanophysicist Michael Roukes outlined his vision for the ultimate limit of nanomechanical systems: force and displacement sensors operating at the level of zero-point fluctuations; nanomechanical quantum devices enabling new functionalities for quantum information processing; and explorations of energy transport at the nanoscale with single-quantum control [120]. This pronouncement followed on the heels of ground-breaking work performed in Roukes' Caltech lab, led by then post-doc Keith Schwab, which demonstrated in an exquisitely sensitive experiment at milli-Kelvin temperatures the quantization of thermal conductance in suspended nanostructures—a universal quantum property of energy transport [121]. The following year, Roukes and Schwab organized a workshop at Caltech dubbed *Quantum-Electro-Mechanics (QEM),* which focused on the properties and applications of mechanical quantum systems. Thus heralded the unofficial launch of quantum nanomechanical systems.

Among the vibrant exchange of ideas at QEM was a prominent motif: elucidation of the experimental challenges facing the development of quantum nanomechanics: cooling structural modes to low thermal occupation numbers; minimizing interactions with uncontrolled degrees of freedom to reduce decoherence; engineering tight-coupling to quantum-limited detectors; and implementing protocols for observing patently quantum mechanical states of motion.

Amidst this panoply of challenges, incipient experimental results soon emerged. Ultra-sensitive nanomechanical displacement transduction schemes were implemented that enabled the achievement of new milestones in sensing: characterization of quantum back-action noise, nanoscale detection of electronic and nuclear spins, integration of nanomechanical elements with superconducting qubits, nanomechanical mass spectrometry of individual bio-molecules, and the demonstration of dynamical back-action cooling [122, 123].

The field flourished, and its rapid acceleration motivated the organization of QEM-2 in 2006. This workshop included researchers from fields that ranged from quantum information to gravitational wave detection to quantum optics to atomic physics; and it filled three jam-packed days of stimulating talks. Importantly, the success of the event helped launch a regular series of Gordon Conferences on mechanical quantum systems, which has occurred every two years since then and has been an important contribution to the field's growth.

As a result of the collective effort and competition, many of the initially identified challenges have been surmounted; and much of Roukes' original vision has become reality: reliable techniques for ground-state cooling, quantum state generation, and quantum-limited displacement detection have all been developed; applications of nanomechanics to quantum computation, communication, and sensing are being pursued;

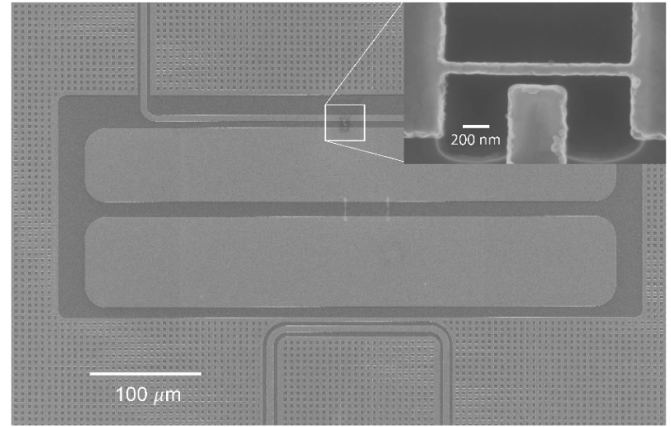

**Figure 19.** Example of an ultra-high frequency nanomechanical device (inset) coupled to a superconducting transmon qubit (main). In this system, the third in-plane flexural mode of the suspended nanostructure, which has a resonance frequency of 3.47 GHz, couples most strongly to the qubit. An applied DC voltage (on the order of volts) between the metallized nanoresonator (aluminum) and the transmon establishes the coupling between motional and charge degrees of freedom [124].

and fundamental explorations of motion, entanglement, and quantum thermodynamics are within reach. At a more general level, quantum nanomechanical devices can now be thought of as veritable tools for integration in hybrid quantum platforms and technologies. Thus, as the new quantum revolution unfolds, quantum nanomechanics will help drive it as well as evolve with it.

*4.2.2. Current and future challenges.* In this section, we outline three of the leading quantum nanomechanical architectures and highlight a primary challenge for each that they face for future development.

Two of the first major benchmarks to be reached on the road to quantum nanomechanics—demonstration of mechanical energy quantization and quantum coherent manipulation of a mechanical resonator's state vector (see ref. [123] and references within)—were demonstrated using a quantum electromechanics architecture that involves integration of mechanical elements on chip with superconducting qubit technology, an idea originally proposed by Schwab nearly 20 years ago (see references within ref. [122]). This approach, which can be thought of as a mechanical analog of certain light-matter interactions explored in CQED and cQED, has been long recognized as a promising and versatile platform for nanomechanical quantum state generation and measurement for both fundamental pursuits and applications in quantum information processing [122, 123]. In fact, in recent years, novel mechanical designs and advances in cQED have catalyzed multiple approaches to improve upon initial qubit-coupled mechanical systems, whose principle short-comings included relatively short device coherence times, which greatly limited their utility for quantum state manipulation, and thus precluded more advanced experiments and functionality. Leading approaches today involve a variety of different nanomechanical elements—flexural devices (figure 19), SAW resonators, bulk acoustic cavities, and phononic crystals—integrated with long-coherence time superconducting qubits (such as the transmon). These advanced designs have pushed qubit-coupled architectures to the brink of the strong-coupling regime, where the mutual interactions between circuit and mechanical degrees-of-freedom dominate over losses to the environment. However, realization of the full potential of these systems will require further increases in interaction strength without comprising device quality.

A second transformative nanomechanical technology is based upon the integration of nanomechanical membranes as compliant parallel-plate capacitors in lumped-element superconducting LC circuits (see ref. [123] and references within). The innovative membrane architecture utilizes radiation pressure in the microwave LC resonator for dynamical back-action cooling of the mechanical modes to temperatures below what standard cryogenic refrigeration can achieve; this enables the cooling of sub-GHz-frequency mechanical modes near their quantum ground states. Additionally, the same electromechanical interactions that are exploited for cooling can be utilized for quantum measurement, entanglement generation, and the transfer of quantum information between electrical and mechanical domains. Consequently, this versatile system has also become a platform for quantum nanomechanics, ushering in a steady string of groundbreaking results including the generation of quantum squeezed states of motion, implementation of mechanical quantum memory, and use in microwave-to-optical transduction techniques. This platform will likely also play a critical role in near-term quantum information technologies, particularly for efficient bidirectional data conversion between superconducting processor elements and photonic interconnects for quantum networking applications. However, the radiation-pressure-mediated interactions are generally very weak, requiring the application of large photon numbers for cooling; this also translates into relatively low conversion efficiency for microwave-optical interfacing. A primary future challenge will thus be to identify clever designs, materials, or measurement strategies to enhance these interactions.

The final cutting-edge quantum nanomechanical systems that we highlight here are optomechanical crystals (see ref. [123] and references within). These suspended nanofabricated structures are patterned to support co-localized optical and phononic modes in confined dimensions. The tight overlap of the optical and mechanical modes allows one to utilize radiation pressure to achieve large optomechanical coupling, which enables the use of the same set of dynamical back-action tools explored in circuit mechanics for ground-state cooling, entanglement generation, and quantum feedback control. These systems have the additional advantage that they can operate at telecom wavelengths, can be integrated on chip with photonic circuitry, and connected via low-loss optical fiber interconnects, which could facilitate their role as quantum interfaces for optical networks and access to efficient and robust quantum optical tools. Thus they are also ideal candidates for an array of functions in quantum networking and sensing. Progress towards this direction has been marked by high profile results in recent years including, perhaps most notably, the remote entanglement of nanomechanical modes of two optomechanical crystals separated by an optical distance of 70 m [125]! One drawback, which needs to be surmounted for further development as quantum interfaces, particularly for interfacing between superconducting and photonic domains, is that photonic wavelengths are anathema for superconducting technologies: the high energy photons excite quasiparticles inside the superconducting films which leads to dissipation and other deleterious effects.

*4.2.3. Advances in science and technology to meet challenges.* The state of quantum nanomechanics today has derived in large part from the cross-pollination of ideas between distinct fields and the proliferation of proposals to implement hybridized systems and techniques. Prominent examples over the arc of the field include [123] the integration of single-charge electronics and Josephson-junction-based devices with nanomechanical systems for transduction; the merger of mechanical elements with superconducting microwave resonators and photonic circuitry; and the mapping of concepts from atomic physics and gravitational-wave detection to the electromechanical-circuit domain; This cross-disciplinary ethos has enabled the field to thrive and it will propel the field over current and future challenges.

In this spirit, several viable routes to explore for the enhancement of electromechanical coupling strengths, both for the case of qubit-mechanics and circuit-mechanics, can be found in recently demonstrated architectures. One example is the *electromechanical crystal* [126], a close cousin of the optomechanical crystal, wherein suspended nanostructures are patterned to engineer a compact lattice with RF phononic modes that can be tightly coupled with superconducting transducer circuitry. The resulting novel dispersion relation contains two important features which enable access to the strong electromechanical coupling regime and quantum level transduction: the existence of a mode where each of the unit cells oscillates in phase ($k = 0$ mode), effectively multiplying the coupling to transducer circuitry by the number of cells; and the implementation of a phononic band gap shield, which acoustically decouples the electromechanical crystal from the surrounding substrate, resulting in a large mechanical mode quality factor. Using a similar design (and reasoning), nano-electromechanical crystals made from piezoelectric materials have very recently been integrated with superconducting qubits demonstrating a likely path to strong electromechanical qubit coupling. It should be noted that phononic band gap shields have recently enabled mechanical quality factors in excess of $10^{10}$ [114], resulting in relaxation times on the order of milli-seconds and seconds—establishing this design technique as a viable tool for nanomechanical quantum memory and on-chip quantum acoustic communication channels

An additional promising route to increasing coupling between circuit and mechanical degrees of freedom, while decreasing channels of dissipation, involves the use of 3D integration. Typically planar architectures have been used for circuit-nanomechanics and qubit-nanomechanics; but a recent demonstration utilizing a nanomembrane in a confined 3D-gap cavity achieved the quantum strong coupling regime of mechanics [127]. This opens the door to new designs to minimize the participation ratio of lossy interfaces, as well as integration with 3D superconducting qubit architectures [128], which would have strong prospects for scalability.

On a final note, both electromechanical crystals and 3D integration techniques—whether using 3D cavities, wafer bonding, or wafer vias—also provide a new parameter space to explore for integrating quantum electromechanics with photonic circuitry, while incorporating the proper isolation to avoid the quasiparticle poisoning of superconducting circuitry from optical photons.

*4.2.4. Concluding remarks.* If the remarkable trajectory of quantum nanomechanical systems is any indication, these systems will continue to become more coherent and more sophisticated; and they will take on prominent roles in heterogenous quantum information, simulation, and sensing platforms; as well, they will provide new opportunities for fundamental studies of entanglement and thermodynamics in complex quantum devices. Some of the challenges on these new horizons can already be identified: the enhancement of optomechanical and electromechanical interactions using metamaterial designs [129] and 3D integration of nanomechanics with superconducting cavities and photonic circuits [130]; exploration of room temperature nanomechanical quantum control; and the development of nanomechanics as a bridge between the atomic world and quantum circuit domains. With the quantum regime of motion outlined by Roukes two decades ago now coming sharply into focus, it is fascinating to ponder what lies beyond this regime and what the next generation of quantum nanomechanical systems will deliver.

## 5. Low-dimensional systems

*5.1. Quantum tunneling devices incorporating 2D magnetic semiconductors*

Hyun Ho Kim and Adam W Tsen

Institute for Quantum Computing and Department of Chemistry, University of Waterloo, Waterloo, Ontario N2L 3G1, Canada

*5.1.1. Status.* Research in 2D materials has experienced rapid growth in the past few years. In particular, various layered compounds exhibiting quantum phenomena, such as superconductivity [131] and magnetism [132], have been isolated in atomically thin form, often in spite of their chemical instability. The nature of the 2D phases can be different than their bulk counterparts, making such systems attractive for fundamental studies. Owing to their high crystallinity and absence of dangling bonds, devices and heterostructures incorporating these materials may also show performance exceeding that of traditional films. In this roadmap article, we focus on a few recent developments in spin-based quantum devices utilizing the 2D magnetic semiconductor, $CrI_3$.

The 2D ferromagnetic (FM) or antiferromagnetic (AFM) compounds that have been reported thus far are: $MPS_3$ (M = Fe, Ni) [133, 134], $Cr_2Ge_2Te_6$ [135], $CrX_3$ (X = I, Br, Cl) [136–140], $Fe_3GeTe_2$ [141], and 1 T-$VSe_2$ [142]. Since it has been rigorously proved that the 2D Heisenberg model can support neither long-range FM nor AFM order at finite temperature [143], magnetism in a monolayer should be anisotropic. Specifically, magnetism survives in monolayer $FePS_3$, $CrI_3$, $CrBr_3$, and $Fe_3GeTe_2$ owing to an out-of-plane easy axis for spin polarization. With the exception of metallic $Fe_3GeTe_2$ and 1 T-$VSe_2$, all exhibit semiconducting or insulating behavior: increasing resistance with decreasing temperature, and all have a magnetic transition temperature ($T_c$) below room temperature. Finally, the magnetic ions in all compounds are FM coupled within the layers, except for AFM $MPS_3$.

$CrI_3$ is particularly intriguing in that bulk and 2D forms exhibit distinct magnetic phases. Due to absence of a structural stacking transition at higher temperature, below $T_c \sim 45$ K, the magnetic coupling between adjacent layers is AFM in thin samples [144], in contrast with FM coupling in the bulk. This property, combined with the Ising-type FM spin ordering within the layers, gives rise to various impressive effects when incorporated in a device heterostructures. Several groups have used few-layer $CrI_3$ as a spin-dependent barrier for quantum tunneling between graphene electrodes [138, 145–148]. The application of a 2 T magnetic field out-of-plane has been found to produce abrupt TMR as large as $10^6$%. As spins in individual layers become aligned with the field, the tunnel barrier is effectively lowered, yielding an exponential rise in tunneling current (see figure 20(a)). While the magnitude of the effect decreases in thinner samples (see figure 20(b)) [145–147, 149], even in bilayer $CrI_3$ the TMR value is over twice the value than that seen from conventional EuS barriers

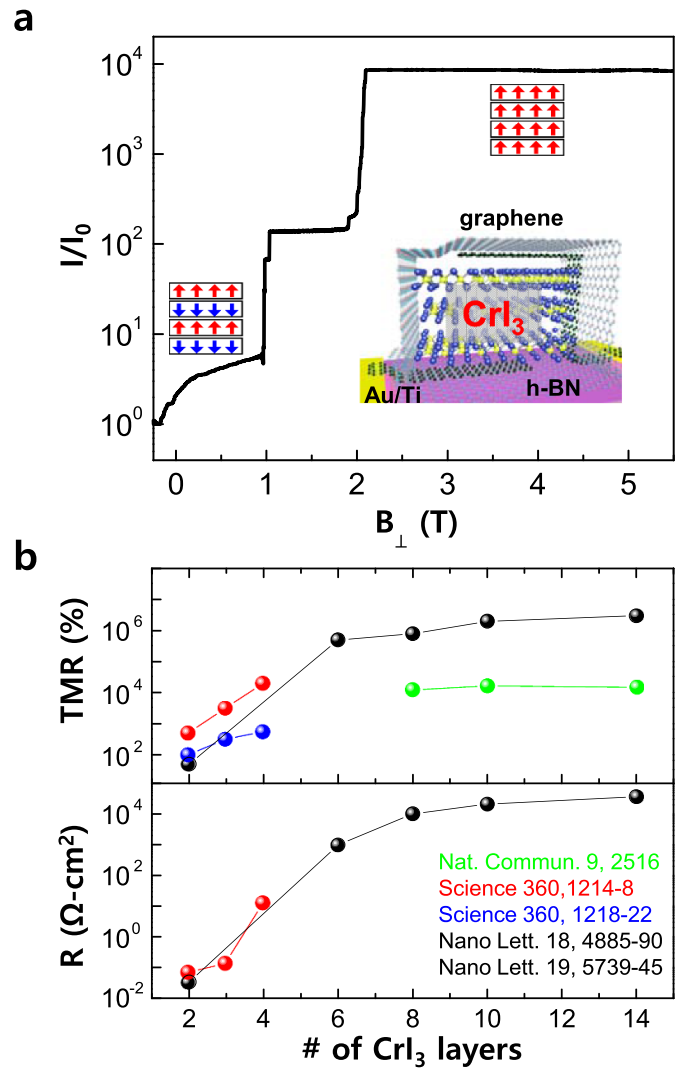

**Figure 20.** Extremely large tunnel magnetoresistance (TMR) across ultrathin $CrI_3$. (a) Change in tunneling current in few-layer $CrI_3$ device as a function of out-of-plane magnetic field at optimal voltage biasing and 1.4 K. Inset shows a schematic illustration of the device. Figure reproduced with permission from: a [148], American Chemical Society. (b) Thickness-dependent TMR (top) and area-normalized DC junction resistance at optimal voltage biasing (bottom).

[150]. Such devices can potentially serve as spin filters or building blocks for magnetic memory.

Using graphene as an electrostatic gate and hexagonal boron nitride (hBN) as the dielectric, two groups have been able to tune the magnetic properties of 2D $CrI_3$ [151–153], thus continuing with a longstanding effort in spintronics to achieve electrical control of magnetism. In monolayer $CrI_3$, doping with hole densities of several $10^{13}$ cm$^{-2}$ can enhance the saturation magnetization and increase $T_c$ by nearly 10%. This is not immediately expected as the exchange interaction in $CrI_3$ is not mediated by itinerant carriers. In bilayers, both doping and a pure electric field generated from two opposing gates can switch the interlayer coupling between AFM and FM ordering [151–153]. Combining this with tunnelling contacts can further allow for gate-tunable TMR characteristics [154, 155]. These experimental highlights demonstrate the versatility and potential of 2D magnetic semiconductors for spin-based quantum devices.

*5.1.2. Current and future challenges.* Despite this recent success, there are a number of challenges to be addressed in order to make such systems more technologically relevant. The most obvious limitation is that all the 2D magnetic semiconductors reported so far have a $T_c$ below room temperature, and so will not yet be suitable for practical devices. Another issue is that the conductance of tunnel devices is rather low, thus limiting their possible switching speeds. In figure 20(b), we have plotted the DC junction resistance at the voltage bias for peak TMR as a function of $CrI_3$ thickness for all the devices reported so far in the literature normalized to their area. There is a clear trade-off in that both the resistance and TMR decrease substantially with decreasing thickness, as is expected for tunneling. Yet, even in bilayers, the resistance is ~$10^{-2}$ $\Omega$ cm$^2$, larger than that for commercial magnetic tunnel junctions (~$10^{-4}$ $\Omega$ cm$^2$). Finally, for memory applications it is desirable to be able to switch between the resistive states with extremely small magnetic fields obtainable from on-chip circuit elements (~1 mT). The interlayer magnetic coupling in the $CrX_3$ family yields much larger critical fields (~1 T), however. While doping can be used substantially reduce this value, it appears that the interlayer AFM ground state is also destabilized [151].

*5.1.3. Advances in science and technology to meet challenges.* While the relative low $T_c$ is an inherent limitation of the material, there have been several reports predicting other 2D magnetic semiconductors at room temperature [156, 157]. As far as we know, they yet await experimental realization. The resistance of the tunnel junctions can be improved, in principle, by selecting metal electrodes with a lower work-function, such as aluminum. Unfortunately, materials such as $CrI_3$ are not directly compatible with conventional fabrication procedures as they will quickly degrade in the air environment. Graphene contacts have been used as a work-around as layered heterostructures can be fully assembled in inert atmosphere. This may actually speak to a larger problem, and thus prompt the development of fabrication tools that are miniaturized in gloveboxes. Finally, the critical switching fields can be lowered by weakening the strength of the interlayer AFM coupling. This may be potentially achieved by chemically changing the interlayer spacing or even re-stacking monolayer samples with a controlled twist angle, similar to what has been demonstrated for graphene and other 2D materials.

*5.1.4. Concluding remarks.* Overall, the relatively young field of 2D magnetism has already led to many exciting results. The TMR physics and electrical control of magnetism may be of significant fundamental interest for the materials and device communities. Nevertheless, it remains to be seen whether the many technical hurdles can be overcome to make such systems more appealing for applications.

*5.2. Topological states*

Dimitrie Culcer[1,2] and Attila Geresdi[3]

[1] School of Physics, The University of New South Wales, Sydney 2052, Australia
[2] Australian Research Council Centre of Excellence in Low-Energy Electronics Technologies, UNSW Node, The University of New South Wales, Sydney 2052, Australia
[3] QuTech and Kavli Institute of Nanoscience, Delft University of Technology, 2600 GA Delft, The Netherlands

*5.2.1. Status.* Topological phases are characterised by a topological invariant that remains unchanged by deformations in the Hamiltonian. Materials exhibiting topological phases include topological insulators (TI), superconductors exhibiting strong spin–orbit coupling, transition metal dichalcogenides, which can be made atomically thin and have direct band gaps, as well as high mobility Weyl and Dirac semimetals. In 2D topological materials, the electron gas on the surface has enabled spectacular phenomena such as the spin–orbit torque: a current through a TI can flip the magnetisation of an adjacent ferromagnet even at room temperature [158, 159]. A power-saving topological transistor harnesses edge states that conduct electricity without dissipation and are responsible for the observed quantised spin and anomalous Hall effects. The first topological transistor design exploits the fact that a top gate drives the system between an insulating normal phase and a topological phase with a quantised conductance [160]. Non-linear electrical and optical effects have taken off, with grand aims including identifying a Hall effect in time-reversal symmetric systems [161, 162] and a direct photocurrent, the *shift current*, which could enable efficient solar cell paradigms.

Topological quantum computing is another motivation to investigate this field. The current physical realisations of quantum bits invariably suffer from finite control fidelity and decoherence due to interaction with the environment. While error correction schemes have been developed to tackle these challenges, topologically protected states can mitigate it by harnessing their degenerate ground state manifold. Topologically protected quantum operations on this manifold require quasiparticles with non-Abelian exchange statistics, which emerge in various engineered nanostructures where electrons are confined to one or two dimensions. Importantly, it is this reduced dimensional behaviour that enables non-Abelian exchange statistics [163]. Specifically, planar semiconductor heterostructures hosting the $\nu = 5/2$ fractional quantum Hall state [163] and spin–orbit coupled nanowires with induced superconductivity in an external magnetic field [164, 165] are widely investigated platforms (figure 21). In these devices, quasiparticle tunnelling [166] and Josephson effect experiments [167] resulted in signatures consistent with the presence of topologically protected electron states. However, no unambiguous experimental confirmation of the non-Abelian exchange statistics exists to date. This next step, which is a prerequisite of topological quantum computation, relies on progress in materials science and the development of readout schemes.

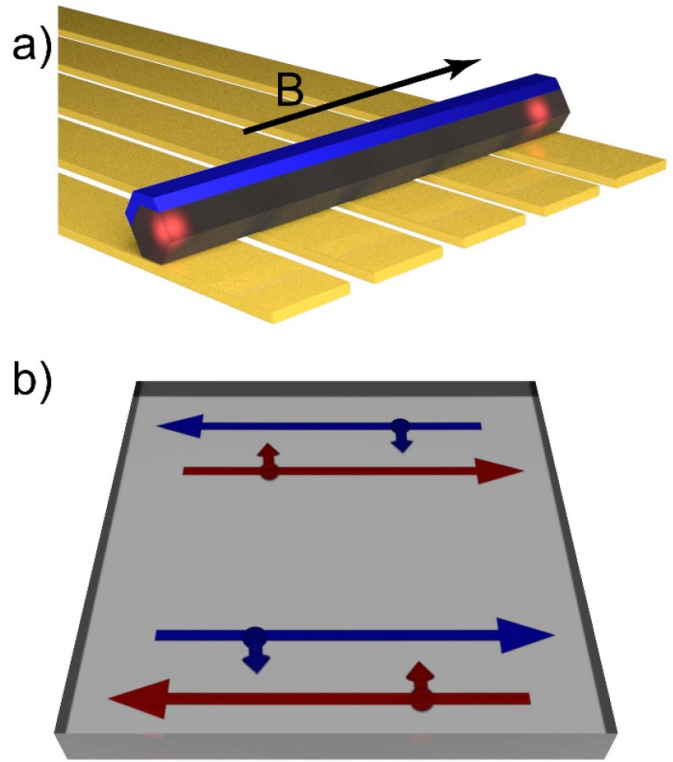

**Figure 21.** Fundamentals of topologically protected electron states. (a) Majorana zero modes (in red) are localised at the ends of a semiconductor nanowire (in grey) with superconductivity induced by thin superconducting layers (in blue). The chemical potential can be adjusted by local electrostatic gates (golden stripes), and an external magnetic field B is applied [164]. (b) The quantum spin Hall (QSH) state, where spin-momentum locking of the edge modes (red and blue arrows) prevents backscattering in the absence of time-reversal breaking perturbations. Note that in the case of the quantum anomalous Hall effect (QAHE), there is only one spin-polarised edge mode (either blue or red) and there is no backscattering whatsoever.

*5.2.2. Current and future challenges.* Because topological transistor research focuses on the dissipationless transport at low temperatures, while spin–orbit torque devices aim for room-temperature operation, significant gaps exist in our overall understanding of topological material devices. The origin of the experimentally observed strong spin–orbit torque, whether from electrically generated spin polarisations on the surface or spin currents in the bulk, is unclear. For topological transistors, the challenge is to make the threshold voltage as small as possible, below that of conventional transistors. Magnetic ordering, critical for certain topological transistors and for room-temperature operation, is also poorly understood in topological materials. For example, known TIs are layered materials, and it is unknown whether intra-layer or inter-layer magnetic interactions dominate, while different mechanisms may be responsible for magnetic ordering on the surface and in the bulk, resulting in different critical temperatures. Moreover, whether the edge states of TIs are topologically protected, remains controversial. Backscattering of edge states, which destroys the conductance quantisation, can be induced by impurities, either directly or by inducing coupling to bulk states, or by size quantisation effects: the edge itself may experience spontaneous time-reversal symmetry breaking due to edge reconstruction. In extreme cases this leads to Anderson localisation of the edge states. The surface states themselves are sensitive to the metallic contacts on the TI.

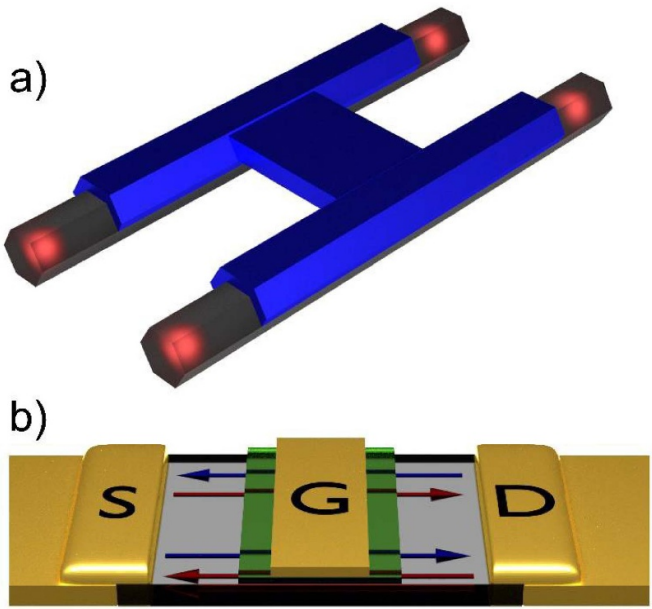

**Figure 22.** Topological device concepts. (a) The Majorana box qubit [171], where the quantum information is encoded in the joint parity of four Majorana states (red spheres). The nanowires are connected via a superconducting bridge (in blue) in order to form a single superconducting island, yet prevent parity leakage. (b) The topological spin transistor, which relies on the gate-tunability of the topological phase transition, recently demonstrated in [160].

Currently, topologically protected electron states are probed by low frequency electronic transport, and high frequency techniques, such as charge sensing, shot noise experiments and the AC Josephson effect. All of these experimental techniques yield signatures of the topologically phase transition, however these signatures were theoretically shown to persist without any topologically protected edge state as well [168]. Future measurements should thus focus on the non-local nature of the topological ordering instead of the local mapping of the edge modes.

A common requirement for topologically protected quantum electronics is the coexistence of a bulk gapped state with the protected edge modes. This requirement demands a careful engineering of materials. Steps in direction have already been made, such as superconductor gap engineering, intentional doping [169] and atomically clean heterointerfaces.

*5.2.3. Advances in science and technology to meet challenges.* While promising for applications, harnessing topological protection with a technological relevance requires further progress to address the limitations and challenges discussed above. Reducing the threshold voltage of a topological transistor is expected to be a matter of identifying the optimal materials, while recent work suggests that the operating temperature of topological transistors can be enhanced by compensated $n-p$ co-doping [170]. Further experiments are needed to resolve the issue of topological protection, since not enough devices exist to determine whether it is fundamental or not, and theoretical predictions of Anderson localisation have not been confirmed or denied. Understanding charge and spin dynamics in the vicinity of interfaces between TIs and ferromagnets is vital in interpreting experiments. Whereas these can be simulated using state-of-the-art computational approaches, they also require conceptual advances in transport theory, considering fundamental issues such as the definition of the spin current when the spin is not conserved.

In materials science, required progress includes the growth of clean materials with tailored band parameters to optimize the topological energy gap separating the ground state manifold and trivial excited states. Heterointerface engineering will enable tuning of proximity effects of superconductivity, magnetic ordering or spin–orbit coupling at interfaces, leading to true designer material systems. These developments, which involve complex charge redistributions and modifications of the electronic structure, will require synergy with advanced numerical modelling methods, where progress is limited due to the numerically expensive calculation of geometries including heterostructures.

While topological device schemes have already been suggested (figure 22), further investigations are required to address connectivity issues in scalable systems. Specifically for topological quantum bit schemes, existing proposals often neglect electrostatic gating and auxiliary structures required for state preparation and readout.

Finally, advances in quantum algorithms are required to consider specific topological quantum hardware and has to optimize the usage of topologically not protected quantum gates, such as non-Clifford gates for Ising anyons [163]. On the other hand, if experimentally observed, more complex topological states, such as Fibonacci anyons can circumvent this requirement, as they enable universal and topologically protected quantum computation [163].

*5.2.4. Concluding remarks.* Topological states form an active field with tremendous promise in fundamental science and conventional and topological quantum computing, which will continue to grow. In transport devices, where attention has focused on ferromagnets, antiferromagnets are evolving at a fast pace, and skyrmions offer new avenues for exploiting topological effects. Spin–orbit coupling offers new functionalities for magneto-resistive devices, such as spin valves, since electrons travelling in a specific direction have a fixed spin orientation determined by their momentum. In non-linear response much work remains to be done in understanding the interplay of topological mechanisms with disorder and phonons, and developing a unified theoretical description in terms of doping, disorder correlations and inter-band coherence. Topologically protected quantum information processing, once demonstrated, will have a major impact on scalable quantum technologies, and will provide new directions for the development of quantum algorithms. Finally, there is a lot of space to investigate layered van der Waals heterostructures where experiments are just beginning.

**Acknowledgments**

D C is supported by the Australian Research Council Centre of Excellence in Future Low-Energy Electronics Technologies (project CE170100039) funded by the Australian Government. A G acknowledges funding by the European Research Council under the European Union's Horizon 2020 research and innovation programme, Grant No. 804988 (SiMS).

# 6. Molecular devices

*Jan A Mol*

Queen Mary University of London

*6.0.5. Status.* First proposed in 1974 [172], the idea of using individual molecules as the functional building block of electronic devices has prompted decades of research into understanding and controlling charge transport down to the single-molecule level. These efforts have not only led the elucidation of fundamental quantum transport phenomena at the atomic and molecular scale, but also to the demonstration of basic electronic components, including diodes and transistors, based on rational molecular design [173]. While molecular device technologies have not yet made the transition from the laboratory to R&D departments, major advances have been made in the underpinning science of single-molecule electronics. Recent studies have focussed on the role of quantum interference in conjugated molecular systems and have highlighted the role of electronic and vibrational degrees of freedom in heat and charge transport. Going beyond electronic transport properties, rational molecular design allows for the engineering of optical, magnetic and quantum effects that are not readily achievable in lithographically defined nanostructures such as solid-state QDs [174]. If these effects can be harnessed in robust and reproducible molecular devices, they will pave the way towards novel information processing, energy harvesting and quantum technologies. Finally, significant progress has been made in the investigation of individual biomolecules. Electronic devices that read the sequence of single-molecule DNA are commercially available, and further advances in single-biomolecule sensing could revolutionise healthcare and data storage [175].

*6.0.6. Current and future challenges.* From the outset, the development of single-molecule electronics has been hampered by the tremendous challenge of making reliable and stable contacts between nanometre-sized molecules and the gold electrodes that connect them the macroscopic outside-world. Without atomically precise molecule-metal contacts, randomness in coupling strength between the molecule and the electrodes lead to large device-to-device variability. Moreover, the high atomic mobility of gold at room temperature limit the long-term stability of single-molecule devices [177]. Traditionally, these issues have been addressed by repeated forming and breaking single-molecule junctions. The resulting ensemble averaged measurements indeed uncover properties that are inherent to the molecule under investigation, however owing to the variability from one measurement to the next they do not constitute any real single-molecule device functionality. Lithographically-defined carbon or platinum-based electrodes have the potential to provide a scalable platform for contacting individual molecules (see figure 23), however they currently suffer from low yield and reproducibility and will require further improvements in coupling chemistry [176].

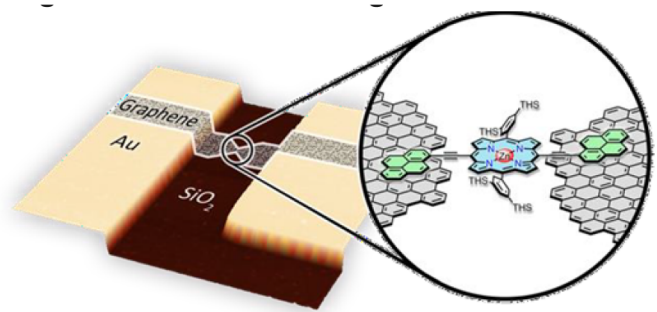

**Figure 23.** Schematic depiction of a graphene-based single-molecule transistor. An individual molecule is anchored to two graphene nanoelectrodes via π-π stacking anchors. Adapted from [176].

In addition to achieving reproducible and stable molecule-electrode coupling, there is a strong need for controlled alignment of the molecular energy levels with respect to the Fermi energy of the electrodes, as this determines the electrical conductance of the molecular device as well other properties including its thermoelectric heat-to-energy conversion rate. Level-alignment can be achieved to a certain extend via rational molecular design, however the most successful approach to date has been to use electrostatic field-control in a transistor-type geometry or gating via an ionic liquid [178]. A key parameter in these single-molecule transistors is the so-called lever arm: the shift of the molecular energy level per volt applied to the gate, which can be improved by the use of high-k gate dielectrics and electrode geometries that reduce screening of the gate electric field. Similar to the molecule-electrode coupling strength, randomness level-alignment leads to device-to-device variability and its origins need to be further understood and mitigated.

*6.0.7. Advances in science and technology to meet challenges.* For a long time, it seemed that the advantages of atomically engineering nanodevices via rational molecular design were being largely negated by the lack of atomically precise electrodes. However, recent advances in top-down and bottom-up nanofabrication are now enabling these kinds of electrodes. Atomically precise nanofabrication has already come to fruition in the field of donor-based quantum computation. Scanning tunnelling microscopy-based nanofabrication techniques could also be leveraged to create electrodes for single-molecule junctions. Similarly, transmission electron microscopy provides a route towards sculpting material with atomic resolution [179]. While these atom-by-atom fabrication techniques do not yet have deliver devices on a technologically relevant scale, they could enable new experiments where the position of each atom in a single-molecule junction is known by design.

Bottom-up approaches, where not only the molecule under investigation but also the electrodes are created via synthetic chemistry methods, offer an alternative route towards atom-by-atom fabrication of single-molecule devices. Solution processable carbon-based electrodes including carbon nanotubes and graphene nanoribbons could provide a scalable means of fabricating large number of chemically identical single-molecule devices that are macroscopic enough to be wired up to the outside world without introducing too much device-to-device variability [180].

Finally, ensemble-averaged single-molecule junction measurements can be achieved not only by repeatedly forming and breaking an individual junction, but also creating structures containing many single-molecule junctions in parallel. Device-functionality in molecular-linked nanoparticle networks arises from the properties of the individual single-molecule junctions linking the network, yet any randomness due to the uncontrolled molecule-metal coupling is averaged out by virtue of having transport through many single-molecule junctions at the same time. In addition to proving the nodes in the network, the nanoparticles can also provide additional device functionality, for example by acting as plasmonic antennas.

*6.0.8. Concluding remarks.* Molecular devices have not had the technological impact once predicted. While rational molecular design enables engineered nanostructures with electronic, magnetic, and vibrational properties that are not otherwise achievable, connecting to these nanostructures has remained challenging. Novel top-down and bottom-up approaches have the potential to bridge the gap between nanometre-size molecular functional building blocks and the macroscopic outside-world. Scalable, robust, and reproducible molecular devices could perform a host of tasks, from rectifying an electrical current as first envisaged by Aviram and Ratner in 1974, to performing quantum computation and determining the amino acid sequence of proteins.

**Acknowledgments**

J A M is a Royal Academy of Engineering Research Fellow. This work was supported by the EPSRC (EP/N017188/1).

## 7. Nanoplasmonics

*Varun Mohan*[1] *and Prashant K Jain*[2,3,4,5]

[1] Department of Materials Science and Engineering, University of Illinois at Urbana-Champaign, Urbana, IL 61801, United States of America
[2] Department of Chemistry, University of Illinois at Urbana-Champaign, Urbana, IL 61801, United States of America
[3] Materials Research Laboratory, University of Illinois at Urbana-Champaign, Urbana, IL 61801, United States of America
[4] Department of Physics, University of Illinois at Urbana-Champaign, Urbana, IL 61801, United States of America
[5] Beckman Institute for Advanced Science and Technology, University of Illinois at Urbana-Champaign, Urbana, IL 61801, United States of America

*7.0.9. Status.* Plasmonics deals with coherent oscillations of loosely-bound electronic charge carriers (surface plasmons) on metal or semiconductor surfaces, usually initiated by a resonant external power source. Nano-plasmonics deals with this phenomenon on the nanometer scale: the coherent oscillations induced by electromagnetic excitation are localized to a nanoscale volume, which amounts to spatial confinement of electromagnetic energy below the diffraction limit of light. Localized surface plasmons (LSPs) enable spatiotemporal concentration, control, and exploitation of optical energy and the orders-of-magnitude amplification of incident electromagnetic fields, naturally lending LSPs to novel applications in communications, signal processing, nanoscale computation, light harvesting, biomedicine, and sensors (figure 24).

With the advent of massively parallel, distributed computing and the increasing volumes of data, information transfer and manipulation in the future will require operations at speeds approaching those of light. Replacement of electronics with photonics is touted to be the answer, but conventional photonics technology is not easily amenable to computation and data manipulation due to the non-interacting nature of photons. Computation would still rely on electronic components; but the components would be connected by fiber optics to allow information transfer at the superior optical bandwidths. However, such a scheme requires the interconversion of photons to-and-from electronic signals, which poses a bottleneck for information processing. A viable strategy for achieving enhanced operation speeds is to utilize intrinsically strong plasmon–plasmon interactions for information processing tasks. This would enable all-optical circuits consisting of nano-plasmonic computation elements [181, 183] connected by photonic buses (figure 25). Speed-limiting conversion of electronic signals to optical ones would no longer be required. In addition, the sub-diffraction localization of fields makes it viable to fabricate optical devices with nanoscale feature sizes on par with state-of-the-art transistor technology, something that is not possible using conventional photonics.

Nano-plasmonics is also naturally suited to quantum information processing. Strong coupling between quantum emitters and the plasmonic modes of nanocavities can be exploited as an efficient, room- temperature source of ultra-bright single photons for quantum circuits [185]. These sources would need neither the expensive cooling infrastructure of currently used photon sources, nor bulky resonator cavities for Purcell enhancement. The potential ability of plasmonic circuits to maintain quantum entanglement during the interconversion of entangled photons to plasmons and back to photons again further enables integrated nanophotonic architectures for cryptography, teleportation and quantum memory [186].

Solar-energy harvesting devices constitute another area where nanoplasmonics is demonstrating promise. While the utility of plasmonic nanostructures for the enhancement of light absorption and photocurrent in silicon photovoltaics has only been modest, nanoplasmonics is gaining new-found attention in photocatalytic conversion. Under visible-light excitation, hot carriers generated in plasmonic metal nanoparticles trigger challenging chemical reactions. Such plasmonic photocatalysis would facilitate the use of concentrated sunlight for eliminating energy-intensive heating required for thermally-activated reactions or for reducing the electrical power needed for driving redox processes. On a more exciting front, plasmonic photocatalysis may allow the harvesting of the free energy of light in the form of excited electron-hole pairs for driving uphill solar-to-fuel conversion [187]. Advances along this front would enable direct solar-to-fuel conversion devices that do not require electrical power.

*7.0.10. Current and future challenges.* The practical realization of proposed nanoplasmonic device functionality requires several challenges to be overcome, broadly categorized as follows:

**Non-ideality**. The performance of nanoplasmonic devices for most photonic and information processing applications relies on high plasmonic quality factors and strongly concentrated fields. Often, experimentally-realized quality and field enhancement factors are orders-of-magnitude smaller than those predicted for ideal architectures. This disparity originates from inadequate experimental control over sensitive factors (e.g. the sharpness of edges in field concentrators) along with our limited modeling capabilities for realistic architectures and complex features such as surface roughness. Moreover, due to the structural sensitivity of these attributes, there is often considerable variation from one fabricated unit to another. There are cases where strong confinement and high-quality plasmon modes have been attained by exquisitely fabricated architectures; but the challenge is to reproducibly achieve such near-ideal characteristics in higher-throughput manufacturing.

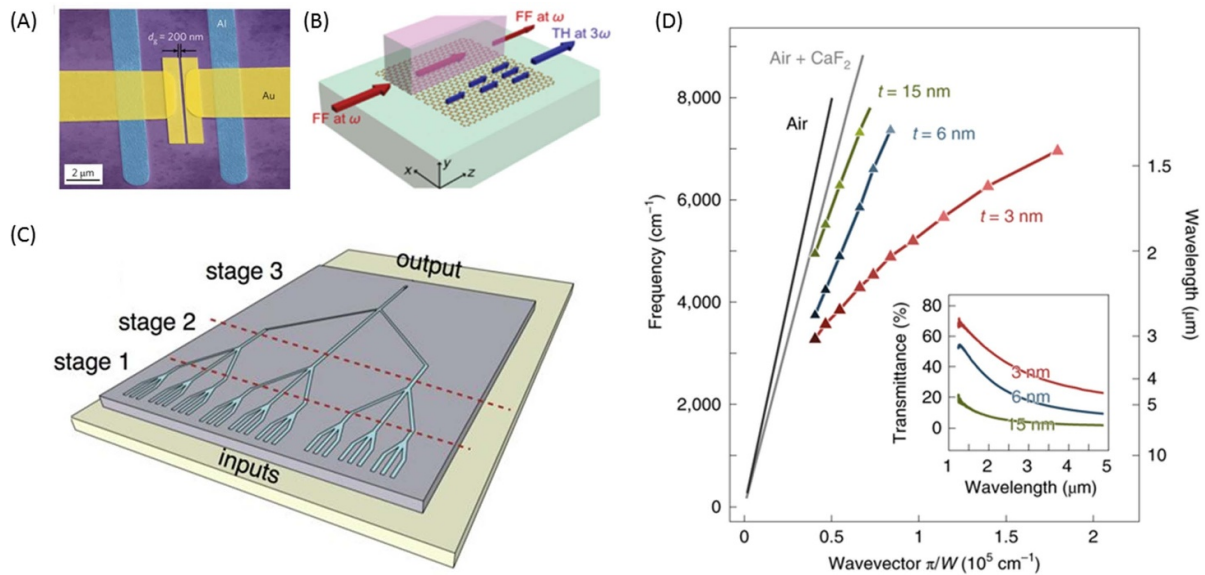

**Figure 24.** Examples of recent nano-plasmonic device designs and applications. Shown in panel (A) is an example of an electronic-plasmonic transducer made using Al-Au tunnel junctions. Reprinted with permission from [181]. Copyright 2017 Nature Publishing Group. This transducer uses inelastic electron scattering to generate surface plasmon polariton (SPP) modes in a coupled waveguide. In panel (B) is a proposed design for a graphene-based third-harmonic generation coupler. Reprinted with permission from [182]. Copyright 2019 American Physical Society. (C) shows a proposed design for a cascaded plasmonic majority gate that uses waveguides to mix travelling SPPs. The phase of the plasmon-polariton at the output waveguide of the majority gate is the variable representing the on/off state. Reprinted with permission from [183]. Copyright 2017 Nature Publishing Group. Panel (D) shows experimental dispersion curves and transmission spectra showing tunability in plasmons excited in ultra-thin metal films (UTMFs) of varying thicknesses. Reprinted with permission from [184]. Copyright 2019 Nature Publishing Group.

**Optical losses.** The most common plasmonic metals in vogue (Au, Cu, and Ag) suffer from high losses at optical frequencies and incompatibility with CMOS technology, which may preclude their use in industrial technologies [188]. Doped semiconductors exhibit lower losses at optical frequencies sufficiently far from the band-gap energy, but carrier concentrations are sub-par. In other words, the general challenge is that large carrier densities go hand-in-hand with high carrier damping rates, which prevents us from approaching idealized levels of plasmonic quality factors or field confinement. Alternative materials with sufficiently high and tunable carrier concentrations and simultaneously low losses and that are also compatible with current industrial manufacturing must be developed and characterized in as much detail as the coinage metals.

**Integrated function**: Often, device functionality relies on the coupling between a plasmonic mode and another optical attribute such as an exciton. Coupling is typically achieved by a hybrid structure (e.g. an assembly of a plasmonic nanoparticle and a QD), but such hybrid structures pose challenges in fabrication and reproducible performance and limit the extent of coupling. It would be desirable to expand the class of materials that integrate plasmonic properties with other optical attributes. As a specific example, the enhanced electric fields inherent in the tightly confined plasmonic modes of metals pave the way for the non-linear optical response of materials to be exploited in unprecedented ways. However, non-linear effects in the current generation of plasmonic materials are weak due to an insufficiently large dielectric response, despite the large field-enhancements achieved. A new generation of plasmonic materials with larger higher-order susceptibilities would fit the need for devices that rely on non-linear phenomena. As another example, for solar-to-fuel technologies, it would be desirable to have material that integrates strong plasmonic light absorption, low carrier relaxation rates, and a surface with the ability to selective catalyse a fuel-forming reaction.

**Mechanistic understanding.** In some cases, a plasmonic device functions as desired; but the mode of operation may either be poorly understood or different from theoretical expectations. For instance, it is debated whether the plasmonic catalysis of some chemical reactions is hot-electron-driven-chemistry or is it simply the manifestation of enhanced kinetics resulting from thermal dissipation of hot electron energy. Depending on each case, one or both or yet other effects may be responsible; but it is necessary to resolve quantitatively the contribution of each mechanism. Such mechanistic understanding is necessary for realistic assessment and techno-economic studies of the industrial potential of the proposed scheme.

*7.0.11. Advances in science and technology to meet challenges.* **New hosts for nanoplasmonic phenomena**: We must look beyond conventional plasmonic metals and explore emerging materials. The burgeoning field of 2D materials is well suited for nano-plasmonics: a 2D heterostructure of graphene and hBN allows confinement length scales two orders-of-magnitude smaller than the incident photon wavelength, along with low losses [189]. These heterostructures show excellent control and localization of the plasmons with wide wavelength-tunability. Non-linear optical effects are also enhanced in 2D heterostructures, which ought to be investigated, as they would enable discovery of optical phenomena and new class of devices. For instance, a recent report predicts third-harmonic generation in doped graphene at midinfrared and terahertz frequencies, which would find utility for signal processing [182]. Further avenues lie in dielectric engineering with doped 2D materials, which would allow further reduction in losses.

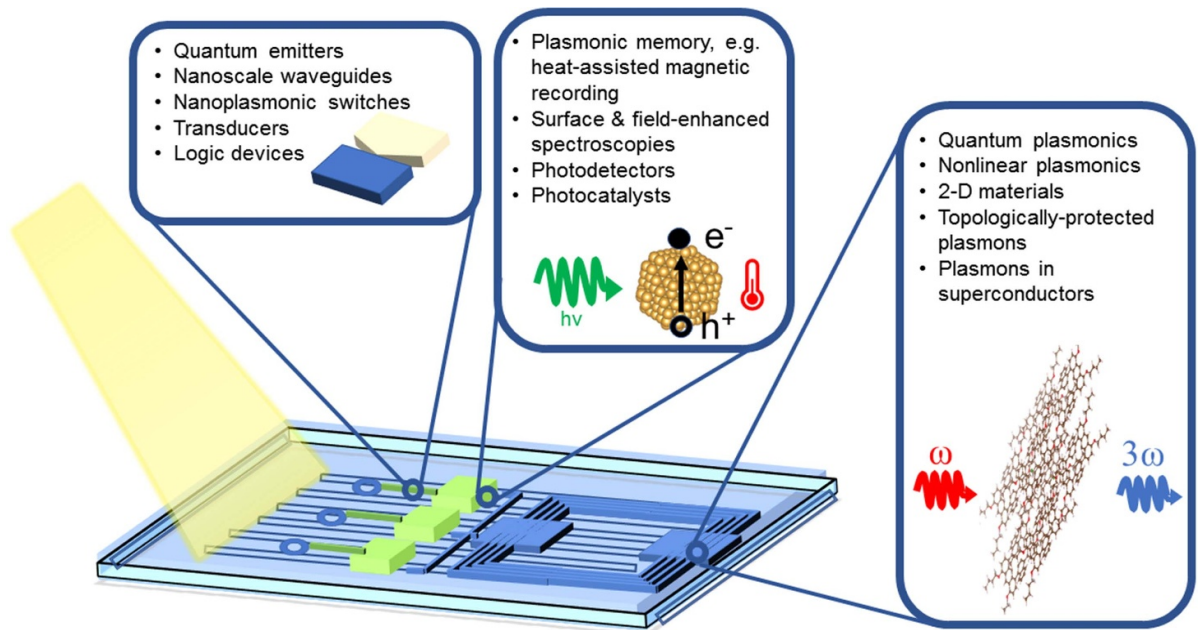

**Figure 25.** Schematic of an integrated, all-optical chip with nano-plasmonic components. Potential future applications of nanoplasmonics are also listed along with avenues for further research.

Attempts to improve the compatibility of plasmonic materials with silicon architectures must be continued. A recent report showed the presence of infrared plasmons in ultra-thin metal films (UTMFs) of Au grown on a $CaF_2$ substrate using a copper-seeded growth, which allowed unprecedented percolation thicknesses of less than 3 nm. The low thickness combined with the low overall sheet resistance made possible by a continuous thin film allowed a reduction in the surface carrier density. This enabled voltage-gated plasmon tunability [184]. On another front, titanium nitride—based materials have recently been shown to be promising due to their CMOS compatibility and a high degree of plasmon tunability [188]. Fabrication of UTMFs of these promising materials would be the next logical step to enable large scale industrial fabricability.

**Advances in quantitative near-field characterization**: While the radiative properties of plasmonic nanostructures are determined by far-field spectroscopy, the nature of nano-plasmonic modes can only be experimentally characterized by near-field probing with high spatial resolution. Further advances are needed in near-field characterization methods. While it has become more common to image the spatial profile of plasmonic fields of nanostructures, field intensities are not typically quantitatively measured. Field probing methods need to become more quantitative [190]. Near-field characteristics are also much more sensitive than far-field attributes to structural deviations from ideality making them valuable for understanding the influence of complex features and heterogeneities. If near-field characterization methods can be made higher-throughput in addition to being quantitative, then these techniques may be utilized for testing fabricated devices—determining whether the architecture meets the design specifications and assessing the nature and degree of heterogeneities that exist from one fabricated unit to another. There is one limitation that also needs attention. High-spatial-resolution probes used in near-field scanning optical microscopy, electron energy loss spectroscopy, and scanning tunneling microscopy likely perturb the electronic and/or electromagnetic response of the plasmonic nanostructure, as a result of which measured near-field characteristics may not be representative. The use of a metamaterial hyperlens substrate to project near-field evanescent waves into the far-field would allow non-perturbative probing and more accurate reconstruction of the near-field characteristics of nanoplasmonic structures and devices. Finally, near-field imaging ought to be advanced such that probing of devices becomes possible in-operando. Such capabilities would enable us to determine feedback responses in nanophotonic devices or decipher the mechanisms of action of plasmonic photocatalysts.

**Improved theoretical models**: With advances in understanding and fabrication methods, increasingly components and devices are being designed where function relies on progressively smaller length-scale features. In fact, we may approach extreme atom-scale field confinement. These advances necessitate all-quantum mechanical descriptions to be adopted for accurate description of nanoplasmonic properties and phenomena. Quantum mechanical effects are now routinely incorporated for gap-plasmons and plasmon resonances of ultrasmall nanoparticles, where semi-classical electrodynamics is not sufficient. Efforts must be made to develop multi-scale modelling methods, wherein nanoscale features and molecular components are treated quantum mechanically, intermeshed with a classical electrodynamics description of the larger scale structure. For systems involving strong coupling, quantum emission, and entanglement, fully QED-based theoretical descriptions of plasmonics at nanometer length scales need to be developed. In addition, *ab-initio* quantum dynamics would be needed for the treatment of phenomena involving structural and electronic dynamics, such as plasmonic photocatalysis, where metal surfaces, hot carriers, and adsorbed molecules interplay. Thus, the opportunities in theoretical plasmonics are as ripe as those in experimental and translational research.

*7.0.12. Concluding remarks.* Nanoplasmonic devices have been demonstrated for myriad applications; but, in general, these proof-of-concept demonstrations are isolated instances of exquisite architectures. The challenge is to translate these laboratory designs to large-scale, high-throughput, integrated chip-based manufacturing. We suggest areas for improvement and avenues for exploration, which would enable large-scale adoption of nano-plasmonics, especially for next-generation all-optical quantum communications and computing.

**Acknowledgments**

Funding for this work was provided by the Energy & Biosciences Institute (EBI) through the EBI-Shell program. V M and P K J co-wrote the manuscript.

## ORCID iDs

Arne Laucht https://orcid.org/0000-0001-7892-7963
Frank Hohls https://orcid.org/0000-0001-9654-3605
Søren Stobbe https://orcid.org/0000-0002-0991-041X
Tim Schröder https://orcid.org/0000-0001-9017-0254
Ferdinand Kuemmeth https://orcid.org/0000-0003-3675-7331
Joe Salfi https://orcid.org/0000-0001-9240-4245
Akira Oiwa https://orcid.org/0000-0001-5599-5824
Juha T Muhonen https://orcid.org/0000-0001-6520-6999
Dimitrie Culcer https://orcid.org/0000-0002-2342-0396
Attila Geresdi https://orcid.org/0000-0003-0123-2922
Prashant K Jain https://orcid.org/0000-0002-7306-3972
Jonathan Baugh https://orcid.org/0000-0002-9300-7134

## References

[1] Sander D et al 2017 J. Phys. D: Appl. Phys. **50** 363001
[2] Adamovich I et al 2017 J. Phys. D: Appl. Phys. **50** 323001
[3] Flensberg K, Odintsov A A, Liefrink F and Teunissen P 1999 Towards single-electron metrology Int. J. Mod. Phys. B **13** 2651
[4] Pekola J P, Saira O-P, Maisi V F, Kemppinen A, Möttönen M, Pashkin Y A and Averin D V 2013 Single-electron current sources: toward a refined definition of the ampere Rev. Mod. Phys. **85** 1421–72
[5] Kaestner B and Kashcheyevs V 2015 Non-adiabatic quantized charge pumping with tunable-barrier quantum dots: a review of current progress Rep. Prog. Phys. **78** 103901
[6] Yamahata G, Giblin S P, Kataoka M, Karasawa T and Fujiwara A 2017 High-accuracy current generation in the nanoampere regime from a silicon single-trap electron pump Sci. Rep. **7** 45137
[7] Giblin S, Fujiwara A, Yamahata G, Bae M-H, Kim N, Rossi A, Möttönen M and Kataoka M 2019 Evidence for universality of tunable-barrier electron pumps Metrologia **56** 044004
[8] Stein F et al 2016 Robustness of single-electron pumps at sub-ppm current accuracy level Metrologia **54** S1–S8
[9] Bureau International des Poids et Mesures 2019 The international system of units (SI) 9th edn (available at: www.bipm.org/)
[10] Devoret M H and Schoelkopf R J 2000 Amplifying quantum signals with the single-electron transistor Nature **406** 1039
[11] Wallraff A, Schuster D I, Blais A, Frunzio L, Huang R S, Majer J, Kumar S, Girvin S M and Schoelkopf R J 2004 Strong coupling of a single photon to a superconducting qubit using circuit quantum electrodynamics Nature **431** 162
[12] Colless J I, Mahoney A C, Hornibrook J M, Doherty A C, Lu H, Gossard A C and Reilly D J 2013 Dispersive readout of a few-electron double quantum dot with fast rf gate sensors Phys. Rev. Lett. **110** 046805
[13] Zheng G, Samkharadze N, Noordam M L, Kalhor N, Brousse D, Sammak A, Scappucci G and Vandersypen L M K 2019 Rapid gate-based spin read-out in silicon using an on-chip resonator Nat. Nanotechnol. **14** 742–6
[14] Karzig T et al 2017 Scalable designs for quasiparticle-poisoning-protected topological quantum computation with Majorana zero modes Phys. Rev. B **95** 235305
[15] Barthel C, Reilly D J, Marcus C M, Hanson M P and Gossard A C 2009 Rapid single-shot measurement of a singlet-triplet qubit Phys. Rev. Lett. **103** 160503
[16] Blais A, Huang R-S, Wallraff A, Girvin S M and Schoelkopf R J 2004 Cavity quantum electrodynamics for superconducting electrical circuits: an architecture for quantum computation Phys. Rev. A **69** 062320
[17] Ahmed I et al 2018 Radio-frequency capacitive gate-based sensing Phys. Rev. Appl. **10** 014018
[18] Reed M D, Johnson B R, Houck A A, DiCarlo L, Chow J M, Schuster D I, Frunzio L and Schoelkopf R J 2010 Fast reset and suppressing spontaneous emission of a superconducting qubit Appl. Phys. Lett. **96** 203110
[19] Yan F et al 2016 The flux qubit revisited to enhance coherence and reproducibility Nat. Commun. **7** 12964
[20] Didier N, Bourassa J and Blais A 2015 Fast quantum nondemolition readout by parametric modulation of longitudinal qubit-oscillator interaction Phys. Rev. Lett. **115** 203601
[21] Schaal S, Rossi A, Ciriano-Tejel V N, Yang T Y, Barraud S, Morton J J and Gonzalez-Zalba M F 2019 A CMOS dynamic random access architecture for radio-frequency readout of quantum devices Nat. Electron. **2** 236
[22] Flamini F, Spagnolo N and Sciarrino F 2018 Photonic quantum information processing: a review Rep. Prog. Phys. **82** 016001
[23] Lodahl P, Mahmoodian S and Stobbe S 2015 Interfacing single photons and single quantum dots with photonic nanostructures Rev. Mod. Phys. **87** 347
[24] Senellart P, Solomon G and White A 2017 High-performance semiconductor quantum-dot single-photon sources Nat. Nanotechnol. **12** 1026
[25] Salter C L, Stevenson R M, Farrer I, Nicoll C A, Ritchie D A and Shields A J 2010 An entangled-light-emitting diode Nature **465** 594
[26] Schwartz I, Cogan D, Schmidgall E R, Don Y, Gantz L, Kenneth O, Lindner N H and Gershoni D 2016 Deterministic generation of a cluster state of entangled photons Science **354** 434
[27] Wang H et al 2017 High-efficiency multiphoton boson sampling Nat. Photon. **11** 361
[28] Atatüre M, Englund D, Vamivakas N, Lee S-Y and Wrachtrup J 2018 Material platforms for spin-based photonic quantum technologies Nat. Rev. Mater. **3** 38
[29] Schröder T, Mouradian S L, Zheng J, Trusheim M E, Walsh M, Chen E H, Li L, Bayn I and Englund D 2016 Quantum nanophotonics in diamond [invited] J. Opt. Soc. Am. B **33** B65
[30] Lee K G, Chen X W, Eghlidi H, Kukura P, Lettow R, Renn A, Sandoghdar V and Götzinger S 2011 A planar dielectric antenna for directional single-photon emission and near-unity collection efficiency Nat. Photon. **5** 166
[31] Saleh B E A and Teich M C 1992 Photon-number-squeezed light Proc. IEEE **80** 451
[32] Gu X, Kockum A F, Miranowicz A, Liu Y-X and Nori F 2017 Microwave photonics with superconducting quantum circuits Phys. Rep. **718** 1–102
[33] Stockklauser A et al 2017 Strong coupling cavity QED with gate-defined double quantum dots enabled by a high impedance resonator Phys. Rev. X **7** 011030
[34] Mi X, Cady J V, Zajac D M, Deelman P W and Petta J R 2017 Strong coupling of a single electron in silicon to a microwave photon Science **335** 156–8
[35] Bruhat L E, Cubaynes T, Viennot J J, Dartiailh M C, Desjardins M M, Cottet A and Kontos T 2018 Circuit QED with a quantum-dot charge qubit dressed by Cooper pairs Phys. Rev. B **98** 155313
[36] Landig A J et al 2018 Coherent spin–photon coupling using a resonant exchange qubit Nature **560** 179–84

[37] Mi X, Benito M, Putz S, Zajac D M, Taylor J M, Burkard G and Petta J R 2018 A coherent spin-photon interface in silicon *Nature* **555** 599–603

[38] Samkharadze N, Zheng G, Kalhor N, Brousse D, Sammak A, Mendes U C, Blais A, Scappucci G and Vandersypen L M K 2018 Strong spin-photon coupling in silicon *Science* **359** 1123–7

[39] Scarlino P *et al* 2019 All-microwave control and dispersive readout of gate-defined quantum dot qubits in circuit quantum electrodynamics *Phys. Rev. Lett.* **122** 206802

[40] van Woerkom D J *et al* 2019 Microwave photon-mediated interactions between semiconductor qubits *Phys. Rev. X* **8** 041018

[41] Scarlino P *et al* 2019 Coherent microwave-photon-mediated coupling between a semiconductor and a superconducting qubit *Nat. Commun.* **10** 3011

[42] Landig A J *et al* 2020 Virtual-photon-mediated spin-qubit–transmon coupling *Nat. Commun.* **10** 5037

[43] Borjans F, Croot X G, Mi X, Gullans M J and Petta J R 2019 Resonant microwave-mediated interactions between distant electron spins *Nature* **577** 195–8

[44] Vandersypen L M K *et al* 2017 Interfacing spin qubits in quantum dots and donors—hot, dense, and coherent *Npj Quantum Inf.* **3** 34

[45] Nichol J M, Orona L A, Harvey S P, Fallahi S, Gardner G C, Manfra M J and Yacoby A 2017 High-fidelity entangling gate for double-quantum-dot spin qubits *Npj Quantum Inf.* **3** 3

[46] Volk C *et al* 2019 Loading a quantum-dot based 'Qubyte' register *Npj Quantum Inf.* **5** 29

[47] Kockum A F *et al* 2019 Ultrastrong coupling between light and matter *Nat. Rev. Phys.* **1.1** 19–40

[48] Thorgrimsson B *et al* 2017 Extending the coherence of a quantum dot hybrid qubit *Npj Quantum Inf.* **3** 32

[49] Malinowski F K, Martins F, Nissen P D, Barnes E, Rudner M S, Fallahi S, Gardner G C, Manfra M J, Marcus C M and Kuemmeth F 2016 Notch filtering the nuclear environment of a spin qubit *Nat. Nanotechnol.* **12** 16

[50] Cerfontaine P, Botzem T, Ritzmann J, Humpohl S S, Ludwig A, Schuh D, Bougeard D, Wieck A D and Bluhm H 2019 Closed-loop control of a GaAs-based singlet-triplet spin qubit with 99.5% gate fidelity and low leakage *Nat. Commun.* **11** 4144

[51] Nakajima T *et al* 2019 Quantum non-demolition measurement of an electron spin qubit *Nat. Nanotechnol.* **14** 555

[52] Teske J D, Humpohl S S, Otten R, Bethke P, Cerfontaine P, Dedden J, Ludwig A, Wieck A D and Bluhm H 2019 A machine learning approach for automated fine-tuning of semiconductor spin qubits *Appl. Phys. Lett.* **114** 133102

[53] Mortemousque P-A, Chanrion E, Jadot B, Flentje H, Ludwig A, Wieck A D, Urdampilleta M, Bauerle C and Meunier T 2018 Coherent control of individual electron spins in a two dimensional array of quantum dots arXiv:1808.06180

[54] Malinowski F K *et al* 2019 Fast spin exchange across a multielectron mediator *Nat. Commun.* **10** 1196

[55] Kandel Y P, Qiao H, Fallahi S, Gardner G C, Manfra M J and Nichol J M 2019 A Heisenberg spin teleport *Nature* **573** 553

[56] Takada S *et al* 2019 Sound-driven single-electron transfer in a circuit of coupled quantum rails *Nat. Commun.* **10** 4557

[57] Fujita T, Kiyama H, Morimoto K, Teraoka S, Allison G, Ludwig A, Wieck A D, Oiwa A and Tarucha S 2013 Nondestructive real-time measurement of charge and spin dynamics of photoelectrons in a double quantum dot *Phys. Rev. Lett.* **110** 266803

[58] Dehollain J P, Mukhopadhyay U, Michal V P, Wang Y, Wunsch B, Reichl C, Wegscheider W, Rudner M S, Demler E and Vandersypen L M K 2019 Nagaoka ferromagnetism observed in a quantum dot plaquette *Nature* **579** 528–33

[59] Sala A and Danon J 2017 Exchange-only singlet-only spin qubit *Phys. Rev. B* **95** 241303(R)

[60] Russ M, Petta J R and Burkard G 2018 Quadrupolar exchange-only spin qubit *Phys. Rev. Lett.* **121** 177701

[61] Malinowski F K, Martins F, Nissen P D, Fallahi S, Gardner G C, Manfra M J, Marcus C M and Kuemmeth F 2017 Symmetric operation of the resonant exchange qubit *Phys. Rev. B* **96** 045443

[62] Kim D, Kiselev A A, Ross R S, Rakher M T, Jones C and Ladd T D 2016 Optically loaded semiconductor quantum memory register *Phys. Rev. Appl.* **5** 024014

[63] Maune B M *et al* 2012 Coherent singlet-triplet oscillations in a silicon-based double quantum dot *Nature* **481** 344–7

[64] Veldhorst M *et al* 2014 An addressable quantum dot qubit with fault-tolerant control-fidelity *Nat. Nanotechnol.* **9** 981–5

[65] Maurand R *et al* 2016 A CMOS silicon spin qubit *Nat. Commun.* **7** 13575

[66] Yoneda J *et al* 2018 A quantum-dot spin qubit with coherence limited by charge noise and fidelity higher than 99.9% *Nat. Nanotechnol.* **13** 102–6

[67] Reed M D *et al* 2016 Reduced sensitivity to charge noise in semiconductor spin qubits via symmetric operation *Phys. Rev. Lett.* **116** 110402

[68] Huang W *et al* 2019 Fidelity benchmarks for two-qubit gates in silicon *Nature* **569** 532–6

[69] Xue X *et al* 2019 Benchmarking Gate fidelities in a Si/SiGe two-qubit device *Phys. Rev. X* **9** 021011

[70] Eng K *et al* 2015 Isotopically enhanced triple-quantum-dot qubit *Sci. Adv.* **1** e1500214

[71] Zajac D M, Hazard T M, Mi X, Nielsen E and Petta J R 2016 Scalable gate architecture for a one-dimensional array of semiconductor spin qubits *Phys. Rev. Appl.* **6** 054013

[72] Betz A C, Wacquez R, Vinet M, Jehl X, Saraiva A L, Sanquer M, Ferguson A J and Gonzalez-Zalba M F 2015 Dispersively detected Pauli spin-blockade in a silicon nanowire field-effect transistor *Nano Lett.* **15** 4622–7

[73] Veldhorst M, Eenink H G J, Yang C H and Dzurak A S 2017 Silicon CMOS architecture for a spin-based quantum computer *Nat. Commun.* **8** 1766

[74] Yang C H *et al* 2019 Silicon qubit fidelities approaching incoherent noise limits via pulse engineering *Nat. Electron.* **2** 151–8

[75] Kane B E 1998 A silicon-based nuclear spin quantum computer *Nature* **393** 133–7

[76] Steger M, Saeedi K, Thewalt M L W, Morton J J L, Riemann H, Abrosimov N V, Becker P and Pohl H-J 2012 Quantum information storage for over 180 s using donor spins in a 28Si 'semiconductor vacuum' *Science* **336** 1280–3

[77] Morello A *et al* 2010 Single-shot readout of an electron spin in silicon *Nature* **467** 687–91

[78] Fuechsle M, Miwa J A, Mahapatra S, Ryu H, Lee S, Warschkow O, Hollenberg L C L, Klimeck G and Simmons M Y 2012 A single-atom transistor *Nat. Nanotechnol.* **7** 242–6

[79] Pla J J, Tan K Y, Dehollain J P, Lim W H, Morton J J L, Zwanenburg F A, Jamieson D N, Dzurak A S and Morello A 2013 High-fidelity readout and control of a nuclear spin qubit in silicon *Nature* **496** 334–8

[80] Muhonen J T *et al* 2014 Storing quantum information for 30 seconds in a nanoelectronic device *Nat. Nanotechnol.* **9** 986–91

[81] Broome M A *et al* 2018 Two-electron spin correlations in precision placed donors in silicon *Nat. Commun.* **9** 980

[82] Usman M, Bocquel J, Salfi J, Voisin B, Tankasala A, Rahman R, Simmons M Y, Rogge S and Hollenberg L C L 2016 Spatial metrology of dopants in silicon with exact lattice site precision *Nat. Nanotechnol.* **11** 763

[83] Hill C D, Peretz E, Hile S J, House M G, Fuechsle M, Rogge S, Simmons M Y and Hollenberg L C 2015 A surface code quantum computer in silicon *Sci. Adv.* **1** e1500707

[84] Tosi G, Mohiyaddin F A, Schmitt V, Tenberg S, Rahman R, Klimeck G and Morello A 2017 Silicon quantum processor with robust long-distance qubit couplings *Nat. Commun.* **8** 450

[85] Pica G, Lovett B W, Bhatt R N, Schenkel T and Lyon S A 2016 Surface code architecture for donors and dots in silicon with imprecise and nonuniform qubit couplings *Phys. Rev.* B **93** 035306, 01

[86] O'Gorman J, Nickerson N H, Ross P, Morton J J L and Benjamin S C 2016 A silicon-based surface code quantum computer *Npj Quantum Inf.* **2** 15019

[87] Ruskov R and Tahan C 2013 On-chip cavity quantum phonodynamics with an acceptor qubit in silicon *Phys. Rev.* B **88** 064308

[88] Salfi J, Mol J, Culcer D and Rogge S 2016 Charge-insensitive single-atom spin-orbit qubit in silicon *Phys. Rev. Lett.* **116** 246801

[89] Abadillo-Uriel J C and Calderon M J 2017 Spin qubit manipulation of acceptor bound states in group IV quantum wells *New J. Phys.* **19** 043027

[90] Golding B and Dykman M I 2003 Acceptor-based silicon quantum computing (arXiv:cond-mat/0309147)

[91] van der Heijden J, Salfi J, Mol J A, Verduijn J, Tettamanzi G C, Hamilton A R, Collaert N and Rogge S 2014 Probing the spin states of a single acceptor atom *Nano Lett.* **14** 1492–6

[92] van der Heijden J, Kobayashi T, House M G, Salfi J, Barraud S, Lavieville R, Simmons M Y and Rogge S 2018 Readout and control of the spin-orbit states of two coupled atoms in a silicon transistor *Sci. Adv.* **4** eaat9199

[93] Stegner A R, Tezuka H, Andlauer T, Stutzmann M, Thewalt M L W, Brandt M S and Itoh K M 2010 Isotope effect on electron paramagnetic resonance of boron acceptors in silicon *Phys. Rev.* B **82** 115213

[94] Kobayashi T *et al* 2020 Engineering long spin coherence times of spin–orbit qubits in silicon *Nat. Mater.* http://doi.org/10.1038/s41563-020-0743-3

[95] Salfi J, Mol J, Rahman R, Klimeck G, Simmons M Y, Hollenberg L C L and Rogge S 2016 Quantum simulation of the Hubbard model with dopant atoms in silicon *Nat. Commun.* **7** 11342

[96] Shearrow A, Koolstra G, Whiteley S J, Earnest N, Barry P S, Heremans F J, Awschalom D D, Shirokoff E and Schuster D I 2018 Atomic layer deposition of titanium nitride for quantum circuits *Appl. Phys. Lett.* **113** 212601

[97] Wehner S, Elkouss D and Hanson R 2018 Quantum internet: a vision for the road ahead *Science* **362** eaam9228

[98] Herbschleb E D *et al* 2019 Ultra-long coherence times amongst room-temperature solid-state spins *Nat. Commun.* **10** 3766

[99] Dobrovitski V V, Fuchs G D, Falk A L, Santori C and Awschalom D D 2013 Quantum control over single spins in diamond *Annu. Rev. Condens. Matter Phys.* **4** 23–50

[100] Hensen B *et al* 2015 Loophole-free Bell inequality violation using electron spins separated by 1.3 kilometres *Nature* **526** 682–6

[101] Vrijen R and Yablonovitch E 2001 A spin-coherent semiconductor photo-detector for quantum communication *Physica* E **10** 569–75

[102] Muralidharan S, Li L, Kim J, Lutkenhaus N, Lukin M D and Jiang L 2016 Optimal architectures for long distance quantum communication *Sci. Rep.* **6** 20453

[103] Oiwa A, Fujita T, Kiyama H, Allison G, Ludwig A, Wieck A D and Tarucha S 2017 Conversion from single photon to single electron spin using electrically controllable quantum dots *J. Phys. Soc. Japan* **86** 011008

[104] Gaudreau L, Bogan A, Korkusinski M, Studenikin S, Austing D G and Sachrajda A S 2017 Entanglement distribution schemes employing coherent photon-to-spin conversion in semiconductor quantum dot circuits *Semicond. Sci. Technol.* **32** 093001

[105] Fujita T, Morimoto K, Kiyama H, Allison G, Larsson M, Ludwig A, Valentin S R, Wieck A D, Oiwa A and Tarucha S 2019 Angular momentum transfer from photon polarization to an electron spin in a gate-defined quantum dot *Nat. Commun.* **10** 2991

[106] Kuroyama K *et al* 2019 Photogeneration of a single electron from a, single Zeeman-resolved light-hole exciton with preserved angular momentum *Phys. Rev.* B **99** 085203-1–085203-5

[107] Kuroyama K, Larsson M, Matsuo S, Fujita T, Valentin S R, Ludwig A, Wieck A D, Oiwa A and Tarucha S 2017 Single electron-photon pair creation from a single polarization-entangled photon pair *Sci. Rep.* **7** 16968

[108] Tajiri T *et al* 2020 Fabrication and optical characterization of photonic crystal nanocavities with electrodes for gate-defined quantum dots *Jpn. J. Appl. Phys.* **59** SGGI05

[109] Aspelmeyer M, Kippenberg T J and Marquardt F 2014 Cavity optomechanics *Rev. Mod. Phys.* **86** 1391

[110] Midolo L, Schliesser A and Fiore A 2018 Nano-opto-electromechanical systems *Nat. Nanotechnol.* **13** 11

[111] Rossi M, Mason D, Chen J, Tsaturyan Y and Schliesser A 2018 Measurement-based quantum control of mechanical motion *Nature* **563** 53

[112] MacCabe G S, Ren H, Luo J, Cohen J D, Zhou H, Sipahigil A, Mirhosseini M and Painter O 2020 Nano-acoustic resonator with ultralong phonon lifetime *Science* **370** 840

[113] Higginbotham A P, Burns P S, Urmey M D, Peterson R W, Kampel N S, Brubaker B M, Smith G, Lehnert K W and Regal C A 2018 Harnessing electro-optic correlations in an efficient mechanical converter *Nat. Phys.* **14** 1038

[114] Balram K C, Davanço M I, Song J D and Srinivasan K 2016 Coherent coupling between radiofrequency, optical and acoustic waves in piezo-optomechanical circuits *Nat. Photon.* **10** 346

[115] Verhagen E and Alù A 2017 Optomechanical nonreciprocity *Nat. Phys.* **13** 922

[116] Chu Y, Kharel P, Renninger W H, Burkhart L D, Frunzio L, Rakich P T and Schoelkopf R J 2017 Quantum acoustics with superconducting qubits *Science* **358** 199–202

[117] Satzinger K J *et al* 2018 Quantum control of surface acoustic-wave phonons *Nature* **563** 661

[118] Lee D, Lee K W, Cady J V, Ovartchaiyapong P and Beszynski Jayich A C 2017 Topical review: spins and mechanics in diamond *J. Opt.* **19** 033001

[119] Flühmann C, Nguyen T L, Marinelli M, Negnevitsky V, Mehta K and Home J P 2019 Encoding a qubit in a trapped-ion mechanical oscillator *Nature* **566** 513–7

[120] Roukes M L 2001 Nanoelectromechanical systems *Transducers '01 Eurosensors XV* pp 658–61

[121] Schwab K C, Henriksen E A, Worlock J M and Roukes M L 2000 Measurement of the quantum of thermal conductance *Nature* **404** 974–7

[122] Schwab K C and Roukes M L 2005 Putting mechanics into quantum mechanics *Phys. Today* **58** 36–42

[123] Aspelmeyer M, Meystre P and Schwab K 2012 Quantum optomechanics *Phys. Today* **65** 29–35

[124] Rouxinol F, Hao Y, Brito F, Caldeira A O, Irish E K and LaHaye M D 2016 Measurements of nanoresonator-qubit

interactions in a hybrid quantum electromechanical system *Nanotechnology* **27** 3642003

[125] Ridinger R, Wallucks A, Marinkovic I, Loschnauer C, Aspelmeyer M, Hong S and Groblacher S 2018 Remote quantum entanglement between two micromechanical oscillators *Nature* **556** 473–7

[126] Kalaee M, Mirhosseini M, Dieterle P B, Peruzzo M, Fink J M and Painter O 2019 Quantum electromechanics of a hypersonic crystal *Nat. Nanotechnol.* **14** 334–9

[127] Noguchi A, Yamazkaki Y, Ataka M, Fujita H, Tabuchi Y, Ishikawa T, Usami K and Nakamura Y 2016 Ground state cooling of a quantum electromechanical system with silicon nitride membrane in a 3D loop-gap cavity *New J. Phys.* **18** 103036

[128] Rosenberg D *et al* 2017 3D integrated superconducting qubits *Npj Quantum Inf.* **3** 42

[129] Wang H *et al* 2019 Mode structure in superconducting metamaterial transmission-line resonators *Phys. Rev. Appl.* **11** 054062

[130] Han X *et al* 2020 Cavity piezo-mechanics for superconducting-nanophotonic quantum interface *Nat. Commun.* **11** 3237

[131] Saito Y, Nojima T and Iwasa Y 2016 Highly crystalline 2D superconductors *Nat. Rev. Mater.* **2** 16094

[132] Gong C and Zhang X 2019 Two-dimensional magnetic crystals and emergent heterostructure devices *Science* **363** 706

[133] Lee J-U, Lee S, Ryoo J H, Kang S, Kim T Y, Kim P, Park C-H, Park J-G and Cheong H 2016 Ising-type magnetic ordering in atomically thin FePS$_3$ *Nano Lett.* **16** 7433–8

[134] Kim K, Lim S Y, Lee J-U, Lee S, Kim T Y, Park K, Jeon G S, Park C-H, Park J-G and Cheong H 2019 Suppression of magnetic ordering in XXZ-type antiferromagnetic monolayer NiPS$_3$ *Nat. Commun.* **10** 345

[135] Gong C *et al* 2017 Discovery of intrinsic ferromagnetism in two-dimensional van der Waals crystals *Nature* **546** 265–9

[136] Huang B *et al* 2017 Layer-dependent ferromagnetism in a van der Waals crystal down to the monolayer limit *Nature* **546** 270–3

[137] Ghazaryan D *et al* 2018 Magnon-assisted tunnelling in van der Waals heterostructures based on CrBr$_3$ *Nat. Electron.* **1** 344–9

[138] Kim H H *et al* 2019 Evolution of interlayer and intralayer magnetism in three atomically thin chromium trihalides *Proc. Natl Acad. Sci. USA* **116** 11131–6

[139] Klein D R *et al* 2019 Enhancement of interlayer exchange in an ultrathin two-dimensional magnet *Nat. Phys.* **15** 1255–60

[140] Cai X *et al* 2019 Atomically thin CrCl$_3$: an in-plane layered antiferromagnetic insulator *Nano Lett.* **19** 3993–8

[141] Deng Y, Yu Y, Song Y, Zhang J, Wang N Z, Wu Y Z, Zhu J, Wang J, Chen X H and Zhang Y 2018 Gate-tunable room-temperature ferromagnetism in two-dimensional Fe$_3$GeTe$_2$ *Nature* **563** 94–99

[142] Bonilla M, Kolekar S, Ma Y, Diaz H C, Kalappattil V, Das R, Eggers T, Gutierrez H R, Phan M-H and Batzill M 2018 Strong room-temperature ferromagnetism in VSe$_2$ monolayers on van der Waals substrates *Nat. Nanotechnol.* **13** 289–93

[143] Mermin N D and Wagner H 1966 Absence of ferromagnetism or antiferromagnetism in one- or two-dimensional isotropic Heisenberg models *Phys. Rev. Lett.* **17** 1133–6

[144] Sivadas N, Okamoto S, Xu X D, Fennie C J and Xiao D 2018 Stacking-dependent magnetism in bilayer CrI$_3$ *Nano Lett.* **18** 7658–64

[145] Wang Z, Gutiérrez-Lezama I, Ubrig N, Kroner M, Taniguchi T, Watanabe K, Imamoğlu A, Giannini E and Morpurgo A F 2018 Very large tunneling magnetoresistance in layered magnetic semiconductor CrI$_3$ *Nat. Commun.* **9** 2516

[146] Song T *et al* 2018 Giant tunneling magnetoresistance in spin-filter van der Waals heterostructures *Science* **360** 1214–8

[147] Klein D R *et al* 2018 Probing magnetism in 2D van der Waals crystalline insulators via electron tunneling *Science* **360** 1218–22

[148] Kim H H, Yang B, Patel T, Sfigakis F, Li C, Tian S, Lei H and Tsen A W 2018 One million percent tunnel magnetoresistance in a magnetic van der Waals heterostructure *Nano Lett.* **18** 4885–90

[149] Kim H H, Yang B, Tian S, Li C, Miao G-X, Lei H and Tsen A W 2019 Tailored tunnel magnetoresistance response in three ultrathin chromium trihalides *Nano Lett.* **19** 5739–45

[150] Miao G-X, Müller M and Moodera J S 2009 Magnetoresistance in double spin filter tunnel junctions with nonmagnetic electrodes and its unconventional bias dependence *Phys. Rev. Lett.* **102** 076601

[151] Jiang S, Li L, Wang Z, Mak K F and Shan J 2018 Controlling magnetism in 2D CrI$_3$ by electrostatic doping *Nat. Nanotechnol.* **13** 549–53

[152] Huang B *et al* 2018 Electrical control of 2D magnetism in bilayer CrI$_3$ *Nat. Nanotechnol.* **13** 544–8

[153] Jiang S, Shan J and Mak K F 2018 Electric-field switching of two-dimensional van der Waals magnets *Nat. Mater.* **17** 406–10

[154] Jiang S, Li L, Wang Z, Shan J and Mak K F 2019 Spin tunnel field-effect transistors based on two-dimensional van der Waals heterostructures *Nat. Electron.* **2** 159–63

[155] Song T *et al* 2019 Voltage control of a van der Waals spin-filter magnetic tunnel junction *Nano Lett.* **19** 915–20

[156] You J-Y, Zhang Z, Gu B and Su G 2019 Two-dimensional room temperature ferromagnetic semiconductors with quantum anomalous Hall effect *Phys. Rev. Appl.* **12** 024063

[157] Fuh H-R, Chang C-R, Wang Y-K, Evans R F L, Chantrell R W and Jeng H-T 2016 Newtype single-layer magnetic semiconductor in transition-metal dichalcogenides VX$_2$ (X = S, Se and Te) *Sci. Rep.* **6** 32625

[158] Wang Y 2017 Room temperature magnetization switching in topological insulator-ferromagnet heterostructures by spin-orbit torques *Nat. Commun.* **8** 1364

[159] Han J *et al* 2017 Room-temperature spin-orbit torque switching induced by a topological insulator *Phys. Rev. Lett.* **119** 077702

[160] Collins J L *et al* 2018 *Nature* **564** 390

[161] Ma Q *et al* 2019 Observation of the nonlinear Hall effect under time-reversal-symmetric conditions *Nature* **565** 337

[162] Kang K *et al* 2019 Nonlinear anomalous Hall effect in few-layer WTe$_2$ *Nat. Mater.* **18** 324

[163] Nayak C, Simon S H, Stern A, Freedman M and Das Sarma S 2008 *Rev. Mod. Phys.* **80** 1083

[164] Lutchyn R, Sau J D and Das Sarma S 2010 *Phys. Rev. Lett.* **105** 077001

[165] Oreg Y *et al* 2010 Helical liquids and Majorana bound states in quantum wires *Phys. Rev. Lett.* **105** 177002

[166] Mourik V, Zuo K, Frolov S M, Plissard S R, Bakkers E P A M and Kouwenhoven L P 2012 *Science* **336** 1003

[167] Laroche D *et al* 2019 Observation of the 4Π-periodic Josephson effect in indium arsenide nanowires *Nat. Commun.* **10** 245

[168] Vuik A *et al* 2019 Reproducing topological properties with quasi-Majorana states *SciPost Phys.* **7** 061

[169] Avila J *et al* 2019 Communications Physics **2** 133

[170] Qi S, Qiao Z, Deng X, Cubuk E D, Chen H, Zhu W, Kaxiras E, Zhang S, Xu X and Zhang Z 2016 High-temperature

quantum anomalous Hall effect in n-p codoped topological insulators *Phys. Rev. Lett.* **117** 056804
[171] Plugge S, Rasmussen A, Egger R and Flensberg K 2017 Majorana box qubits *New J. Phys.* **19** 012001
[172] Aviram A and Ratner M A 1974 Molecular rectifiers *Chem. Phys. Lett.* **29** 277–83
[173] Evers F, Korytár R, Tewari S and van Ruitenbeek J M 2020 Advances and challenges in single-molecule electron transport *Rev. Mod. Phys.* **92** 035001
[174] Gehring P, Thijssen J M and van der Zant H S J 2019 Single-molecule quantum-transport phenomena in break junctions *Nat. Rev. Phys.* **1** 381–96
[175] Heerema S J and Dekker C 2016 Graphene nanodevices for DNA sequencing *Nat. Nanotechnol.* **11** 127–36
[176] Limburg B *et al* 2018 Anchor groups for graphene-porphyrin single-molecule transistors *Adv. Funct. Mater.* **28** 1803629
[177] Lörtscher E 2013 Wiring molecules into circuits *Nat. Nanotechnol.* **8** 381–4
[178] Perrin M L, Burzurí E and van der Zant H S J 2015 Single-molecule transistors *Chem. Soc. Rev.* **44** 902–19
[179] Kalinin S V, Borisevich A and Jesse S 2016 Fire up the atom forge *Nature* **539** 485–7
[180] Zhu J, McMorrow J, Crespo-Otero R, Ao G, Zheng M, Gillin W P and Palma M 2016 Solution-processable carbon nanoelectrodes for single-molecule investigations *J. Am. Chem. Soc.* **138** 2905–8
[181] Du W, Wang T, Chu H S and Nijhuis C A 2017 Highly efficient on-chip direct electronic-plasmonic transducers *Nat. Photon.* **11** 623–7
[182] Tingting W, Yu L, Maier S A and Lei W 2019 Phase-matching and peak nonlinearity enhanced third-harmonic generation in graphene plasmonic coupler *Phys. Rev. Appl.* **11** 014049
[183] Dutta S, Zografos O, Gurunarayanan S, Radu I, Soree B, Catthoor F and Naeemi A 2017 Proposal for nanoscale cascaded plasmonic majority gates for non-Boolean computation *Sci. Rep.* **7** 1–10
[184] Maniyara R A, Rodrigo D, Yu R, Canet-Ferrer J, Ghosh D S, Yongsunthon R, Baker D E, Rezikyan A, Abajo F J G D and Pruneri V 2019 Tunable plasmons in ultrathin metal films *Nat. Photon.* **13** 328–33
[185] Bogdanov S I *et al* 2018 Ultrabright room-temperature sub-nanosecond emission from single nitrogen-vacancy centers coupled to nanopatch antennas *Nano Lett.* **18** 4837–44
[186] Fasel S, Robin F, Moreno E, Erni D, Gisin N and Zbinden H 2005 Energy-time entanglement preservation in plasmon-assisted light transmission *Phys. Rev. Lett.* **94** 149901
[187] Yu S and Jain P K 2019 Plasmonic photosynthesis of $C_1$–$C_3$ hydrocarbons from carbon dioxide assisted by an ionic liquid *Nat. Commun.* **10** 1–7
[188] Naik G V, Shalaev V M and Boltasseva A 2013 Alternative plasmonic materials: beyond gold and silver *Adv. Mater.* **25** 3264–94
[189] Woessner A *et al* 2015 Highly confined low-loss plasmons in graphene–boron nitride heterostructures *Nat. Mater.* **14** 421–5
[190] Deeb C *et al* 2010 Quantitative analysis of localized surface plasmons based on molecular probing *ACS Nano* **4** 4579–86